\documentclass[twocolumn,rmp,showpacs,floatfix,aps]{revtex4}

\usepackage{epsfig}
\usepackage{amsmath}
\usepackage{longtable}

\newcommand{\ea}{{\em et al.}}

\newcommand{\ti}[1]{``{#1},''}

\begin{document}

\parbox{7in}{\hfill{TRI-PP-02-08}} 

\title
{Induced pseudoscalar coupling \\
of the proton weak interaction.}

\author
{Tim Gorringe \footnote{Electronic address: gorringe@pa.uky.edu}}
\affiliation
{Dept. of Physics and Astronomy,
Univ. of Kentucky,
Lexington, 
KY, 40506, U.~S.~A.}

\author
{Harold W. Fearing \footnote{Electronic address: fearing@triumf.ca}}
\affiliation
{TRIUMF, 
Vancouver, 
British Columbia, V6T 2A3, Canada}

\begin{abstract}
The induced pseudoscalar coupling $g_p$ is the least well known of the
weak coupling constants of the proton's charged--current interaction.
Its size is dictated by chiral symmetry arguments, and its measurement
represents an important test of quantum chromodynamics at low
energies.  During the past decade a large body of new data relevant to
the coupling $g_p$ has been accumulated. This data includes 
measurements of radiative and non radiative muon capture on targets
ranging from hydrogen and few--nucleon systems to complex nuclei.  Herein
the authors review the theoretical underpinnings of $g_p$, the
experimental studies of $g_p$, and the procedures and uncertainties in
extracting the coupling from data.  Current puzzles are highlighted
and future opportunities are discussed.
\end{abstract}

\pacs{23.40.-s, 11.40.Ha, 13.10.+q, 11.30.Rd, 25.30.-c, 11.40.-q}

\maketitle
\tableofcontents

\newpage


\section{Introduction}
\label{s: general intro}

It is well known that the weak interactions of leptons are governed by
a current-current interaction, where the currents are given by a
simple $V-A$ form, \makebox{$\gamma_\mu (1-\gamma_5)$}. When hadrons
are involved, the interaction is still current-current and still $V-A$
but the individual vector and axial vector currents become more
complicated, picking up both form factors and new structures involving
the momentum transfer $q$. The couplings associated with these new
structures are the so called `induced' couplings which have been a
topic of investigation, both theoretical and experimental, for a long
time. We are concerned in this review with one of these, the induced
pseudoscalar coupling, $g_p$.

The most general weak current, actually the matrix element of that
current, for say a neutron and proton can be written as
\begin{eqnarray}
\label{e: general current}
\overline{u}_n(&+&g_v \gamma^\mu+\frac{i g_m}{2 m_N}\sigma^{\mu\nu}
q_{\nu}+\frac{g_s}{m_\mu} q^\mu \nonumber \\
&-&g_a \gamma^\mu
\gamma_5-\frac{g_p}{m_\mu}q^\mu \gamma_5-\frac{i g_t}{2
m_N}\sigma^{\mu\nu} q_\nu \gamma_5 ) u_p
\end{eqnarray}
where the notation for gamma matrices, spinors, etc.\ is that of
\citet{Bj64} and $m_\mu$ and $m_N$ are respectively the
muon and nucleon masses.  The momentum transfer $q=p_n-p_p$, with
$p_p$, $p_n$ respectively the proton and neutron momenta. This most
general form for the current contains six coupling `constants', which
are actually functions of $q^2$, namely $g_v$ and $g_a$, the vector
and axial vector couplings, $g_m$, the weak magnetism coupling, $g_p$,
the induced pseudoscalar coupling, and $g_s$ and $g_t$, the second
class induced scalar and induced tensor couplings.\footnote{There is
an unfortunate confusion about the signs of the axial couplings which
has arisen as conventions have changed over the years. In very early
calculations in a different metric, the signs were such that $g_a$ was
positive. With the widespread use of the metric of \citet{Bj64} it
became conventional to write the weak current like Eq.~(\ref{e:
general current}) only with all signs positive. This was the
convention used in \citet{Fe80, Be87, Be89} and most other modern
papers, and implies that $g_a$ and $g_p$ are negative. With the advent
of chiral perturbation theory, for which $g_a$ is normally taken to be
positive, the convention changed again. We have adopted this latter
convention. This means that the axial current of Eq.~(\ref{e: general
current}) has an explicit overall minus sign and that $g_a$ and $g_p$
are positive numbers.}

Of these six, the second class terms $g_s$ and $g_t$ transform
differently than the others under G-parity, which is a combination of
charge conjugation and a rotation in isospin space. In the standard
model they are generated only via quark mass differences or
electromagnetic effects and are thus predicted to be small
\citep{Gr85, Sh96}. Thus although the experimental evidence is not
unequivocal \citep{Wi00a, Wi00b, Mi01}, they are normally assumed to
be zero and will be ignored in this review. The vector terms $g_v$ and
$g_m$ are related by the Conserved Vector Current (CVC) theory to the
isovector electromagnetic current. Thus their static values are
related to the charge and anomalous magnetic moments of the nucleons
and their $q^2$ dependence is given by measurements of the isovector
electromagnetic form factors of the proton and neutron. The axial
vector coupling $g_a$ can be determined precisely from neutron beta
decay and its $q^2$ dependence from neutrino scattering \citep{Ah88}
or from pion electroproduction
\citep{De76, Es78, Ch93, Be92}.

This leaves the induced pseudoscalar coupling, $g_p$, which is by far
the least well known.  In principle it can be predicted using the
Partially Conserved Axial Current (PCAC) theory via a relation known
as the Goldberger-Treiman relation \citep{Go58}. The details of this
derivation will be discussed in the next section. Physically the
dominant diagram is one in which the nucleon emits a pion which
propagates and then couples to the $\mu-\nu$ or $e-\nu$ vertex. The
coupling $g_p$ thus contains a pion pole. In practice however it has
been difficult to verify this relation and there are some situations,
notably radiative muon capture in liquid hydrogen, where experiment
and theory do not agree.

Thus the main purpose of this review is to discuss in detail the
information available, both theoretical and experimental, regarding
the value of $g_p$. In Sec.~\ref{s: gptheory} we discuss the
theoretical prediction for $g_p$ arising from PCAC and from the more
modern approaches using chiral perturbation theory. Sections \ref{s:
sources} and \ref{s: muchem} describe in general some of the sources
of experimental information and some of the atomic and molecular
physics required to understand muon processes in hydrogen or
deuterium. The remainder of the review is divided into three main
parts which discuss the three main areas which have led to information
on $g_p$. The first of these, Secs.~\ref{s: mcproton}-\ref{s:
otherfb}, deals with ordinary, non radiative, muon capture (OMC) and
radiative muon capture (RMC) on the proton, deuteron, and $^3$He. Here
the experiments are difficult, but the results presumedly are not
obscured by unknown details of nuclear structure. In the next major
part, Sec.~\ref{s: exclnuclOMC}, exclusive OMC on complex nuclei is
discussed, with particular emphasis on spin observables in partial
transitions.  Here the sensitivity to $g_p$ is high, but the
experiments are often difficult and their interpretation is model
dependent.  Then in the final part, Sec.~\ref{s: nuclRMC}, inclusive
RMC on complex nuclei is examined. RMC is quite sensitive to $g_p$ and
nowadays measurements of inclusive RMC on $Z>2$ nuclei are relatively
straightforward. However model dependence in the theory makes the
extraction of a unique value of $g_p$ very difficult. Finally in
Sec.~\ref{s: summary} we will summarize the current situation with
regard to $g_p$ and make some suggestions for further work.

Related review articles include those of
\citet{Mu77}, \citet{Gr85}, \citet{Gm87}, \citet{Me01}, and \citet{Be01a}.

\section{Theoretical predictions for $g_p$}
\label{s: gptheory}

\subsection{PCAC, $g_p$, and the Goldberger-Treiman relation}
\label{s: theory-PCAC}

Historically the measurement of $g_p$ has been considered interesting
and important because there is a very definite prediction for its
value derived many years ago \citep{Go58} based on the very
fundamental notion of the Partially Conserved Axial Current
(PCAC). Detailed derivations are given in textbooks,
e.g.~\citet{Bj64}, so we simply outline the basic ideas here. The
underlying assumption is that the divergence of the axial current is
proportional to the pion field. Applying this idea to the divergence
of the axial current of Eq.~(\ref{e: general current}), taking matrix
elements, and evaluating at four-momentum transfer $q^2=0$,  gives the
relation known as the Goldberger-Treiman relation:
\begin{equation}
\label{e: G-T relation}
g_{\pi NN}(0) F_\pi=m_N g_a(0)
\end{equation}
where $g_{\pi NN}$ is the pion nucleon coupling constant,
$F_\pi=92.4\pm0.3$ MeV is the pion decay constant, and $m_N$ is the
nucleon mass. This equation is rather well satisfied when one uses
modern values of $g_{\pi NN}$. A measure of the difference, which is
known as the Goldberger-Treiman discrepancy, is given by the equation
\begin{equation}
\label{e: GT discrepancy}
\Delta_{GT}=1-\frac{m_N g_a(0)}{g_{\pi NN}(m^2_{\pi}) F_\pi}.
\end{equation}
The value of $\Delta_{GT}$ was about 6\% using the older (larger)
value of $g_{\pi NN}=13.4$, and there were a number of papers
discussing possible sources of this discrepancy. See
e.g. \citet{Jo75, CoS81, Co90} and references cited therein. With the
newer and somewhat smaller value of $g_{\pi NN}(m_\pi^2)=13.05\pm0.08$
\citep{De97, St93, Arn95}, and updated values of $g_a$ and $F_\pi$
the discrepancy is now 2\% or less. See e.g. \citet{Na00, Go99} 
and references cited therein. Thus at $q^2=0$ the 
Goldberger-Treiman relation is quite well satisfied.

Now consider the matrix element of the divergence of the axial current
for non-zero $q^2$. This leads, using Eq.~(\ref{e: G-T relation}), to
an expression for $g_p$ given by
\begin{equation}\label{e: GT-gp}
g_p(q^2)=\frac{2 m_\mu m_N}{m_\pi^2-q^2}g_a(0)
\end{equation}
with $m_\pi$ the charged pion mass.  This is known as the
Goldberger-Treiman expression for $g_p$. Observe the explicit presence
of a pion pole. At $q^2=-0.88 m_\mu^2$, which is the relevant momentum
transfer for muon capture on the proton, this formula gives $g_p(-0.88
m_\mu^2)=6.77 g_a = 8.58$, where we have used the latest value of
$g_a(0)=1.2670 \pm 0.0035$ from the \citet{Pa00} and taken for $m_N$
the average of neutron and proton masses. This is what we shall refer
to in subsequent sections as the lowest order PCAC value
$g_p^{PCAC}$. At $q^2=-m_\mu^2$ the result is $g_p(-m_\mu^2)=6.47 g_a
= 8.20$.

Somewhat later the first order correction to this, proportional to the
derivative of $g_a$, was derived using current algebra techniques
\citep{Ad66}.  Numerically however the correction is
rather small, as described in the next section.

\subsection{ChPT derivations of $g_p$}
\label{s: theoryChPT}
In the years since the original derivations of the Goldberger-Treiman
relation there have been major advances in our understanding of the
way to include chiral symmetry in such analyses, particularly in the
framework of what is known as chiral perturbation theory (ChPT), or,
when nucleons are involved, heavy baryon chiral perturbation theory
(HBChPT). This approach provides a way of incorporating the symmetries
of QCD into a systematic low energy expansion, where the expansion
parameter is something of the order $m_\pi/m_N$. Thus one can
reproduce all the old current algebra results, but more importantly
calculate in a systematic way the corrections to these results.

This approach was applied \citep{Be94} to obtain for
$g_p$ the result
\begin{equation}
\label{e: gp-gpiNN}
g_p(q^2)=\frac{2 m_\mu g_{\pi NN}(q^2) F_\pi}{m_\pi^2-q^2}-
\frac{1}{3}g_a(0) m_\mu m_N r_A^2,
\end{equation}
where $r_A^2$ is the axial radius of the nucleon, defined in the usual
way via $g_a(q^2)=g_a(0)(1+q^2 r_A^2/6+{\cal O}(q^4))$. This is
essentially the result obtained much earlier by \citet{Ad66} and by
\citet{Wo70}, but the systematic approach allows us to be confident
that the corrections are of higher order.

Using the value of the axial radius $r_A^2=0.42 \pm 0.04$ fm obtained
from anti neutrino-nucleon scattering \citep{Ah88}, this formula leads
to $g_p(-0.88 m_\mu^2)=8.70-0.45=8.25$, so the correction term is
indeed rather small.

An alternative approach, still within HBChPT, is simply to calculate
the amplitude for muon capture, as was done by \citet{Fe97}, 
and identify $g_p$ by comparing with Eq.~(\ref{e: general
current}).
This gives, up to corrections of ${\cal O}(p^4)$,
\begin{equation}
\label{e: chptgp}
g_p(q^2) = \frac{2m_{\mu}m_N}{(m_\pi^2-q^2)}\left[g_a(q^2)
	   - \frac{m_\pi^2}{(4{\pi}F_\pi)^2}(2b_{19}-b_{23})\right]
\end{equation}
where $b_{19}$ and $b_{23}$ are low energy constants (LEC's) of the
basic Lagrangian. At first glance this appears to be different than
the result above, but can be put in the form of Eq.~(\ref{e: gp-gpiNN})
by noting that in the same approach $g_{\pi NN}$ and the axial radius
squared $r_A^2$ are given by
\begin{equation}
\label{e: gpiNN}
g_{\pi NN}(q^2) = \frac{m_N}{F_\pi}\left(g_a(0)-
\frac{m_\pi^2 b_{19}}{8 \pi^2 F^2_\pi}
\right).
\end{equation}
and
\begin{equation}
\label{e: axial radius}
r_A^2 = - 6 \frac{b_{23}}{g_a(0) (4{\pi}F_\pi)^2}
\end{equation}

If one wants to express $g_p$ in terms of $g_a$, as is conventional,
rather than $f_\pi g_{\pi NN}$, some simple manipulation of 
Eq.~(\ref{e: gp-gpiNN}) using Eq.~(\ref{e: G-T relation}) gives 
\begin{equation}
\label{e: gp-ga}
 g_p(q^2) = \frac{2m_{\mu}m_N}{(m_\pi^2-q^2)} g_a(0) (1+\tilde{\epsilon})-
 \frac{m_\mu m_N g_a(0) r_A^2}{3}
\end{equation}
where $(1+\tilde{\epsilon})=[g_{\pi NN}(q^2)/g_{\pi NN}(0)]$.  Note
that $(1+\tilde{\epsilon})=1$ if one neglects the $q^2$ dependence of
$g_{\pi NN}$, which we will do in this review. This gives $g_p(-0.88
m_\mu^2)=8.58-0.45=8.13$ where the slight difference from the result
of Eq.~(\ref{e: gp-gpiNN}) originates in the difference between left
and right hand sides of Eq.~(\ref{e: G-T relation}) when experimental
values are used. It is this latter value, 8.13, which we will take as 
$g_p^{PCAC}$ when the constant term is included.

It is important to note that the specific results in terms of the
LEC's of ChPT depend on the specific choice of starting Lagrangian and
the details of the calculation. The expressions of Eqs.~(\ref{e:
chptgp}), (\ref{e: gpiNN}), and (\ref{e: axial radius}) come from
\citet{Fe97}, but similar results were subsequently obtained by
\citet{Be98} and \citet{An00}. However the expressions in terms of the
physically measurable quantities, as given in Eqs.~(\ref{e: gp-gpiNN})
or (\ref{e: gp-ga}) are independent of the detailed conventions of the
approach.

Finally we can summarize this section by observing that theoretical
prediction for $g_p$ as given in Eqs.~(\ref{e: gp-gpiNN}) or (\ref{e:
gp-ga}) and based on chiral symmetry seems to be quite robust. The
original Goldberger-Treiman relation is understood as the first term
in an expansion and correction terms have been evaluated and are
understood via a ChPT calculation carried out through the first three
orders. Thus a test of this prediction should be an important test of
our understanding of chiral symmetry and of low energy QCD.

\section{Sources of information on $g_p$}
\label{s: sources}

At the simplest level $g_p$ appears as a phenomenological parameter in
the fundamental definition of the weak nucleon current, Eq.~(\ref{e:
general current}), so it should be obtainable from any process which
directly involves this current.  This would include beta decay, muon
capture, radiative muon capture, and in principle any of the crossed
versions of these reactions, as for example processes initiated by
neutrinos. The $g_p$ term is proportional to the momentum transfer
however, so in practice beta decay is not a useful source of
information since the momentum transfer is so small. Neutrino
processes are of course extremely difficult to measure.  This leaves
ordinary muon capture and radiative muon capture as the two main
sources of information on $g_p$.

Thus for ordinary, i.~e., non radiative, muon capture, $g_p$ is purely
a phenomenological parameter which appears in the most general weak
current. PCAC is not needed and it is quite reasonable to simply fit
to data, treating $g_p$ as a free parameter. This is what has
typically been done, with the result looked upon as a test of the PCAC
prediction.

At a somewhat deeper level the $g_p$ coupling is understood to arise
from the diagram in which a pion, emitted from a nucleon, couples to
the weak leptonic current. Since this pion-nucleon coupling is a
component of the axial current, other processes involving this current
such as pion electroproduction for example, also in principle allow
one to obtain information on $g_p$.  The interpretation of information
from such processes must be somewhat different, however, than that
obtained from processes like muon capture which contain the
phenomenological weak current explicitly. For processes like
electroproduction the direct information available is really
information about the $\pi NN$ vertex. Thus the connection to $g_p$ is
only via the theoretical infrastructure of chiral symmetry and of PCAC
and the interpretation that $g_p$ originates in a pion exchange
diagram.  If PCAC were in fact wrong, the whole connection would break
down and there would be no convincing physical reason why fitting data
using for $g_p$ a multiple of the PCAC expression containing the pion
pole should work. 

It is interesting to note that RMC is somewhere in between the
situations corresponding to OMC and pion electroproduction. The
dominant diagrams contain the phenomenological weak current directly,
albeit with one leg off shell, but there is also a diagram which
explicitly contains pion exchange.

Finally there is a third level where the weak current and one pion
exchange do not appear directly but where, in the context of HBChPT,
some of the LEC's needed for $g_p$ do appear.  For such processes it
is, at least in principle, possible to determine those LEC's, and thus
determine $g_p$, at least indirectly, via an equation analogous to
Eq.~(\ref{e: chptgp}). Again such information must be interpreted in a
context which accepts the validity of PCAC.

For all of these processes it may be that measurable quantities such
as correlations relative to some of the particle spins, or capture
from hyperfine states, or capture to or from a specific nuclear state,
or some such more detailed observable may provide more information
than just the overall rate, so these should be considered.

We will begin by looking at these various processes, starting with the
simplest, muon capture on the proton, and working up through reactions
on nuclei to see what has been learned and what potentially could be
learned with regard to $g_p$.

\section{Muon chemistry and the initial spin state}
\label{s: muchem}

When a negative muon is stopped in matter a muonic atom is formed.
Unfortunately in hydrogen and deuterium such atoms undergo a
complicated sequence of chemical processes, changing the spin--state
populations with time, as the muons eventually reach the level from
which they are captured.  As discussed elsewhere, the capture rates
for muonic hydrogen and muonic deuterium are strongly dependent on the
spin state of the $\mu$--nucleus system.  Thus a detailed knowledge of
this muon chemistry is required in order to determine $g_p$ from
H$_2$/D$_2$ experiments and it is appropriate to discuss this
chemistry before considering the capture process. Below we denote the
two hyperfine states of the muonic atom by $F_{\pm} = I \pm 1/2$ where
$I = 1/2$ for the proton and $I = 1$ for the deuteron, so that $F$ is
the total spin of the atom.

Before we discuss the details we make a few general comments.  One
important aspect of muon chemistry is $\mu$--atom scattering from
surrounding molecules which results in the hyperfine depopulation of
the upper F--state into the lower F--state.  Another important aspect
is collisional formation of muonic molecules which results in
additional arrangements of $\mu$--nucleus spin states.  Also muon
recycling from molecular states to atomic states via $\mu$-catalyzed
fusion occurs for $d \mu d$ molecules and $p \mu d$ molecules.  How
these effects unfold for muons in hydrogen and deuterium, as a
function of the density, is the focus of our discussion in this
section.

In Secs. \ref{s: pmup} and \ref{s: dmud}, respectively, we discuss the
chemistry of $\mu$p atoms in pure H$_2$ and $\mu$d atoms in pure
D$_2$.  In Sec. \ref{s: pmud} we describe the chemistry of muons in
H$_2$/D$_2$ mixtures. Related review articles are those of
\citet{Fr92, Po73, Br89, Br82}.

\subsection{Muons in pure hydrogen}
\label{s: pmup}

\begin{figure}
\begin{center}
\epsfig{figure=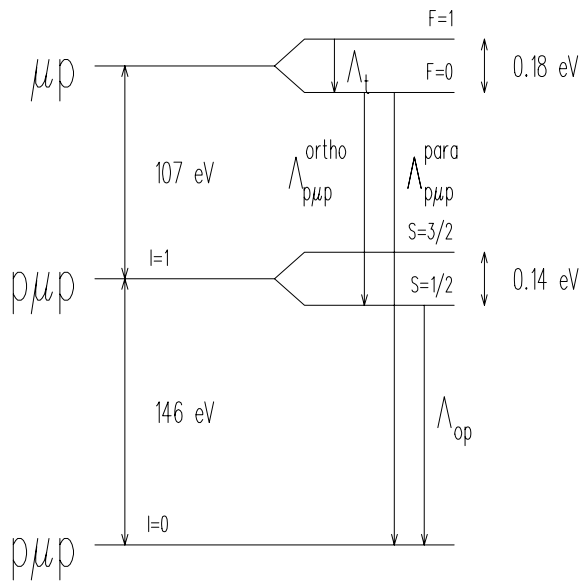, width=7.5cm,angle=0}
\caption{A simplified  diagram of the various atomic and 
molecular states and transitions relevant to muon capture in pure
H$_2$.}
\label{f: mup chemistry}
\end{center}
\end{figure}

To assist the reader a simplified diagram of muon chemistry in pure
H$_2$ is given in Fig. \ref{f: mup chemistry}.  The figure shows the
F$_+$ and F$_-$ states of the $\mu$p atom, the ortho (I=1) and para
(I=0) states of the p$\mu$p molecule, and relevant atomic and
molecular transitions.

The $\mu$p atom is initially formed in a highly excited state with a
principal quantum number $n \sim 14$ and a kinetic energy $\sim$ 1 eV.
The excited atom then rapidly de--excites via combinations of Auger
emission
\begin{equation}
(\mu p)_n + e \rightarrow (\mu p)_{n'} + e,
\label{e: auger}
\end{equation}
radiative decay
\begin{equation}
(\mu p)_n  \rightarrow (\mu p)_{n'} + \gamma,
\label{e: radiative}
\end{equation}
and Coulomb de--excitation
\begin{equation}
(\mu p)_n + p \rightarrow (\mu p)_{n'} + p.
\label{e: coulomb}
\end{equation}
Note that in Auger emission and radiative decay the $\mu p$ recoil has
a relatively small kinetic energy, since the $e$ or $\gamma$ carry the
released energy, while in Coulomb de--excitation the $\mu p$ recoil
has a relatively large kinetic energy, since the $\mu p$ and $p$ share
the released energy.  Consequently when formed the ground state atom
has kinetic energies of typically about 1 eV but occasionally up to 
100 eV. For recent experiments on energy distributions of ground state
$\mu p$/$\pi p$ atoms see \citet{Si96} and \citet{Sc99}.  Note that at
formation of the ground state atom the hyperfine states are
statistically populated, {\it i.e.}  3:1 for $F_+$:$F_-$.

This `hot' ground state atom is rapidly thermalized by elastic
scattering, 
\begin{equation}
(\mu p)_{\uparrow \downarrow} + p \leftrightarrow (\mu p)_{\uparrow
\downarrow} + p,
\label{e: mup elastic}
\end{equation}
and spin--flip collisions,
\begin{equation}
(\mu p)_{\uparrow \downarrow} + p \leftrightarrow (\mu p)_{\uparrow
\uparrow} + p,
\label{e: mup spinflip}
\end{equation}
from the atomic nuclei of the neighboring molecules. Here $\uparrow
\uparrow$ denotes the triplet state and $\uparrow \downarrow$ denotes
the singlet state.  Once the $\mu$p kinetic energy falls below the
0.18 eV $\mu$p hyperfine splitting the singlet--to--triplet
transitions are energetically forbidden and triplet--to--singlet
transitions depopulate the upper F$_+$ state.  The resulting $F_+$
state lifetime is about $0.1$~ns in liquid H$_2$ and about $10$~ns in
10 bar H$_2$ gas. The short lifetime arises from the large $\mu$p
scattering cross sections due to a near--threshold $\mu p + p$ virtual
state.  For further details of experimental studies of $\mu$p
scattering in H$_2$ environments see \citet{Ab97} and references
therein.

At sufficient densities, {\it i.e.} at pressures exceeding 10 bar, the
formation of $p \mu p$ molecules is important.  The molecule (see
Fig.~\ref{f: mup chemistry}) comprises a para--molecular ground state,
with total orbital angular momentum $\ell = 0$ and nuclear spin $I =
0$, and an ortho--molecular excited state, with total orbital angular
momentum $\ell = 1$ and nuclear spin $I = 1$.  Note spin--spin and
spin--orbit interactions produce a five--fold splitting of the
ortho--molecular state with two $S = 1/2$ sub--states and three $S =
3/2$ sub--states, where $S$ is the total spin angular momentum of the
$p \mu p$ complex.  Importantly the different states have different
make--ups in terms of $\mu p$ components with parallel spins ({\it
i.e.} $F_+$) and anti--parallel spins ({\it i.e.} $F_-$).
Specifically the para--state is 3:1 triplet--to--singlet, the $S =
1/2$ ortho--states are 1:3 triplet--to--singlet, and the $S = 3/2$
ortho--states are pure triplet.  For further details see \citet{Ba82}.

The $p \mu p$ molecules are formed by Auger emission
\begin{equation}
\mu p + H \rightarrow p \mu p + e.
\label{e: pmup}
\end{equation}
Calculations of the rates for the process have been performed by
\citet{Fa99}, \citet{Fa89}, \citet{Po76}, and \citet{Ze59}.  Formation of 
the ortho--state involves an E1 transition with a predicted rate
$\Lambda^{ortho}_{p \mu p} \simeq n/n_o \times 1.8 \times 10^6 s^{-1}$
and formation of the para--state involves an E0 transition with a
predicted rate $\Lambda^{para}_{p \mu p} \simeq n/n_o \times 0.75
\times 10^4 s^{-1}$, where $n/n_o$ is the H$_2$ target number density
normalized to the liquid H$_2$ number density.  Recent measurements of
the summed, {\it i.e.} ortho and para, rate are typically $30$\%
greater than the calculated rate.  See \citet{Mu96} for further
details.

Note that the E1 transition feeding the ortho--molecular state 
populates only the $S = 1/2$ sub--states ({\it i.e.} yielding a 1:3 ratio
of $F_+$:$F_-$ spin--states).  \citet{We60} and \citet{An00} have
discussed the possible mixing of the $S=1/2, 3/2$ levels which would
lead to changes in the 1:3 triplet--to--singlet ratio for the
ortho--molecule.  However the available calculations of
\citet{Ha64a, Ha64b, We64} and \citet{Ba82} have suggested such effects are
negligible.\footnote{In addition the relative rates for $\mu$ capture
in H$_2$ gas (mostly singlet--atom capture) and H$_2$ liquid (mostly
ortho--molecule capture) are consistent with a 1:3 triplet:singlet
make--up of the ortho--molecule.  For details see
\protect{Sec. \ref{s: comp expt}}.}

At first glance the $\Delta I = 0$ selection rule for E1 transitions
forbids decay of the ortho excited state to the para ground state.
However as discussed by \citet{Ba82}, via the small components of the
relativistic wave functions, the ortho state contains $I = 0$
admixtures and the para state contains $I = 1$ admixtures.  Therefore
ortho--to--para E1 transitions occur via cross--combinations of the
small components and the large components of the molecular wave
functions.  Note the rate $\Lambda_{op}$ for this Auger process is a
function of the electron environment of the $p \mu p$ molecule.
\citet{Ba82} obtained $\Lambda_{op} = 7.1 \pm 1.2
\times 10^4 s^{-1}$ assuming an environment consisting of 75\% $ [ (p
\mu p)^+ 2p 2e]^+ $ and 25\% $ [ (p \mu p)^+ e] $. 
These proportions are a consequence of the Hirshfelder reaction
\citep{Hi36} involving $p \mu p$ complexes and H$_2$ molecules as
discussed by \citet{Fa89, Fa99}.  Note that the only published
experimental value of $\Lambda_{op} =(4.1 \pm 1.4) \times 10^4 s^{-1}$
from \citet{Ba81b, Bar82} is in marginal disagreement with the
calculation at the $2\sigma$ level.

In summary, in H$_2$ gas at pressures $0.1 < P < 10$~bar, where the $F
= 1$ atoms disappear very quickly and the $p \mu p$ molecules form
very slowly, the capture process is essentially dominated by singlet
atoms.  However in liquid H$_2$ the molecular formation rate
$\Lambda_{p \mu p}$ and ortho--to--para transition rate $\Lambda_{op}$
are important.  Here the rate is a superposition of singlet atomic
capture, ortho--molecule capture and para--molecule capture and
depends on $\Lambda_{p \mu p}$, $\Lambda_{op}$, and the measurement
time window.

\subsection{Muons in pure deuterium}
\label{s: dmud}

The atomic capture and cascade processes for muons in pure D$_2$ and
pure H$_2$ are very similar. Most importantly for muons in D$_2$ the
ground state $\mu d$ atoms are rapidly formed in a statistical mixture
of the hyperfine states, {\it i.e.}  2:1 for $F_+$:$F_-$.

The $\mu d$ atoms are then thermalized by elastic scattering
\begin{equation}
(\mu d)_{\uparrow \downarrow} + d \leftrightarrow (\mu d)_{\uparrow
\downarrow} + d
\end{equation}
and spin--flip collisions
\begin{equation}
(\mu d)_{\uparrow \downarrow} + d \leftrightarrow (\mu d)_{\uparrow
\uparrow} + d
\end{equation}
on surrounding nuclei.  When the $\mu d$ kinetic energy falls below
the 0.043 eV hyperfine splitting the spin--flip collisions then
depopulate the higher lying F$_+$ state.  However the cross sections
are considerably smaller for $\mu d + d$ than $\mu p + p$ and
consequently the $F_+$ lifetime is considerably longer in deuterium
than hydrogen. For example in liquid D$_2$ at $34~K$, the hyperfine
depopulation rate is $\Lambda = 42.6 \times 10^6 s^{-1}$.  For further
details see \citet{Ka82, Ka83}.

An interesting feature of $\mu d$ chemistry is resonant formation of
$d \mu d$ molecules.  For example in liquid D$_2$, while $d \mu d$
formation from $F=1/2$ $\mu d$ atoms involves a non--resonant Auger
process, the $d \mu d$ formation from $F=3/2$ $\mu d$ atoms involves a
resonant excitation process, {\it i.e.} one where the $d \mu d$ binding
energy is absorbed by D$_2$ vibro--rotational modes.  In liquid D$_2$
at $34~K$ the effective rates are $\Lambda^{1/2}_{d \mu d} \sim 5
\times 10^4 s^{-1}$ for non--resonant, {\it i.e.} doublet, formation
and $\Lambda^{3/2}_{d \mu d}\sim 4 \times 10^6 s^{-1}$ for resonant,
{\it i.e.} quartet, formation.  Note that the resonant formation is
temperature dependent.  For further details see \citet{Br89}.

When $d \mu d$ molecules are formed, the two deuterons immediately fuse
via $\mu$--catalyzed fusion
\begin{equation}
d + d \rightarrow ^{3}\!\!He + n ~~~~~[ B.R. \sim 55\% ]
\label{e: dd -> 3He}
\end{equation}
\begin{equation}
d + d \rightarrow  ^{3}\!\!H + p ~~~~~[ B.R. \sim 45\% ]
\label{e: dd -> 3H}
\end{equation}
which quickly recycles the muon from the molecular states to the
atomic states.  Consequently for muons in pure D$_2$ the deuterium
capture is from $\mu d$ atoms and not $d \mu d$ molecules, independent
of density and temperature.  However the $\mu$ sticking probability in
$\mu$ catalyzed fusion, 13\% in Eq.~(\ref{e: dd -> 3He}) and 1\% in
Eq.~(\ref{e: dd -> 3H}), is non--negligible.  Consequently with
increasing D$_2$ target density an increasing $\mu$$^3$He capture
background is unavoidable.

In summary, for muons in pure D$_2$ the capture is a superposition of
the rates from the doublet atom and the quartet atom and $\mu$ capture
from $d \mu d$ molecules is completely negligible.  Actually for the
particular conditions of the $\mu d$ experiments by \citet{Ba86} and
\citet{Ca89} the deuterium capture is almost entirely from doublet
atoms. See Sec. \ref{s: D experiment} for details.  However due to
muon sticking in the $d \mu d$ fusion process, a correction for
backgrounds from $\mu$$^3$He capture is necessary in liquid D$_2$.

\subsection{Muons in hydrogen-deuterium mixtures}
\label{s: pmud}

Our interest in hydrogen-deuterium mixtures is two--fold.  First
`pure--H$_2$' experiments and `pure--D$_2$' experiments must
inevitably be concerned with contamination from other isotopes.
Second some early experiments on deuterium capture used
hydrogen-deuterium mixtures, {\it e.g.}
\citet{Wa65c} used 0.3\% $D_2$ in H$_2$ liquid and \citet{Be73} used
5\% $D_2$ in H$_2$ gas.

Muon transfer from
$\mu p$ atoms to $\mu d$ atoms 
\begin{equation}
\mu p + d \rightarrow \mu d + p
\end{equation}
occurs with an energy release of 135 eV and a transfer rate of
$n/n_o \times c_d \times 1.7 \times 10^{10} s^{-1}$,
where $c_d$ is the D$_2$ concentration in the H$_2$ target and $n/n_o$
the target number density relative to liquid H$_2$ \citep{Ad92}.
Consequently, in H$_2$ liquid a $10^{-3}$ D$_2$ concentration and in
$P > 10$~bar H$_2$ gas a $10^{-2}$D$_2$ concentration, is sufficient
to engineer the transfer in roughly $100$~ns.

Following transfer, the $\mu d$ atom is thermalized via collisions
with H$_2$ molecules.  An interesting feature of $\mu d + p$
scattering is the Ramsauer--Townsend minimum at a kinetic energy
$1.6$~eV.  The tiny $\mu d + p$ cross section means slow
thermalization and large diffusion of $\mu d$ atoms in $H_2$ gas.  In
addition the slow thermalization and small deuterium concentration
makes hyperfine depopulation via spin--flip collisions
\begin{equation}
(\mu d)_{\uparrow \downarrow} + d \leftrightarrow (\mu d)_{\uparrow
\uparrow} + d
\end{equation}
very slow by comparison to $\mu p$ atoms in pure H$_2$ and $\mu d$
atoms in pure D$_2$.  

At sufficient densities the formation of $p \mu d$ molecules
occurs by Auger emission 
\begin{equation}
\mu d + H_2 \rightarrow p \mu d + H + e
\label{e: pmud formation}
\end{equation}
with a rate in liquid H$_2$ of $\Lambda_{p
\mu d} = 5.6 \times 10^6 s^{-1}$
\citep{Pe90}.  The formation of molecules is
important as (i) capture is consequently a superposition of $\mu d$
reactions and $\mu p$ reactions, and (ii) the various $p \mu d$ states
have different decompositions into $\mu d$ $F$--states.  Further the
parent distribution of $\mu d$ atom $F$--states effects the resulting
distribution of $p \mu d$ molecule states, making the relative
population of $p \mu d$ states a complicated function of target
density and deuterium concentration.

Muon catalyzed fusion from $p \mu d$ molecules occurs via 
both radiative reactions
\begin{equation}
p \mu d  \rightarrow \mu ^{3}\!He + \gamma
\label{e: fusion radiative}
\end{equation}
and non--radiative reactions
\begin{equation}
p \mu d  \rightarrow \mu + ^{3}\!\!He.
\label{e: fusion nonradiative}
\end{equation}
The radiative rates \citep{Pe90} are $\Lambda_{1/2} = 0.35 \times
10^{6} s^{-1}$ and $\Lambda_{3/2} = 0.11 \times 10^{6} s^{-1}$ and the
non--radiative rates are $\Lambda_{1/2} = 0.056 \times 10^{6} s^{-1}$
and $\Lambda_{3/2} = 0$, where the subscripts $1/2, 3/2$ denote the
p--d spin states.  The slow rates make $\mu$ capture from $p \mu d$
molecules an important contribution at high target densities.  Further
the 100\% sticking probability for the radiative reaction makes $\mu$
capture in muonic $^3$He a troublesome background.

In summary, both $\sim 10^{-3}$ D$_2$ admixtures in H$_2$ liquid and
$\sim 10^{-2}$ D$_2$ admixtures in H$_2$ gas have been used in the
study of $\mu d$ capture.  Unfortunately the hyperfine depopulation of
$\mu d$ atoms in H$_2$/D$_2$ mixtures is slow and therefore the
doublet--quartet make--up in H$_2$/D$_2$ experiments is dependent on
target density and deuterium concentration.  Further at densities
where $p
\mu d$ molecules are formed, the observed rate of muon capture is a
complicated superposition of the $\mu p$, $\mu d$ and $\mu$$^{3}$He
rates and thus disentangling the outcome is difficult.

\section{Muon capture in hydrogen}
\label{s: mcproton}

\subsection{Theory of ordinary muon capture}
\label{s: omcproton}

\subsubsection{Standard diagram calculations}

The simplest of the muon capture reactions is OMC on the proton, $\mu +
p \rightarrow n + \nu$, which has been studied theoretically for many
years. Some of the early work included that of \citet{Pr59} and
\citet{Fu59}. \citet{Op64} evaluated amplitudes for both
OMC and RMC in an expansion in powers of $1/m_N$.  Many other authors
have performed similar evaluations of the OMC rate.

The basic physics is completely determined by the weak nucleon current
given in Eq.~(\ref{e: general current}), which is the most general
possible form consistent with the known current-current form of the
weak interaction. Given this current, the OMC amplitude is determined
by its product with the leptonic weak current. One then uses standard
diagrammatic techniques to square the amplitude, put in phase space
and thus obtain an expression for the rate in terms of the coupling
parameters $g_v, g_m, g_a$, and $g_p$. This expression is the same
whether the couplings are obtained from purely phenomenological
sources or from some detailed fundamental model.

Fortunately a lot is known about the couplings. The weak vector
current is completely determined by the well established Conserved
Vector Current theory which tells us that the weak vector current is
simply an isospin rotation of the isovector electromagnetic
current. In practice this means that $g_v(q^2)= F_1^p(q^2)-F_1^n(q^2)$
and $g_m(q^2)=\kappa_p F_2^p(q^2) - \kappa_n F_2^n(q^2)$ where
$\kappa_p=1.79285$, $\kappa_n=-1.91304$ are the proton and neutron
anomalous magnetic moments and where $F_1^p(q^2)$, $F_1^n(q^2)$,
$F_2^p(q^2)$, and $F_2^n(q^2)$ are the usual proton and neutron
isovector electromagnetic form factors. The coupling $g_a$ is well
determined from neutron beta decay, $g_a(0)=1.2670 \pm 0.0035$
\citep{Pa00}. The momentum dependence is known, at least at low
momentum transfers, from neutrino scattering \citep{Ah88} or from pion
electroproduction \citep{De76, Es78, Ch93}. For a long time there was
some disagreement between these two sources, but that has now been
resolved by a more careful analysis of the corrections necessary in
pion electroproduction \citep{Be92}. Thus all of the ingredients for a
theoretical calculation of the OMC rate on the proton are well
determined by general principles which have been verified in many
other situations, except for the value of $g_p$, which is given
primarily by the theoretical predictions discussed in Section \ref{s:
gptheory} above.

\subsubsection{ChPT calculations}

The amplitude for OMC on the proton has also been evaluated in the
context of HBChPT by \citet{Fe97}. Such calculations start with the
general ChPT Lagrangian, in this case through $O(p^3)$, and evaluate
the amplitude consisting of tree and one loop diagrams. The outcome of
such calculations are expressions for couplings appearing in the most
general amplitude of Eq.~(\ref{e: general current}) in terms of the
LEC's appearing in the Lagrangian. This approach thus provides a
systematic way of calculating couplings and their corrections, as was
described for $g_p$ in Section \ref{s: theoryChPT} above. However the
vector part of the amplitude must still satisfy CVC, $g_a$ must still
reproduce neutron beta decay, etc., so in actual fact the couplings to
be used are exactly the same as those which have always been used in
the phenomenological approach and there is no new information arising
from a ChPT calculation, except perhaps for the (small) correction
term appearing in $g_p$, Eq.~(\ref{e: gp-gpiNN}). What is accomplished
by such a calculation however is to evaluate some of the LEC's which
are needed for other calculations, such as RMC.

Similar ChPT evaluations of OMC were subsequently carried out by
\citet{An00} and \citet{Be98}. In the latter case the so called small
scale expansion was used, which is a way of including the $\Delta$ as an
explicit degree of freedom in the ChPT formalism, rather than
absorbing its effects in the LEC's.

\subsubsection{Spin effects}
When a muon is stopped in hydrogen it proceeds via a rather
complicated series of atomic and molecular processes, as was described
above, to a low level state in either a $\mu p$ atom or a $p \mu p$
molecule from which the muon is captured. The details of this cascade
process and relative probabilities for population of the various
states depend on the density of the target and the time at which one
starts detecting the captures. For now it suffices to note that the
capture from any initial state will be a linear combination of
captures from the singlet and triplet $\mu p$ states. Thus it is
important to calculate separately the rates from these two spin
states.

\begin{figure}
\begin{center}
\epsfig{file=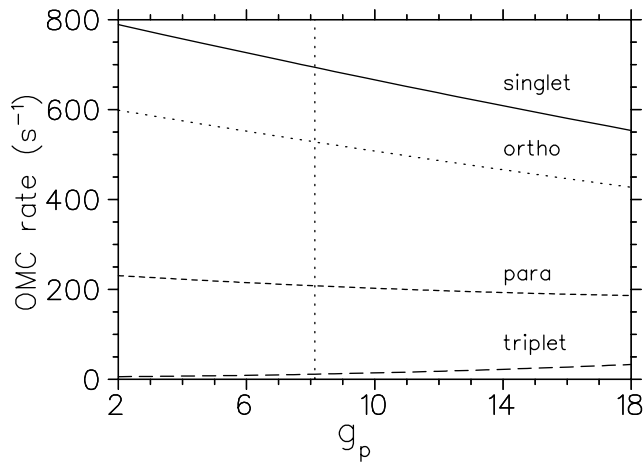,width=6cm, angle=90}
\caption{Capture rates for OMC on a proton plotted
versus $g_p(-0.88 m_\mu^2)$ for the muon initially in the singlet or
triplet state of the $\mu p$ atom or in the ortho or para state of the
$p\mu p$ molecule. The vertical line corresponds to the PCAC value of
$g_p = 8.13$, as defined in Eq.~(\ref{e: gp-ga}), which includes the
constant term. Details of the model used are described in Sec.~\ref{s:
comp expt}}
\label{f: singlet/triplet capture}
\end{center}
\end{figure}

Figure \ref{f: singlet/triplet capture} shows the individual singlet
and triplet captures rates for OMC, using a standard diagram
calculation\footnote{The calculation has been updated to include form
factors and modern values of the couplings, particularly $g_a$. The
general table of numerical values, Table \ref{t: parameters}, gives the
values of parameters and constants used.}
\citep{Fe80}, plotted versus the value of $g_p$. Clearly the
capture rate from singlet state is much larger than that from the
triplet state. However it is also much less sensitive to the value of
$g_p$. Unfortunately, although one can increase the sensitivity to
$g_p$ to some extent by choosing conditions that enhance the triplet
contribution, the singlet rate is so much larger that it dominates in
essentially all circumstances.

\begingroup
\begin{table*}
\caption{Numerical values of the parameters and derived quantities 
used in the text and in our evaluations of rates for comparison with
experiment.}
\label{t: parameters}
\begin{ruledtabular}
\begin{tabular}{llcl}
 & & &  \\
Symbol & Description & Value & Reference\\
 & & & \\
\hline
 & & & \\
$F_\pi$ & pion decay constant & $92.4 \pm 0.3 $\ MeV & \citet{Pa00} \\
$g_{\pi NN}(m_\pi^2)$ & pion nucleon coupling & 
     $13.05 \pm 0.08$ & \citet{De97} \\
$G_F V_{ud}$ & Fermi constant for $\beta$ decay &
     $1.13548 \times 10^{-5} \rm{GeV}^{-2}$ & \citet{Pa00} \\
$g_a$(0) & axial coupling from $\beta$-decay & 
     $1.2670 \pm 0.0035$ & \citet{Pa00} \\
$r_A^2$ & rms radius squared for $g_a$ & 
     $0.42 \pm 0.04 \ {\rm fm}^2$ & \citet{Ah88} \\
$g_p^{PCAC}$ & PCAC value, $g_p(-0.88 m_\mu^2)$ & 
     $6.77\, g_a(0) =8.58$ & Eq.~(\ref{e: GT-gp}) \\
             & PCAC value, including constant term & 
     $6.42\, g_a(0) =8.13$ & Eq.~(\ref{e: gp-ga}) \\
$\Lambda_{p\mu p}$ & $p \mu p$  molecular formation rate & 
     $2.5 \times 10^6 \ {\rm s}^{-1}$ & average, \citet{Wr98}\\
$\Lambda_{p\mu p}^{ortho}/\Lambda_{p\mu p}^{para}$ & ratio 
of ortho to para molecular formation  &  $240:1$ & \citet{Fa99}\\
$\Lambda_{op}$ & ortho to para transition rate & 
     $4.1\pm 1.4 \times  10 ^4 \ {\rm s}^{-1}$ & \citet{Ba81a} \\
$ 2 \gamma^{ortho}$ &ortho-molecular overlap factor & 
     $1.009\pm 0.001$ & \citet{Ba82}  \\
$ 2 \gamma^{para}$ &para-molecular overlap factor & 
     $1.143\pm 0.001$ & \citet{Ba82}  \\
$g_m(0)$ & weak magnetism coupling, $\kappa_p-\kappa_n$ & 
     $3.70589$ & \citet{Pa00} \\
$r_m^2$ & rms radius squared for $g_m$ & $0.80 \ {\rm fm}^2$ & 
\citet{Me96} \\
$r_v^2$ & rms radius squared for $g_v$ & $0.59 \ {\rm fm}^2$ & 
\citet{Me96} \\
 & & & \\
\end{tabular}
\end{ruledtabular}
\end{table*}
\endgroup

\subsection{Theory of radiative muon capture}
\label{s: rmcproton}

We now want to consider the radiative muon capture process $\mu + p
\rightarrow n + \nu + \gamma$. The basic ingredients are the same as
for OMC and the additional coupling of the photon is known. However
the presence of the photon changes the range of momentum transfers
available and leads to contributions coming from regions much closer
to the pion pole than for OMC. More specifically, for OMC on the
proton the momentum transfer is fixed at $q^2=-0.88 m_\mu^2$. For RMC
however the momentum transfer for some of the diagrams can approach
$+m_\mu^2$. These diagrams, all of which involve radiation from
hadronic legs, are not the dominant ones. The muon radiating diagram
dominates, at least in the usual gauge, and it involves similar
momentum transfer as for OMC. However the other diagrams contribute
enough in the measurable photon energy region k$>$60 MeV that the
overall sensitivity of RMC to $g_p$ is significantly increased as
compared to OMC.

\subsubsection{Standard diagram calculations}

\begin{figure}
\begin{center}
\epsfig{figure=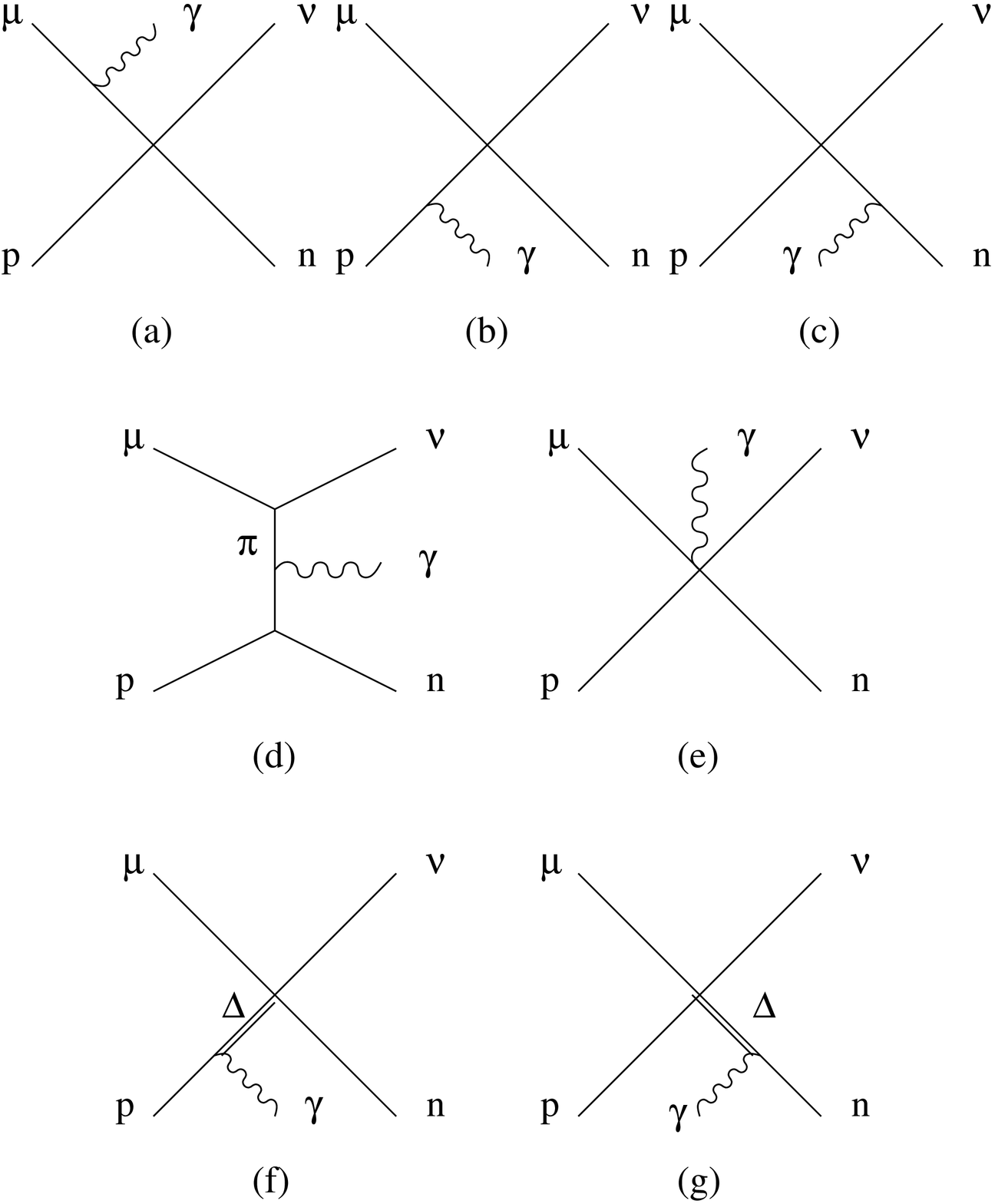, width=7.5cm,angle=0}
\caption{Diagrams contributing to the standard diagrammatic approach
to RMC: (a)-(e) are the standard diagrams, (f) and (g) are those
involving $\Delta$ contributions}
\label{f: RMCdiag}
\end{center}
\end{figure}

The standard approach to RMC on the proton has been a Feynman diagram
approach, Fig.~\ref{f: RMCdiag}(a)-(e), which includes the diagrams
involving radiation from the muon, the proton, the neutron via its
magnetic moment, and from the exchanged pion which generates the
induced pseudoscalar term. The fifth diagram makes the result gauge
invariant using a minimal substitution. \citet{Op64} performed one of
the earliest such calculations, using however a $1/m_N$ expansion of
the amplitude. The completely relativistic calculation of \citet{Fe80} 
is an example of a more modern calculation using this 
approach.\footnote{\citet{Hw78} also evaluated RMC in hydrogen
using what they called a linearity hypothesis. This was shown to be
incorrect however by \citet{Wu79},
\citet{Fe80}, and \citet{GmO81}.}

There are some possible enhancements to this basic approach. For
example the intermediate nucleon, between weak and electromagnetic
vertices, can in principle be a $\Delta$. Such effects, 
Fig.~\ref{f: RMCdiag}(f)-(g), were considered by \citet{Be87, Be89}. 
They increase the photon spectrum by amounts ranging from 2-3\%
at 60 MeV to 7-8\% at the upper endpoint. \citet{TK01}
found a similar sized effect.

Some additional terms, higher order in an expansion of the RMC
amplitude in powers of the photon momentum $k$ or the momentum
transfer $q$, were obtained originally by \citet{Ad66}. These arise 
via the requirements of gauge invariance and PCAC
for the full amplitude. Similar terms were obtained by \citet{Ch77} 
and by \citet{Kl85}. They however seem to be fairly
small.

This basic diagrammatic approach clearly contains most of the
important physics. However there are some things it does not
contain. In particular in the simplest approach gauge invariance is
imposed only via a minimal substitution $p \rightarrow p-eA$ on the
explicit momentum dependence of the operators in the weak vertex,
Eq.~(\ref{e: general current}). Thus one picks up only two terms, one
from the explicit $q$ in the $g_m$ term and one from the $q$ in the
$g_p$ term. In principle there may be many other terms, gauge
invariant by themselves, for example coming from situations where the
photon couples to some internal loop structure of the $\pi NN$ vertex.

Also for RMC it is difficult to put form factors in at the various
vertices in a general way, though lowest order form factor effects can
be included via a low energy expansion as done by \citet{Ad66}. This
is because the momentum transfer at the various vertices in different
diagrams is different and so putting in form factors evaluated at
these different momenta would destroy the delicate cancellation needed
for gauge invariance. This is not a problem unique to RMC. It appears
in any electromagnetic process described by more than one diagram if
the momentum transfers are different. Various prescriptions have been
proposed to address this problem, but all are pretty much ad hoc.  In
OMC introducing form factors reduces the rate, but only by a few
percent. This is because the relevant momentum transfer is small
compared to the scale relevant for the form factor. In RMC all
momentum transfers are comparable or smaller than that for OMC, so
hopefully the form factors introduce only a small correction in RMC as
well. This turns out to be the case in the simple approach described
below.

\subsubsection{ChPT calculations}
\begin{figure}
\begin{center}
\epsfig{figure=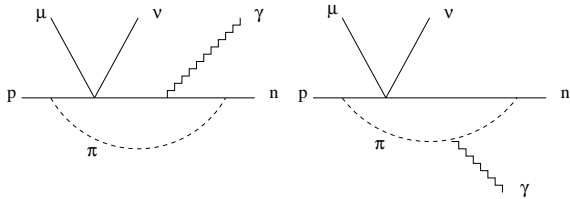, width=7.5cm,angle=0}
\caption{Several examples of diagrams which appear in a HBChPT
calculation but which are not included in the usual minimal
substitution gauge contribution of the diagrammatic approach. }
\label{f: gauge example}
\end{center}
\end{figure}

Just as for OMC, it is possible to calculate the amplitude for RMC
using HBChPT. Unlike OMC however such a calculation introduces some
new physics. The ChPT Lagrangian is constructed to satisfy gauge
invariance and also CVC and PCAC. Thus the calculated amplitude will
satisfy all of these general principles. This means for example that
such a calculation may contain explicitly gauge terms beyond the
simple minimal substitution used in the standard diagrammatic
approach. Figure \ref{f: gauge example} shows some examples of such
terms present in a ChPT calculation which are not present in the
diagrammatic calculation. A second advantage of such a calculation is
that it automatically incorporates form factors, at least to the order
of the calculation. One can calculate weak and electromagnetic form
factors explicitly as was done by \citet{Fe97} or
\citet{Be98} within ChPT. The same sub-diagrams
responsible for these form factors will appear in the RMC calculation,
so the form factors will be present in such a calculation, and in a
gauge invariant way.

The ChPT approach has one disadvantage however, not shared by the
diagrammatic approach. It is an expansion in a small parameter. Thus
the calculations are done at best to ${\cal O}(p^3)$ which is
sufficient to include tree level and the lowest contribution to one
loop diagrams. In principle a diagrammatic calculation keeps all
orders of whatever diagrams are included. In practice however
diagrammatic calculations are limited to tree level graphs, so do not
include the loop contributions of ChPT and probably at most have only
a few higher order terms coming from relativistic corrections which
would not be included in the ChPT approach.

The most complete of the HBChPT calculations of RMC has been done by
\citet{An98}, who worked to third order, {\it i.e.} ${\cal O}(p^3)$, 
or in their terminology NNLO (next to next to leading order). They
found that the loop contributions contributed less than 5\% correction
to the tree level amplitude and that the result was in reasonable
agreement with the standard diagrammatic approach.

A similar calculation, but just involving tree level diagrams, and
working only to ${\cal O}(p^2)$ was performed by \citet{Me98}. They
initially found a photon spectrum harder by 10\% or so than that of
the diagrammatic approach in the region of photon energies greater
than 60 MeV, but this was later attributed \citep{My99} to unwarranted
approximations made in the phase space evaluation. A similar ${\cal
O}(p^2)$ calculation was done by \citet{Be01b} who however used the so
called small scale expansion which allows one to put the $\Delta$ in
explicitly. Such an approach is not fundamentally different from the
usual HBChPT approach. It just allows one to extract explicitly the
contribution of the $\Delta$ to the various LEC's. In the usual
approach the $\Delta$ degrees of freedom are integrated out and their
effects absorbed in the LEC's. These authors found results similar to
those of earlier diagrammatic calculations and specifically that the
$\Delta$ contribution was only of order of a few percent, consistent
with the findings of
\citet{Be87, Be89}.

\subsubsection{Model calculations}
\label{s: RMCmodel}
Recently there has been a somewhat different model calculation of RMC
\citep{TK01} based on a $\pi \rho \omega a_1$ Lagrangian
which also includes $\Delta$'s \citep{Sm99}. This is basically a
diagrammatic approach, but based on a Lagrangian which has a number of
additional pieces, and which thus results in a explicitly gauge
invariant expression, which also satisfies CVC and PCAC and which
generates some of the higher order terms analogous to those derived by
\citet{Ad66}. For the best values of the parameters the $\Delta$
contributions range from about 3-7\% which can be increased to
roughly 7-11\% at the extreme edge of the allowed parameter
space. Thus their results are very similar to those obtained by
\citet{Be87, Be89}, and so this model would seem to give a result only
a few percent different from the ChPT result or from the standard
diagrammatic approach including the $\Delta$.

\subsubsection{Spin effects}
\begin{figure}
\begin{center}
\epsfig{file=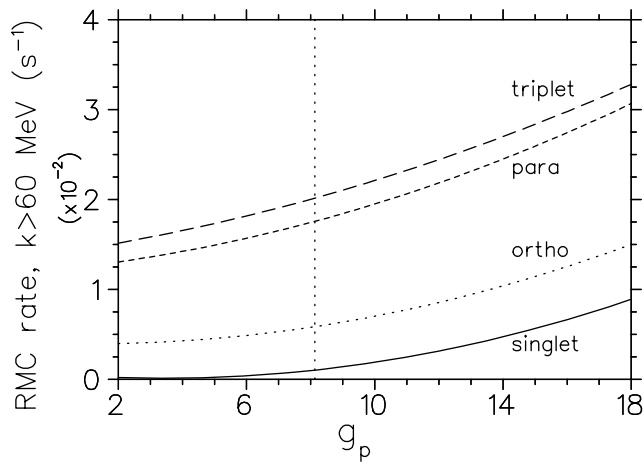,width=6cm,angle=90}
\caption{The photon spectrum for RMC on a proton, integrated from 60
MeV to the upper end point, shown as a function of $g_p(-0.88
m_\mu^2)$.  Shown separately are the rates for the muon initially in
the singlet or triplet state of the $\mu p$ atom or in the ortho or
para state of the $p\mu p$ molecule. The vertical line corresponds to
the PCAC value of $g_p = 8.13$, as defined in Eq.~(\ref{e: gp-ga}),
which includes the constant term. Details of the model used are
described in Sec.~\ref{s: comp expt}}
\label{f: RMCsingtrip}
\end{center}
\end{figure}

For RMC, just as for OMC, there are fairly dramatic differences in the
capture rate from singlet and triplet initial states, and likewise in
sensitivities to $g_p$. Figure \ref{f: RMCsingtrip} shows the
integrated photon spectrum above 60 MeV plotted versus $g_p$ for RMC
on the proton. Now it is the triplet state which dominates and the
singlet which is most sensitive to $g_p$, just the opposite situation
from OMC. Furthermore, if one looks at the RMC photon spectrum before
integration \citep{Be89} one sees that variations 
in $g_p$ about the PCAC value tend to affect the singlet rate more
at the upper end of the spectrum than at the lower end, whereas for
the triplet state the change is more uniform across the spectrum. This
further enhances the sensitivity to $g_p$, since it is the upper end
of the spectrum which is experimentally accessible.

Finally one should note that the effects which change the relative
proportion of singlet and triplet states work in different directions
for OMC and RMC. This is because triplet capture is most important for
RMC whereas singlet capture is most important for OMC. Thus for
example, some effect which increases the proportion of the triplet
state relative to the singlet will decrease the OMC rate and increase
the RMC rate, for a given value of $g_p$.

\subsection{Experimental methods for hydrogen and deuterium}
\label{s: exptmethods}
Herein we discuss the experimental methods for muon capture on
hydrogen and deuterium.  They include measurements of ordinary capture
via neutron detection, Sec. \ref{s: neutronmeth}, Michel electron
detection, Sec. \ref{s: lifetimemeth}, and radiative capture via
$\gamma$--ray detection, Sec. \ref{s: radcapt}.

\subsubsection{Bubble chamber studies}
\label{s: bubcham}
The first observations of OMC on hydrogen were reported by \citet{Hi62},
\citet{Be62} and \citet{HiD62}.  In these
early bubble chamber experiments the muons were identified by range
and curvature and neutrons were identified via their knock--on
protons.  An attractive feature of bubble chamber studies is the
determination of the neutron energy from the measurement of the proton
kinematics.  This is helpful in distinguishing the neutrons from OMC
in H$_2$ from combinatorial backgrounds of uncorrelated muons and
knock--on protons. Recall the neutrons from OMC in H$_2$ are
mono--energetic $E_n =$ 5.2 MeV.  However the bubble chamber studies
were limited by statistics.

\subsubsection{Lifetime method}
\label{s: lifetimemeth}
The `lifetime method' involves comparing the negative muon lifetime
and positive muon lifetime when stopped in hydrogen or deuterium.  For
positive muons the lifetime is the inverse of the $\mu$ decay rate,
whereas for negative muons the lifetime is the inverse of the sum of
the $\mu$ decay rate and the $\mu$ capture rate.  The lifetime
difference thus determines the capture rate.  This approach was used
at Saclay to determine the $\mu$ capture rate in liquid H$_2$
\citep{Ba81a} and liquid D$_2$ \citep{Ba86}.

The Saclay experiments used a pulsed beam with typically a $3000$~Hz
repetition rate and typically a $3.0$~$\mu$s pulse width.  The beam
entered via a lead collimator and a copper degrader, stopping in a
$24$~cm diameter Cu--walled target.  The target was filled with LH$_2$
or LD$_2$ with high elemental and isotopic purity.  The target was
viewed by six Michel electron telescopes each comprising a three--ply
plastic scintillator sandwich.  A quartz oscillator was used to
determine the time between the beam pulse start signal and the decay
electron stop signal.  Note that the stop signals were recorded in
time windows of about 1 ${\mu}$s $< t <$ 16 ${\mu}$s following the end
of the beam pulse.
 
The major challenge in the lifetime experiment is the required
accuracy.  Since the $\mu$ decay rate is about 1000 times the $\mu$
capture rate, the $\mu^+$--$\mu^-$ lifetime difference is only about 
$0.1$\%.  Consequently a determination of the capture rate $\Lambda$
to about $5\%$ requires a measurement of the lifetime $\tau$ to about
$5 \times 10^{-5}$.  Therefore both high statistics and low
systematics are important.

One concern is muon stops in non--hydrogen materials.  Detection of
decay electrons from extraneous materials will alter the time spectrum
and distort the measured lifetime.  Note that the degrader,
collimator, and target cell were all constructed from high--Z
materials, so muon stops were rapidly absorbed.  Also special care was
taken to avoid the occurrence of muon stops in plastic scintillator.
Lastly, H$_2$ or D$_2$ of ultra--high elemental and isotopic purity
was used.  For $\mu p$ experiments a small quantity of D$_2$ in H$_2$
will result in $\mu$--transfer (see Sec. \ref{s: pmud}).  For $\mu d$
experiments, where muons are quickly recycled from $d \mu d$ molecules
to $\mu d$ atoms (see Sec. \ref{s: dmud}), the effects of muon
transfer to high--Z contaminants are especially troublesome.

Another concern is a time distortion due to rate effects, {\it i.e.}
distortion arising from finite resolving time and electron pulse
pile--up. \citet{Ba81a} found the main effect of pulse pile--up is an
additional time component with an effective lifetime $\tau /2$.  They
corrected for this effect by measuring lifetimes at different rates
and extrapolating to zero. The correction to $\tau$ was roughly
$\sim$~$0.1$~ns.  Additionally time dependent and time independent
physical backgrounds were accounted for.

In $\mu d$ experiments a correction is also necessary for
$\mu$ capture on $^{3}$He nuclei.  The $\mu$$^{3}$He atoms are
produced via $d \mu d$ formation and $d d$ fusion (see Sec. \ref{s:
dmud}).  The $^{3}$He background, which was monitored by the detection
of the 2.5 MeV neutrons from the $d d$ fusion, was a 10\% correction
in the Saclay $\mu d$ experiment.

Finally we note that although the Saclay group measured both the
$\mu^+$ and $\mu^-$ lifetimes they employed the then available world
average values for $\tau_+$ in order to extract the capture rates in
LH$_2$ and LD$_2$. The earlier hydrogen experiment used $\tau_+ =
2197.148 \pm 0.066$~ns \citep{Ba81a}, though this was later updated by
\citet{Ma84, Ma82}. The later deuterium measurement \citep{Ba86} 
used $\tau_+ = 2197.03 \pm 0.04$~ns.  This point is discussed in
detail in Sec. \ref{s: comp expt}.

\subsubsection{Neutron method}
\label{s: neutronmeth}

The `neutron method' involves directly detecting the recoil neutrons
from muon capture in H$_2$ or D$_2$.  For hydrogen capture the
neutrons are mono--energetic with $E_n =$ 5.2 MeV while for deuterium
capture the neutrons are peaked at $E_n \sim$ 1.5 MeV.  Hydrogen data is
available on liquid targets from \citet{Bl62} and
\citet{Ro63} and on gas targets from \citet{Al69} and \citet{By74}.  
Deuterium data is available in pure D$_2$ from \citet{Ca89} and H$_2$/D$_2$
mixtures from \citet{Wa65c} and \citet{Be73}.

The `neutron method' involves (i) counting the incoming muons and
outgoing neutrons and (ii) determining their corresponding detection
efficiencies.  In a typical set--up the $\mu$--beam is directed into
the target vessel via a scintillator telescope and the neutrons are
detected in a liquid scintillator counter array.  Pulse--shape
analysis enables the separation of neutrons from $\gamma$--rays and
veto counters enable the separation of neutrons from electrons.
Both electrons  and gammas from $\mu$--decay are intense backgrounds.
Typically the detection of neutrons is initiated about 1~$\mu$s after
$\mu$ arrival.

The low yield of capture neutrons from OMC in H$_2$ or D$_2$ means
neutron backgrounds from $\mu$ capture in surrounding materials, $\mu$
transfer to target impurities, and other sources, are troublesome.  By
using high--Z materials for the collimator, vessel, etc., the neutron
backgrounds from $\mu$ stops in extraneous materials are
short--lived.\footnote{Further, the experiments of \citet{Al69},
\citet{Be73}, and \citet{By74} used a counter arrangement 
in the target vessel to define the $\mu$--stops in hydrogen.}  By
using high--purity gas/liquid with small $Z > 1$ contamination the
problem of transfer is minimized.  Additionally a neutron background
generated by the combination of Michel bremsstrahlung and ($\gamma$,
n) reactions is observed.  These photo--neutrons have the 2.2 $\mu$s
lifetime of the $\mu$ stops.  Therefore studies with $\mu^+$ stops,
yielding photo neutrons without capture neutrons, are necessary to
subtract this background.

Also the accurate determination of the neutron detection efficiency is
a difficult problem.  Calibration via the 8.9 MeV neutrons from the
$\pi^- p \rightarrow \gamma n$ reaction is helpful, but careful
simulations of neutron interactions in counters, target, etc., are
necessary.

Note that the deuterium experiment using the neutron method is
especially challenging.  First the deuterium neutrons form a
continuous distribution peaking at 1.5 MeV.  Second a large background
is produced by $d d$ fusion following $d \mu d$ formation.

\subsubsection{Radiative capture}
\label{s: radcapt}

RMC on H$_2$ has a yield per muon stop of $\sim 10^{-8}$ and a
continuum gamma--ray spectrum with $E_{\gamma} \leq$ 99.2 MeV.  The
first measurement of RMC on H$_2$ was recently accomplished at 
TRIUMF \citep{Jon96, Wr98}.

The experiment detected photons from $\mu$ stops in liquid H$_2$.  An
ultra--pure muon beam ($\pi / \mu = 10^{-3}$) was directed into a
liquid H$_2$ target via a scintillator telescope.  The target flask
and vacuum jacket were constructed from Au and Ag in order to ensure
wall stops in high--Z materials.  The flask contained pure hydrogen
with a D$_2$ contamination of about 1 ppm and a $Z > 1$ contamination
of less than 1 ppb.  The $\gamma$--ray detector was a high acceptance,
medium resolution, pair spectrometer and comprised a lead cylinder for
$\gamma$--ray conversion, cylindrical multiwire and drift chambers for
$e^+ e^-$ tracking, and axial B--field for momentum analysis. The
photon detection efficiency was $\epsilon \Omega \sim 10^{-2}$ and
photon energy resolution was $\Delta E / E \sim 10\%$.

The major difficulty in RMC on H$_2$ is the tiny yield.  Consequently
photon backgrounds from pion stops in liquid hydrogen, muon capture in
nearby materials, and $\mu$ decay are dangerous.

Pion stops in liquid H$_2$ undergo both charge exchange, $\pi^- p
\rightarrow \pi^o n$, and radiative capture, $\pi^- p \rightarrow
\gamma n$.  The resulting photons comprise a Doppler spectrum of
55--83 MeV and monoenergetic peak at 129 MeV, thus endangering the
region of interest for RMC on H$_2$.  \citet{Wr98} suppressed this
background via an ultra--high purity muon beam, prompt photon timing
cut, and the difference in the $\pi$/$\mu$ ranges.  Additionally the
authors determined the residual background from the 55--83 MeV photons
via the residual signal from the 129 MeV $\gamma$--ray peak.

The bremsstrahlung of electrons from $\mu$--decay yields a continuum
background with $E_{\gamma} < m_{\mu} / 2$.  It prevents the
measurement of the RMC spectrum below 53 MeV and threatens the
measurement of the RMC spectrum above 53 MeV, because of the finite
resolution of the pair spectrometer.  \citet{Wr98} designed their
spectrometer to minimize the contribution of the high--energy tail in
the response function.  Additionally the remaining background from
Michel bremsstrahlung was measured by stopping a $\mu^+$ beam in
liquid H$_2$, yielding the photon background from $\mu$ decay without
the photon signal from $\mu$ capture.

In addition the backgrounds from $\mu$ stops in extraneous materials
and $\mu$ transfer to target impurities were minimized by (i) using
high--Z materials in the target vicinity and (ii) using high isotopic
purity H$_2$ as the target material.  Also ancillary measurements were
employed to determine the photon backgrounds from accelerator sources
and cosmic--ray interactions.

The experiment recorded $397 \pm 20$ photons with energies $E_{\gamma}
>$ 60 MeV and times $t >$ 365 ns.  After subtracting the backgrounds
from $\mu$--decay bremsstrahlung ($48 \pm 7$), Au/Ag radiative $\mu$
capture ($29 \pm 11$), and other sources, a total of $279 \pm 26$
photons from RMC in liquid H$_2$ was obtained.  The resulting partial
branching ratio for RMC on liquid H$_2$, {\it i.~e.} the integrated
photon spectrum for $k > 60$ MeV divided by the sum of the muon decay
and capture rates, was R$_{\gamma}$$( k > 60 MeV ) = ( 2.10 \pm 0.21 )
\times 10^{-8}$.  Note this value corresponds to the particular
occupancy of the $\mu$p spin--states in the TRIUMF experiment, {\it
i.e.}\ with $t >$ 365 ns at liquid H$_2$ densities.

\subsection{Comparison of experiment and theory}
\label{s: comp expt}

In Table \ref{t: hydrogen summary} we summarize the results of the
various measurements of muon capture on hydrogen.  For each experiment
we have shown the density of the target $n/n_0$, relative to liquid
hydrogen, the approximate time delay from muon stop until counting
starts $\Delta t$, the time averaged proportion of singlet, ortho and
para states relevant for the experiment, and the rate obtained under
these conditions.  Note that the singlet/ortho/para ratios depend on
the muon chemistry, and the underlying parameters determining that
chemistry, and the delay time $\Delta t$. In Table
\ref{t: hydrogen summary} we used a $p \mu p$ molecular formation rate
$\Lambda_{p \mu p} = 2.5 \times 10^6$~s$^{-1}$, a ratio of
ortho-state formation to para-state formation of $240$:$1$
\citep{Fa99, Po76, Ze59},  an ortho-to-para transition rate $\Lambda_{op}
= 4.1 \times 10^4$~s$^{-1}$ \citep{Ba81a}, and gamma factors of $2
\gamma^{ortho} = 1.009$ and $2 \gamma^{para} = 1.143$ 
\citep{Ba82}. Furthermore the rates tabulated correspond to 
different experimental conditions and so are not directly comparable.

\begingroup
\begin{table*}[tpb]
\caption{Summary of world data for OMC and RMC on hydrogen. The 
columns correspond to the target density in units of liquid hydrogen
density $n_o$, the time delay between $\mu$ stop and start of
counting, the effective, time averaged, singlet/ortho/para ratio
corresponding to the particular experimental conditions, the capture
rate corresponding to these ratios, and the value of $g_p$ implied
using the calculation described in the text. For RMC the value given
in the rate column corresponds to $R_\gamma(k>60 MeV)$, the partial
branching ratio obtained by integrating the photon spectrum above
$k=60 MeV$ and dividing by the sum of the $\mu$ decay and capture
rates. Note that the ratios of molecular states in the S:O:P column
depend on the parameters, e.~g. $\Lambda_{op}$, used, and are given
only for those experiments that measure the neutron or gamma yield. For
the Saclay experiment \citep{Ba81a} the relationship between the
measured rate and the ortho capture rate depends on details of the
experiment and is given in the original paper.}
\label{t: hydrogen summary}
\begin{ruledtabular}
\begin{tabular}{lccccc}
 & & & & & \\ Ref.  & n/n$_o$ & $\Delta t$($\mu s$) & S:O:P & rate
($s^{-1}$) & $g_p(-0.88 m_\mu^2)$\\ & & & & & \\
\hline
 & & & & & \\
~~~~~~ OMC & & & & &\\
\citet{Hi62}		 & 1.0     & 0.0 & 0.15:0.77:0.07 & 
$420 \pm 120$  & $19.5 \pm 11.6$ \\
\citet{HiD62}		 & 1.0	   & 0.0 & 0.15:0.77:0.07 & 
$428 \pm 85$	  & $18.7 \pm 8.2$\\
\citet{Be62}		 & 1.0	   & 0.0 & 0.15:0.77:0.07 & 
$450 \pm 50$	& $16.4 \pm 4.9$\\
\citet{Bl62}		 & 1.0	   & 1.0 & 0.01:0.88:0.11 & 
$515 \pm 85$	& $6.3 \pm 8.7$\\
\citet{Ro63}		 & 1.0	   & 1.2 & 0.01:0.88:0.12 & 
$464 \pm 42$	& $11.4 \pm 4.2$\\
\citet{Al69}	         & 0.014   & 0.9 & 1.00:0.00:0.00 & 
$651 \pm 57$    & $11.0 \pm 3.8$\\
\citet{By74} 	         & 0.072   & 1.4 & 1.00:0.00:0.00 & 
$686 \pm 88$	& $8.7 \pm 5.7$\\
\citet{Ba81a}(original $\tau_+$)  & 1.0	   & 2.5 & -- 
& $460 \pm 20$	  & $7.9 \pm 3.0$\\
\phantom{\citet{Ba81a}}(new $\tau_+$) &    &  &  & $435 \pm 17$	  
& $10.6 \pm 2.7$\\
~~~~~~ RMC & & & & \\
\citet{Wr98}(original theory)   & 1.0    & 0.365  & 0.06:0.85:0.09     & 
$(2.10 \pm 0.21) \times 10^{-8}$           & $12.4 \pm 0.9 \pm 0.4$\\ 
\phantom{\citet{Wr98}}(new theory)  &    &  & & &$12.2 \pm 0.9 \pm 0.4$\\
 & & & & & \\
\end{tabular}
\end{ruledtabular}
\end{table*}
\endgroup

First we stress that the various experiments are sensitive to
different combinations of the $\mu$ atomic and molecular states, {\it
i.e.} the triplet, singlet, ortho and para states.  The situation is
simplest in the H$_2$ gas experiments of \citet{Al69} and \citet{By74}
where muon capture is almost exclusively from the $F = 0$ atomic state
and $p \mu p$ molecule formation is a few percent correction.  However
for experiments in liquid H$_2$, while capture from the ortho state is
the largest, capture from other states is significant.  The precise
blend of states is determined by the delay $\Delta t$ between the muon
arrival time and the counting start time.  The greater the delay the
smaller the contribution from the $F = 0$ atomic state and the larger
the contribution from the para molecular state, as the muon has
additional time to form the $p\mu p$ molecular state and convert from
ortho to para state, in accord with the processes described in
Sec. \ref{s: muchem}. Thus for example the bubble chamber experiments,
where $\Delta t = 0$, have therefore the largest contribution of
singlet atom capture and smallest contribution of para molecule
capture.  In contrast the Saclay lifetime experiment, where $\Delta t
\sim $ 2.5 $\mu$s, had the smallest contribution of singlet atom
capture and, though still dominated by ortho capture, the largest
contribution of capture from the para state.\footnote{$\Delta t \sim $
2.5 $\mu$s is an approximate average value for the Saclay
experiment. In analyzing the data the detailed time structure had to
be explicitly treated.}

In the last column of this table is given the value of $g_p$
corresponding to the experimental rate. To get these numbers the
singlet and triplet capture rates were calculated and combined as
appropriate for the experimental conditions of the individual
experiment. The model used was the standard diagrammatic approach of
\citet{Fe80} updated to the extent that modern values of the
couplings, particularly $g_a=1.267$, were used and form factors were
included. These form factors were taken to be of the form
$f(q^2)=1+q^2 <r^2>/6$ with the values of the rms radii squared
$<r^2>$ taken as 0.59 fm$^2$and 0.80 fm$^2$ for $g_v$ and $g_m$
\citep{Me96} respectively and 0.42 fm$^2$ for $g_a$ \citep{Ah88}. At
the momentum transfer appropriate for OMC these form factors are
typically $0.98-0.96$ so that including them reduces the theoretical
OMC rate by 2-4\%. The constant term appearing in $g_p$, as
in Eq.~(\ref{e: gp-gpiNN}) or (\ref{e: gp-ga}), was included and $g_p$
was parameterized as
\begin{equation}
\label{e: gp-parm}
 g_p(q^2) = R \frac{2m_{\mu}m_N}{(m_\pi^2-q^2)} g_a(0) -
 \frac{m_\mu m_N g_a(0) r_A^2}{3}.
\end{equation}
Here $R$ is not intended to have physical significance, but just be a
conventional way of parametrizing the variation of $g_p$ from the PCAC
value. At the PCAC point, $R=1$ and $g_p(-0.88
m_\mu^2)=8.58-0.45=8.13$.

For RMC form factors were included also using the $1+q^2 <r^2>/6$ form
and the necessary gauge terms were generated via a minimal
substitution. This gives a gauge invariant result which includes most
of the terms found by \citet{Ad66}. For RMC these form factors make
essentially no difference, e.g. $<$ 1\% in the rate corresponding to
the TRIUMF experiment, presumedly because both spacelike and timelike
momentum transfers contribute and, with the linear approximation for
the form factors, tend to cancel.  The $\Delta$ was included as in
\citet{Be87, Be89}.

The OMC results obtained from this calculation are in very good
agreement with other modern OMC calculations such as those of
\citet{An00} and
\citet{Be01b}.  It is interesting to observe however that the values of
$g_p$ quoted in Table \ref{t: hydrogen summary} are 0.3-0.8 higher
than those in a similar table given in \citet{Ba81b}. This can be
traced to two main effects, namely the increase in $g_a$ to its modern
value, 1.254$\rightarrow$1.267, and the use of more modern form
factors which fall somewhat less rapidly with $q^2$ than those used in
\citet{Ba81b}. Both of these effects lead to a larger theoretical 
rate for a given value of $g_p$ and thus, as can be seen from
Fig.~\ref{f: singlet/triplet capture}, to a larger $g_p$ to fit a
given experimental rate.
 
A further comment is required concerning the value of $g_p$ obtained
from the Saclay experiment
\citep{Ba81a, Ba81b}.  In analyzing their data the authors extracted the
capture rate from the difference between their measured value for
$\tau_-$ and the world average value for $\tau_+$.  They used the
world average for comparison because their measured value for
$\tau_+$, although consistent with the world average, had a larger
uncertainty.  Since the publication of \citet{Ba81a} the world average
of $\tau_+$ has changed from $2197.15 \pm 0.07$~ns to $2197.03 \pm
0.04$~ns \citep{Pa00}.  In determining $g_p$ for Table
\ref{t: hydrogen summary} we decided it best to update the $\mu^+$
lifetime in extracting the $\mu^-$ capture rate,\footnote{This was
also done, in a conference proceedings, by one of the members of the
original experimental group. See \citet{Ma84}.} which now becomes $435
\pm 17 s^{-1}$ instead of $460 \pm 20 s^{-1}$.  Therefore the value 
of $g_p(-0.88 m_\mu^2) = 10.6 \pm 2.7$ in Table
\ref{t: hydrogen summary} is one standard deviation larger than
$g_p(-0.88 m_\mu^2) = 7.9 \pm 3.0$ which is obtained from the original
published rate with our theoretical calculation, and even larger than
the value $g_p(-0.88 m_\mu^2) = 7.1 \pm 3.0$ given in the original
paper \citep{Ba81b}.

The values of $g_p(-0.88 m_\mu^2)$ obtained from the five more recent,
electronic, OMC experiments are all in general agreement within their
errors and result in a world average value $10.5 \pm 1.8$. The single
determination with the smallest uncertainty is that of
\citet{Ba81b} which gives $10.6 \pm 2.7$.  Both of these
values are somewhat larger than the PCAC prediction $g_p(-0.88
m_\mu^2) = 8.13$, which includes the constant term, though are
consistent with it at the 1 to 1.3 standard deviation level.  Note
that if one does not update the $\mu^+$ lifetime in the \citet{Ba81b}
analysis the Saclay result is $g_p(-0.88 m_\mu^2) = 7.9 \pm 3.0$ and
world average is $g_p(-0.88 m_\mu^2) = 9.4 \pm 1.9$, both of which are
also consistent with PCAC, but both still about 0.7 larger than the
corresponding numbers, $7.1 \pm 3.0$ and $8.7 \pm 1.9$ quoted in the
original paper, because of the updates to the parameters of the theory.

However the result $g_p(-0.88 m_\mu^2) = 12.4 \pm 0.9 \pm 0.4$, or
$12.2 \pm 1.1$ using the updated theory, from the TRIUMF radiative
capture experiment \citep{Wr98} is 50\% larger than, and clearly
inconsistent, with PCAC. Note that the RMC result is quite consistent
with the world average from OMC or the Saclay result for $g_p$ if one
uses the updated $\mu^+$ lifetime.  However, with the older $\mu^+$
lifetime that \citet{Ba81b} have used, the Saclay result and RMC
result do not overlap within their uncertainties.
  
We can thus summarize the situation in hydrogen as follows. The RMC
result is several standard deviations larger than the PCAC prediction
and clearly inconsistent with it. The OMC results have always been
considered to be in agreement with PCAC, based on the results of the
Saclay experiment. We have seen though that, as a result of updates to
the parameters in the theory and to subsequent measurements of the
$\mu^+$ lifetime, the value of $g_p$ obtained from the OMC result has
increased. The increase is only about one standard deviation, so the
result is still marginally consistent with PCAC. The central value
however is now also high, and actually somewhat closer to the RMC
result than to the PCAC value.

\subsection{Attempts to resolve the discrepancy}

\subsubsection{General comments}
In the previous section we saw that the results for $g_p$ obtained
from RMC were significantly higher than the PCAC prediction and in
definite disagreement. The new interpretation of the OMC results
suggest that they are too high also, being in good agreement with the
RMC result, but perhaps only marginally consistent with PCAC.  Since
the older interpretation of the OMC data seemed to be in good
agreement with PCAC, there has been essentially no consideration of
possible difficulties with OMC. However the discrepancy between the
RMC results and PCAC has generated a lot of discussion and a number of
attempts to find additional effects to explain it, some of which will
be now discussed.

A first question one asks is it logical to have a different $g_p$ for
RMC and OMC?  At a fundamental level the value of $g_p$ should be the
same. However given the present state of analysis, an apparent
difference could arise simply because RMC is much more complicated
than OMC and involves a lot of additional diagrams. Thus if something
is left out of the RMC analysis, a fit to the data using the standard
approach may require a different value of $g_p$ to compensate for the
piece left out. In this view a difference in the values of $g_p$
extracted from OMC and RMC using current theory may reflect something
wrong or missing in the theory rather than a failure of the PCAC
relation of Eq.~(\ref{e: gp-gpiNN}). Note however that while a
difference between the values of $g_p$ extracted from OMC and RMC
might be rationalized this way, this would not explain any differences
between the OMC and PCAC results since for OMC $g_p$ is a parameter in
the most general weak amplitude. It can thus be determined from data
largely independently of the details of the weak capture part, as
opposed to the molecular and atomic part, of the theory.

A second question to ask is what does it take to bring the RMC result
closer to PCAC, and what does that do to OMC? As discussed earlier the
rates for both RMC and OMC depend strongly on the proportion of
singlet versus triplet components in the initial state. To get the RMC
result to agree better with PCAC we need to increase the predicted
rate for a given $g_p$, which can be done by increasing the amount of
triplet capture and reducing the singlet capture. Since OMC is
dominated by the singlet rate, this reduces the OMC rate, which also
moves the extracted $g_p$ toward the PCAC value. However the
sensitivity of OMC and RMC to increasing triplet is different, and
thus it becomes difficult to completely fix one without destroying the
agreement of the other.
\footnote{To get simultaneous agreement of both the TRIUMF RMC
result and Saclay OMC result with PCAC a larger increase in triplet
occupancy is needed for RMC than for OMC.  Since the TRIUMF RMC
experiment detected the majority of its photons with muon capture
times less than a few $\mu$s, whereas the Saclay OMC experiment
detected the majority of its electrons with muon decay times more than
a few $\mu$s, in principle such circumstances are possible. For
example a long-lived, i.e. $\sim$1~$\mu$s lifetime , triplet atom
component could increase triplet occupancy for the TRIUMF RMC
experiment much more than for the Saclay OMC experiment. However, such
a long-lived triplet atom is completely at odds with our current
knowledge of muon chemistry in liquid hydrogen.}

In any case, in the subsequent paragraphs we will discuss several
aspects of the muon chemistry which change the average
singlet/ortho/para ratio for each experiment and thus in principle
affect the comparison of the results with PCAC. We will also discuss
two other suggestions, dealing specifically with the RMC calculation
which have been put forward as potentially resolving the discrepancy
implied by the RMC results.

\subsubsection{Value of the ortho-para transition rate $\Lambda_{op}$}

\begin{figure}
\begin{center}
\epsfig{figure=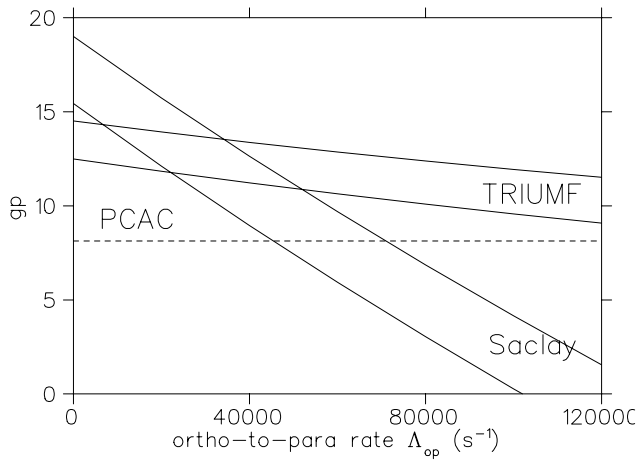, width=6cm,angle=90}
\caption{The region in the \protect{$g_p$ - $\Lambda_{op}$} plane
which is allowed by the TRIUMF RMC experiment \protect{\citep{Wr98}}
and the Saclay OMC experiment \protect{\citep{Ba81a}}. The dashed line
indicates the PCAC value, $g_p^{PCAC}=8.13$, as defined in
Eq.~(\ref{e: gp-ga}), which includes the constant term. Details of the
model used are described in Sec.~\ref{s: comp expt}. Note that in
determining the error bands the uncertainty included in the
experimental results due to the uncertainty in $\Lambda_{op}$ has not
been included, as the results are being plotted against
$\Lambda_{op}$.}
\label{f: gpvslop}
\end{center}
\end{figure}

As has been noted, the experimental combination of singlet and triplet
states is important, and this is determined by the various transition
rates governing the chemical processes the muon undergoes between
stopping and capture. The least certain of these rates is
$\Lambda_{op}$, the ortho-para transition rate. There is a single
theoretical value $\Lambda_{op} = (7.1 \pm 1.2) \times 10^4 s^{-1}$
\citep{Ba82} which however is nearly a factor of two
larger than the experimental value $(4.1 \pm 1.4) \times 10^4 s ^{-1}$ 
\citep{Ba81b, Bar82},\footnote{This group also 
obtained a value $7.7 \pm 2.7 \times 10^4 s^{-1}$ via a measurement of
the electron time spectrum. See \citet{Ma82}.} though another
experiment is in progress
\citep{Ar95}. A larger value of $\Lambda_{op}$
increases the relative amount of para state, and thus increases the
amount of the triplet component contained in the initial state. Figure
\ref{f: gpvslop} shows the region in the $g_p$ - $\Lambda_{op}$ plane
allowed by the TRIUMF RMC experiment \citep{Wr98} and
the latest OMC experiment \citep{Ba81a}.

For these experiments an increase in the value of $\Lambda_{op}$,
which leads to a greater triplet component in the initial state,
increases the predicted RMC rate and decreases the OMC rate for a
given value of $g_p$. In the vicinity of the PCAC value of $g_p$, the
RMC rate increases and the OMC rate decreases with increasing
$g_p$. This means that for fixed value of the rate the extracted value
of $g_p$ decreases with increasing $\Lambda_{op}$ for both OMC and
RMC, as can be seen from the figure. However the Saclay OMC experiment
is much more sensitive to $\Lambda_{op}$ than the TRIUMF RMC
experiment.

The updates and improvements we have made in the theory, which raise
the OMC band in this figure, make it a little easier to accommodate a
larger value of $\Lambda_{op}$ and still have consistency between OMC
and RMC. In fact now a modest increase in $\Lambda_{op}$ from the
experimental value which has normally been used improves the agreement
of both OMC and RMC with PCAC while keeping them consistent. However
increasing $\Lambda_{op}$ enough to get agreement of OMC with PCAC or
increasing it further to the theoretical value still leaves RMC
several standard deviations above the PCAC prediction.  Increasing it
sufficiently to produce the PCAC value of $g_p$ from the RMC data
would lead to a catastrophic disagreement between the Saclay OMC
experiment and the PCAC prediction.  Thus while one can envision a
value of $\Lambda_{op}$ which slightly improves the agreement with the
PCAC prediction for both OMC and RMC, there still appears to be no
value which will simultaneously result in good agreement with PCAC for
both cases.

\subsubsection{Admixtures of a $J=3/2$ ortho-molecular state}

\begin{figure}
\begin{center}
\epsfig{figure=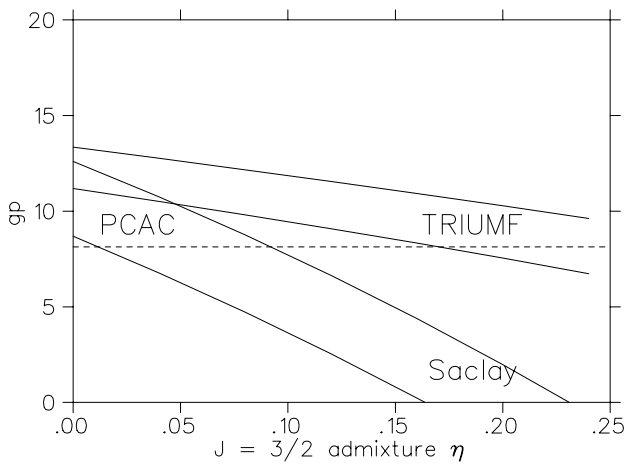, width=6cm,angle=90}
\caption{The region in the \protect{$g_p$ -- $\eta$} plane which is
allowed by the TRIUMF RMC experiment \protect{\citep{Wr98}} and the
Saclay OMC experiment \protect{\citep{Ba81a}}. The dashed line
indicates the PCAC value, $g_p^{PCAC}=8.13$, as defined in
Eq.~(\ref{e: gp-ga}), which includes the constant term. Details of the
model used are described in Sec.~\ref{s: comp expt}. $\Lambda_{op}$
has been taken as $4.1 \times 10^4 s^{-1}$ and the uncertainty in its
value has not been included in the plotted error bands. }
\label{f: gpvsxi}
\end{center}
\end{figure}

It was pointed out in the very early work of \citet{We60} that the
ortho state could also include a component of total spin angular
momentum $S=3/2$. Subsequent theoretical calculations
\citep{Ha64a, Ha64b, We64, Ba82} seemed to indicate that this 
component was zero or very small. Nevertheless if there were such a
component it would effectively increase the amount of triplet state in
the initial state.  This has been suggested as a possible explanation
by \citet{An00, An01a}. Their original numbers were not quite right,
due to a misinterpretation of the appropriate initial state in the OMC
experiment, but the idea is worth examining. Figure \ref{f: gpvsxi}
shows the region in the $\eta$-$g_p$ plane allowed by the TRIUMF RMC
experiment and the most recent OMC experiment. Here $\eta$ is the
fraction of the spin $3/2$ state present in the ortho state. The OMC
calculation here differs slightly from that of \citet{An00} by virtue
of the fact that form factors have now been included. The theoretical
prediction, $\eta=0$ corresponding to no additional spin $3/2$
component, gives the most consistent result. The updates to the OMC
theory, which again raise the band from the Saclay experiment, make it
possible to slightly improve both RMC and OMC results with a small
non-zero value of $\eta$. However analogous to the previous case, a
value of $\eta$ which makes OMC agree with PCAC leaves RMC too high
and a value which makes RMC agree leads to dramatic disagreement of
OMC. Thus this effect does not appear to resolve the discrepancy.

\subsubsection{Direct singlet - para transitions}

As shown in Fig.~\ref{f: mup chemistry} it is in principle possible
for the $\mu p$ singlet atom to make a direct transition to the para
$p \mu p$ molecule. This also would increase the proportion of the
triplet component in the initial state. This transition is governed by
the rate $\Lambda^{para}_{p \mu p}$ which is predicted to be $\simeq
0.75 \times 10^4 s^{-1}$ for liquid hydrogen, which is  two
orders of magnitude smaller than the transition rate to the ortho
state, $\Lambda^{ortho}_{p \mu p} \simeq 1.8 \times 10^6 s^{-1}$,
\citep{Fa99}, and so should be a negligible effect.

\subsubsection{Large $\Delta$ resonance effects}

In a recent calculation of RMC in a Lagrangian based model, described
in Section \ref{s: RMCmodel} above, \citet{TK01} argue that the
discrepancy is at least partially resolved by a contribution coming
primarily from the $\Delta$. The first observation to be made however
is that using their best parameters the $\Delta$ contribution they
obtain is essentially the same size as the contribution obtained
earlier \citep{Be87, Be89} which was already included in the analysis
of the TRIUMF RMC experiment. Other values of the parameters, but
still within the region allowed by other experiments, can raise this
$\Delta$ contribution by 3-4\% in the most important part of the
spectrum, where the discrepancy is more than 40\%.  It would appear,
as they in fact note, that their primary results really basically
agree with previous ones and so do not provide an explanation of the
discrepancy. They go on to observe however that one can change one of
the parameters corresponding to off-shell properties of the $\Delta$
and thus fit the experimental spectrum. That is certainly possible, as
one can usually arbitrarily change one parameter to get a fit to
another. However it requires the $\Delta$ contribution to be an order
of magnitude larger than what they consider most reasonable.
Furthermore one would have to reconcile such an explanation with the
general result that changing off-shell properties can not change
physically measurable quantities. See e.~g.\ \citet{Fe98b, Fe00,
FeS00, Sc01} and references cited therein. Although it is clear that
there are uncertainties in any attempt to include a $\Delta$ in a
calculation such as this, their best estimate of the $\Delta$
contribution is quite consistent with the (small) contribution found
in the two previous independent calculations which included the
$\Delta$ \citep{Be87, Be89, Be01b}. An enhancement obtained by varying
off-shell parameters which is large enough to explain the full
discrepancy does not seem too likely.\footnote{There was also a
suggestion by \citet{Ch98, Ch99} of an additional term in the
Lagrangian which could contribute, but it was shown that such a term
was already included in standard calculations, and so was not an
additional effect \citep{Fe98a, Sm98}. }

\subsubsection{Other possibilities}

Another suggestion was made by \citet{Be01b}, namely that the
discrepancy between PCAC and the RMC results could be resolved by a
series of small effects in RMC which happen to add up. Such effects
might include for example those discussed above or perhaps higher
order loop effects which have not been included, radiative
corrections, or variations in some of the parameters. Clearly one can
not test this suggestion without calculating all such effects, which
has not been done. We note however that in so far as the molecular
effects we have discussed are concerned, they all serve to increase
the amount of triplet capture, at least for current liquid H$_2$
experiments.  However in each of these cases, or in some combination
of them, as they are somewhat equivalent, it appears that changes
sufficient to make RMC agree, lead to drastic disagreement for OMC.

Another particular effect they proposed was an isospin breaking
effect. This proposal was based on the observation that the RMC rate
changed quite a bit when the charged pion mass in the pion propagator
was replaced by the neutral pion mass, thus supposedly showing a
sensitivity to the kinds of presumedly electromagnetic isospin
breaking effects responsible for the pion mass splitting. This
sensitivity clearly originates in the fact that some of the momentum
transfers in RMC can get quite close to the pion pole. Thus changing
the position of the pole, even by a few MeV, can change the results
significantly. One should note however that the pion exchanged is in
fact a charged one. Thus in an ideal theory which properly
incorporated all electromagnetic effects, after all the
renormalizations, and after all the terms which lead to mass
splittings were included, the pion mass appearing in the propagator
would in fact be the physical charged pion mass, not the neutral
one. Thus this particular argument for sensitivity to isospin effects
in spurious. In such a complete theory one would probably expect terms
in the numerator of the amplitude proportional to the pion mass
splitting, and likewise the neutron-proton mass splitting. But the
scale in the denominator would likely be something like the pion mass,
so these terms should be small.  Nevertheless, there could be
accidental enhancements, so this is a calculation which should be
done.

One might also want to consider radiative corrections to both OMC and
RMC. Typically these might be expected to be \cal{O}($\alpha/\pi$),
and this was what was found by the one such calculation we are aware
of \citep{Go72}. This is too small an effect to be relevant now, but
might be important in the interpretation of future, high precision,
OMC experiments.

\subsection{Summary and outlook for hydrogen}

It is thus possible to summarize the situation in hydrogen as follows.
{}From the theory side, all calculations of OMC in hydrogen are in
essential agreement, as they must be as all couplings and form factors
except $g_p$ are fixed by generally well accepted principles such as
CVC or by other experiments. The several calculations of RMC, which in
principle could differ, e.g. via the extra loop contributions included
in the ChPT calculation relative to the standard diagrammatic
calculation, are in fact in general agreement at the few percent
level.

On the experimental side, all modern OMC experiments now generally
lead to values of $g_p$ which are consistent with each other within
the errors.  The perception that OMC agrees with PCAC has changed
somewhat however with updates to the theoretical calculations and to
the $\mu^+$ lifetime. These updates have increased the world average
from OMC by only about one standard deviation, but result in a value
which is now higher than the PCAC value by about 1.3 standard
deviations and which is in fact in better agreement with the RMC
result than with PCAC. The one existing RMC experiment gives a value
of $g_p$ which is 50\% larger than the PCAC prediction and with
uncertainties small enough that it is in clear disagreement with that
prediction. Though there have been a number of suggestions and
attempts to explain the discrepancy indicated by the RMC result, none
so far have been successful, and the difference remains a puzzle.

So what needs to be done next to resolve this situation?  Specifically
a high precision measurement of the OMC rate in gaseous hydrogen would
be very helpful. By using gas rather than liquid the muon chemistry
uncertainties are greatly reduced.  Indeed a new measurement of the
singlet capture rate for OMC in H$_2$ gas at $P \simeq 10$~bar using
the lifetime method is currently under way at PSI \citep{Ka00}.  The
goal is a precision of about 1\% in the singlet rate and about 6\% in
the coupling $g_p$.  Note that the experiment will use a novel
hydrogen time projection chamber that enables direct monitoring of
muon stops, muon transfer and $\mu$-atom diffusion.

Also useful  would be a new measurement of OMC in liquid hydrogen with
significantly improved uncertainties. Such a measurement might
distinguish between two possible scenarios. In one, both OMC and RMC
may give values of $g_p$ which are too high and inconsistent with
PCAC, which would suggest that the problem is something common to OMC
and RMC, perhaps some difficulty with our understanding of the
molecular or atomic effects. In the other OMC might clearly agree with
PCAC and disagree with RMC, which would suggest that the problem is
something wrong with our understanding of RMC.

It would also help to confirm some aspects of the muon chemistry,
particularly the value of $\Lambda_{op}$ which is important in
determining the initial muonic state and for which the existing
experimental \citep{Ba81b, Bar82} and theoretical \citep{Ba82} values
differ by almost a factor of two. Such an experiment is under way
\citep{Ar95} and should have results soon.

Another alternative is to try to measure some other quantity which is
especially sensitive to $g_p$. One such quantity is the triplet
capture rate for OMC in H$_2$. Here the central problem is that the
triplet rate is about thirty times smaller than the singlet rate.
Further during the time evolution of the $\mu$p system following
atomic capture, the singlet state is always present, so the singlet
capture dominates the OMC rate at all times.  However some suggestions
have surfaced for isolating the triplet capture from the much larger
singlet capture.  One possibility suggested by \citet{De83} is to
measure the time dependence of the neutron polarization. Neutrons
following capture from the singlet state have polarization -1, while
those from the triplet state have a polarization depending on the weak
dynamics and $g_p$.  Measuring as a function of time allows one to
determine a relative neutron polarization, rather than an absolute
one, which is more difficult.  Such a measurement of the neutron
polarization could, in principle, help in identifying the triplet
capture and isolating the $g_p$ contribution.  Another suggestion
\citep{Ba83} notes that the triplet $\mu$p atoms, unlike the singlet
$\mu$p atoms, will precess in a magnetic field.  Consequently the time
spectra of decay electrons or capture neutrons will be modulated by
this precession when muons are stopped in H$_2$ gas at low pressures,
where triplet atoms are relatively long-lived, thus enabling isolation
of triplet capture.  While interesting and worth reconsidering, such
experiments are of course extremely difficult.

For RMC, it is clear that a new experiment in liquid hydrogen, where
triplet capture is largest, would also be both interesting and useful
to check the TRIUMF result.  Further a new RMC experiment in gaseous
hydrogen, where singlet capture is largest, would be very
interesting. However an RMC experiment in gaseous hydrogen is
extraordinarily difficult since the muon stopping rate is much less
and the singlet RMC rate is much smaller than the triplet RMC rate.

Since for RMC the triplet rate is by far the largest, it is possible,
by utilizing both a gas target, where capture is primarily from the
singlet state, and a liquid target, where there can be significant
captures from the triplet state, to obtain as the dominant
contribution to the RMC rate either singlet capture or triplet
capture.  This is different from the situation for OMC since there the
singlet rate is the largest and so in both gas and liquid experiments
the singlet capture will dominate the total rate.

In principle it is possible to measure spin degrees of freedom in RMC
as well and a suggestion of this type has also been made recently by
\citet{An02}. Such experiments would be extremely difficult because of
the low rates and the necessity of either starting with a polarized
muonic atom or measuring the polarization of the outgoing photon, but they
are more sensitive to $g_p$ than the rate and are sensitive to a
different combination of amplitudes and so give independent
information on the process \citep{Fe75}.

\section{Muon capture in deuterium}
\label{s: deuterium}

We next want to consider OMC on deuterium, {\it i.e.} the reaction
$\mu + d \rightarrow n + n + \nu$, which involves many of the same
ingredients as OMC on the proton, but also some additional
complications.

\subsection{Theory of muon capture in deuterium}
\label{s: D theory}

Since there are now two nucleons involved in the capture process,
several new ingredients appear which were not relevant for capture on
the proton. In particular the process now depends on the two nucleon
interaction, which is reflected in the properties of the initial state
deuteron, such as the wave function, percentage D state, etc., and
also in the final state interaction between the two outgoing neutrons.
Furthermore now meson exchange corrections (MEC's) can contribute.

There have been many calculations of the OMC rate in
deuterium.\footnote{Some representative earlier ones include those of
\citet{Wa65b}, \citet{Pa72}, \citet{Tr72}, \citet{So74}, \citet{Dog79}, 
\citet{Iv79b}, \citet{Ng75}, \citet{Ho76}, \citet{Sv78}, \citet{Sv79}, 
\citet{Sv80}, \citet{Go82}, and \citet{Mi73,Mi83}.} Some of 
the more recent ones include \citet{Ta90}, \citet{Ad90},
\citet{Do90,Do91}, \citet{Mo92},
\citet{Hw99} and \citet{An01b}. 
Originally the main emphasis was on using this process to extract the
neutron-neutron scattering length $a_{nn}$.  In the kinematics in
which the two neutrons have low relative momentum, the neutron
spectrum peaks at a value several orders of magnitude larger than that
for kinematics with large relative momentum. Such a measurement of
this low energy neutron spectrum, while extremely difficult, is still
interesting, but perhaps less so than originally in view of the
advances in our knowledge of the nucleon-nucleon force from other
sources. However the sensitivity to $a_{nn}$ means that it must be
well known to have any hope of using this process to extract $g_p$, at
least if one focuses on the region of low relative neutron momentum
where the rate is largest.

Furthermore the total OMC rate, which is dominated by the doublet
rate, is only moderately sensitive to $g_p$. A 50\% change in $g_p$
produces only about a 10\% change in the rate \citep{Do91}.

The total rate is also affected by MEC's which have been calculated
using specific diagrams involving pion or rho exchange \citep{Do90} or
using a more elaborate hard pion Lagrangian involving $\pi$'s,
$\rho$'s, $\omega$'s and $a_1$'s \citep{Iv79b, Ad90, Ta90}. Generally
these MEC's are about a 10\% correction to the total rate.

The results of modern calculations now generally agree. For example
the doublet capture rate obtained by \citet{Ta90} is 398-400 s$^{-1}$
of which 31-33 s$^{-1}$ comes from MEC's and where the range
corresponds to different nucleon-nucleon potentials. \citet{Do90}
obtained 402 s$^{-1}$ and \citet{Ad90} found 416$\pm$7 s$^{-1}$.

It it interesting to note that since this is a three body final state,
just like RMC in the proton case, it is possible to get timelike
momentum transfers approaching $+m_\mu^2$ and thus get near the pion
pole, where in principle the sensitivity to $g_p$ is enhanced. To our
knowledge sensitivity to $g_p$ in that region has not been
investigated, although \citet{Mi83} considered some effects of
timelike form factors and \citet{Do90} and \citet{Go82} showed that
meson exchange currents are an important effect there. Unfortunately
this kinematic region occurs when the two neutrons go out back to back
and the $\nu$ has relatively low energy, which is a region which is
strongly suppressed relative to the region where the neutron-neutron
final state scattering is important, thus making experiments
difficult.

There is a very strong hyperfine effect in the capture rate on
deuterium. The quartet rate is only 10-15 s$^{-1}$
\citep{Do90, So74, Ho76}. It has been suggested \citep{Mo92, Mo93, Do91} 
that the ratio of quartet to doublet rate is much more sensitive to
$g_p$ than the doublet rate and thus that a measurement of the quartet
rate, or directly of the ratio, would give a better value of $g_p$.

In principle RMC in deuterium should also be an interesting process,
giving information on $g_p$. However one might expect that MEC's would
be even more important here, since at the least the photon would have
to couple to all of the intermediate states which generate the MEC's
in OMC. To our knowledge there have not been calculations of RMC in
deuterium.

\subsection{Experiments on muon capture in deuterium}
\label{s: D experiment}

A summary of available data for $\mu d$ total capture rates is given
in Table \ref{tab: dobservables}.  It lists experiments with mixed
H$_2$/D$_2$ targets by \citet{Wa65c} and \citet{Be73} and pure D$_2$
targets by \citet{Ba86} and
\citet{Ca89}. A H$_2$/D$_2$ target is helpful in reducing
the neutron background from $d \mu d$ fusion.  The \citet{Ba86} 
result was obtained using the `lifetime method' (see
Sec. \ref{s: lifetimemeth}) and the other results were obtained using
the `neutron method' (see Sec. \ref{s: neutronmeth}).  Note the
neutron measurements give partial rates for muon capture with neutron
energies $> E_{thres}$ and require some input from theory to
determine the total rate of $\mu d$ capture.

\begin{table*}[tbp]
\caption{Summary of world data for the $\mu$d doublet capture rate. The 
columns correspond to the target density in units of liquid hydrogen
density $n_o$, target temperature, the time delay between $\mu$ stop
and start of counting, the neutron energy threshold, the deuterium
concentration, and the doublet capture rate. }
\label{tab: dobservables}
\begin{ruledtabular}
\begin{tabular}{lcccccc}
 & & & & & & \\ Ref.  & n/n$_o$ & T(K) & $\Delta t (\mu s)$ &
E$_n$(MeV) & $c_{d}$ & rate ($s^{-1}$) \\ & & & & & & \\
\hline
 & & & & & & \\
\citet{Wa65c}	 & 1.0	   & 18	  & 0.8	    &	1.4	     & 0.0032 
& $365 \pm 96$   \\
\citet{Be73}	 & 0.013   & 293  & 0.5	    &	1.5	     & 0.05   
& $445 \pm 60 ^a$   \\
\citet{Ba86}	 & 1.0	   & 18	  & 2.5	    &	 -	     & 1.00   
& $470 \pm 29$   \\
\citet{Ca89}	 & 0.04	   & 40	  & 1.9	    &	1.5, 2.5     & 1.00   
& $409 \pm 40$   \\
 & & & & & & \\
\end{tabular}
\end{ruledtabular}
$^a$ Note that the \citet{Be73} doublet capture rate was obtained assuming
a much faster hyperfine depopulation rate than is consistent with
our current understanding of muon chemistry in H$_2$/D$_2$ mixtures.
See text for details.
\end{table*}

The measurement by \citet{Wa65c} used a liquid H$_2$ target with a
0.32\% D$_2$ admixture.  Under these conditions the $\mu$ capture is
dominantly from $p \mu d$ molecules, since muon transfer yields $\mu$d
atoms in roughly 20 ns and proton capture yields $p\mu d$ molecules in
roughly 200 ns. See Sec. \ref{s: pmud} for details.  Consequently the
neutron signal in \citet{Wa65c} involves a superposition of capture on
protons, deuterons and $^3$He, the latter being produced via $p d$
fusion in $p \mu d$ molecules.  Note that in analyzing their results the
authors assumed a statistical mixture of the $\mu d$ spin--states in
the $p \mu d$ molecules, as would be expected if hyperfine
depopulation by spin--flip collisions is much slower than $p \mu d$
formation.  Also the different time dependence of $\mu$$^3$He neutrons
and $\mu$d neutrons was employed in separating their
contributions.\footnote{The neutrons from $\mu$d capture build--up
quickly with the nanosecond timescale of the $\mu$ transfer process.  The
neutrons from $\mu$$^3$He capture build--up slowly with the microsecond
timescale of the $p d$ fusion process.}  Their final result for
doublet capture was $\Lambda_{1/2} = 365 \pm 96$s$^{-1}$.

The measurement by \citet{Be73} used a 7.6~bar H$_2$ gas target with a
5\% D$_2$ admixture.  Under these conditions the $\mu$d atom formation
is rapid but $p \mu d$ molecule formation is negligible, thus avoiding
the complication of contributions from protons, deuterons, and $^3$He.
Strangely they found that under the assumption of a statistical mix of
the $\mu$d spin--states their extracted rate for doublet capture was
1100--1500~s$^{-1}$, {\it i.e.} three times the theoretical value.
Therefore, in order to understand their results, they postulated a
much faster hyperfine depopulation rate than assumed by \citet{Wa65c}.
Thus assuming a pure doublet mix of $\mu$d spin states the authors
obtained a doublet rate of $445 \pm 60$~s$^{-1}$.  However the
subsequent studies of hyperfine depopulation in H$_2$/D$_2$ systems,
{\it e.g.} \citet{Br81}, have failed to support this claim and
therefore the \citet{Be73} result is extremely puzzling.

The more recent experiments of \citet{Ca89} and \citet{Ba86} were
conducted in pure deuterium.  In these circumstances the $\mu d + d$
collision rate is sufficient to fully depopulate the F$_+$ state. Note
any $d \mu d$ formation is followed by prompt fusion and muon
recycling into $\mu$d atoms.  The resulting rates for doublet capture
were $\Lambda_{1/2} = 409 \pm 40$~s$^{-1}$ from \citet{Ca89}, using a
gas target and the neutron method, and $\Lambda_{1/2} = 470 \pm
29$~s$^{-1}$ from \citet{Ba86}, using a liquid target and the lifetime
method.  Using the calculation of \citet{Do91}, the \citet{Ca89}
result implies $g_p = 8 \pm 4$ and the \citet{Ba86} result implies
$g_p = 2 \pm 3$, which are in disagreement at the level of about 1-1.5
$\sigma$. Although one experiment used a gas target and the other
experiment used a liquid target, in both cases it is believed that
capture is from the doublet atomic state, so that muon chemistry in
pure deuterium is unlikely to account for the different results.
However both experiments were challenging and faced different
experimental difficulties.  Therefore at present we conclude the two
results simply indicate a rather large uncertainty in the value of the
coupling $g_p$ extracted from the $\mu$d system.

There has also been one measurement of the high energy neutron
spectrum following muon capture in deuterium \citep{Le87} using
neutron time of flight methods. In this kinematic situation the
neutrons come out back to back, so the direction is well defined, but
there is no start signal for time of flight measurements. This group
used a novel technique to overcome this, by using the detection of one
of the neutrons in a counter placed close to the target as a start
signal for the time of flight measurement of the other neutron in a
counter placed further away from the target. The results suggest the
enhancement due to meson exchange corrections found by \citet{Go82}
and are in qualitative agreement with the calculations of \citet{Do90}
at least as far as the shape of the spectrum is concerned, but are not
precise enough to give any real information on $g_p$. In view of the
potentially increased sensitivity to $g_p$ however, such a measurement
of high energy neutrons is worth re-examining.

A new higher precision measurement of the $\mu$d doublet capture rate
would be useful in resolving the possible discrepancy between the
current results of \citet{Ba86} and \citet{Ca89}.  However in
deuterium, unlike in hydrogen, the uncertainties in calculating the
contributions of exchange currents may ultimately limit the
achievable accuracy in extracting $g_p$.  We also remark that in pure
deuterium and hydrogen-deuterium mixtures the hyperfine depopulation
rate is much slower than in pure hydrogen.  Consequently measuring
either the quartet rate or hyperfine dependence in $\mu$d capture,
which are strongly dependent on $g_p$, may be experimentally easier in
$\mu$d.  It would be worthwhile to reconsider this source of
information on $g_p$, as has been suggested by \citet{Mo92} and
\citet{Do91}.

\section{Muon capture in $^3$He}
\label{s: mcHe3}

We now want to consider both OMC and RMC on $^3$He. Since it is
possible to break up the final state there are several processes
possible,
\begin{equation}
\mu^- + ^{3}\!\!He \rightarrow ^{3}\!\!H + \nu
\label{e: triton chan}
\end{equation}
\begin{equation}
\mu^- + ^{3}\!\!He \rightarrow ^{2}\!\!H + n + \nu
\label{e: breakup1 chan}
\end{equation}
\begin{equation}
\mu^- + ^{3}\!\!He \rightarrow p + n + n  + \nu
\label{e: breakup2 chan}
\end{equation}
with analogous processes involving a photon in the case 
of RMC.

Of these, most work has been done on the triton final state,
Eq.~(\ref{e: triton chan}). Like the deuteron, since there is more
than one nucleon, the complications of exchange currents must be
addressed. Additionally three-body forces are now possible. However
there has been a tremendous amount of work done on three-body wave
functions in recent years and these wave functions are now extremely
well known. Thus the wave function complications which plague
interpretation of capture in heavier nuclei really do not enter
here. Furthermore the rate is significantly higher than in capture on
protons and deuterons and the atomic and molecular processes which
make the interpretation of capture in hydrogen or deuterium difficult
are not present here. Thus $^3$He is an ideal compromise, and we will
see that one of the best measurements of $g_p$ to date comes from OMC
in $^3$He.

\subsection{Theory of ordinary muon capture in $^3$He}

\label{s: 3He theory}
\label{s: he3theory}
There have been two major approaches to the theory of muon capture in
$^3$He, the elementary particle model (EPM) and the impulse
approximation (IA), supplemented by explicit calculations of meson
exchange corrections. The EPM exploits the fact that the $^3$He -
$^3$H system is a spin $1/2$ isodoublet just like the proton-neutron
system. Thus one can carry over directly the calculations done for
capture on the proton. The couplings and form factors now apply to the
three-body system as a whole, rather than the nucleon. However they
can be obtained in many cases directly from other processes such as
electron scattering on $^3$He or tritium $\beta$-decay or from theory,
i.e., by applying PCAC to the nuclear pseudoscalar form factor.  This
approach has the advantage that it already includes effects of MEC's,
which are buried in the form factors, and that it can be done
relativistically. However it is difficult to relate the nuclear form
factors to the fundamental couplings and form factors of the weak
current, e.~g. $g_p$. Some sort of microscopic model is necessary to
make this connection.

One of the first such EPM calculations was that of \citet{Ki65}. The
idea was later applied to $^3$He by \citet{Be76}, \citet{Fe80},
\citet{Kl81}, and \citet{Kl82}. A more modern and very careful
calculation with particular attention to the form factors and the
various uncertainties was carried out by \citet{Co93}, and their
results were confirmed by \citet{Go00} and by \citet{Ho02}. The final
result is a prediction from the EPM for the statistical capture rate
of $1497 \pm 21$s$^{-1}$ \citep{Co93}, using the standard PCAC value
for $g_p$.

The IA on the other hand treats the capture process as capture on an
essentially free proton. This one nucleon interaction is then expanded
in powers of $1/m$ and used as an effective interaction in a matrix
element between nuclear wave functions. The information specific to
the nucleus comes in only via the nuclear wave functions. Early
applications of this approach applied to $^3$He include \citet{Pe68},
\citet{Ph75}, \citet{Kl81}, and \citet{Kl82}. Again a very 
careful calculation was done by \citet{Co93} using the variational
wave functions of \citet{Ka89} with the result for the rate 1304
s$^{-1}$. This was essentially confirmed by \citet{Ho02} using for the
wave functions momentum space Faddeev solutions for a variety of
potentials. They found however about a $\pm 3\%$ spread among the
results using the various wave functions.  The fact that the IA result
is about 15\% lower than the EPM is characteristic of the IA and
indicates the effect of the MEC's. \citet{Co96} refined these IA
calculations by including explicitly the MEC's, based on a specific
Lagrangian involving $\pi, \rho, a_1, \Delta$ couplings. The end
result, which also included some effects of three-body forces, was
1502 $\pm$ 32 s$^{-1}$. This is in excellent agreement with the EPM
result. Finally, very recently \citet{Ma01} repeated the calculation
of \citet{Co96} using wave functions arising from the Argonne $v_{14}$
or $v_{18}$ potentials with some three-body forces. They also made
some refinements to the weak current and reduced the size of the
uncertainty by fixing the $\Delta$ contribution to the exchange
current corrections via a fit to tritium beta decay. They obtained
1494 $\pm$ 9 s$^{-1}$.

Thus to summarize, the capture rate for OMC on $^3$He is well
determined by both EPM and IA+MEC calculations with an accuracy of
1-3\%. We note that at this level of accuracy radiative corrections
might be important and should be evaluated.

Finally it is worth mentioning that there have also been calculations
of various spin correlations in OMC \citep{Co93, Co94, Co96, Go00,
Ma01}.  Generally these correlations are more sensitive to $g_p$ and
less sensitive to the MEC's than the rate is. There have also been
calculations of some breakup channels
\citep{Wa65a, Ph75, Sk99}.

\subsection{Experiments on OMC in $^3$He}
\label{s: 3He experiment}

The $^3$H channel in OMC on $^3$He yields a muon neutrino with
momentum 100 MeV/c and triton recoil with energy 1.9 MeV. The triton
yield per $\mu$ stop is roughly 0.3\%.  Contrary to muon capture on
complex nuclei, the recoil triton has a relatively high energy and a
relatively small charge, and is therefore directly detectable.  In
Sec. \ref{s: overview rate} below we discuss the current status of the
$^3$He $\rightarrow$ $^3$H capture rate experiments and in
Sec. \ref{s: overview asym} the $^3$He $\rightarrow$ $^3$H recoil
asymmetry experiments.

\subsubsection{$^3$He capture rate experiments}
\label{s: overview rate}
So far several measurements of the $^3$He $\rightarrow$ $^3$H rate
have been performed.  They comprise the early work of \citet{Au67},
\citet{Cl65}, \citet{Fa63}, \citet{Za63}, and recent work of \citet{Ac98}.

The basic method involves counting the numbers of $\mu$--stops and
$^3$He--recoils when a beam of muons is stopped in $^3$He.  In
\citet{Au67} and \citet{Ac98} a $^3$He gas ionization chamber was used
whereas in \citet{Fa63} and \citet{Cl65} a $^3$He liquid scintillator
was used.  One advantage of a liquid scintillation counter is the
higher muon stopping rate in the liquid helium target.  However one
advantage of a gas ionization chamber is the better separation of the
$\mu$ and $^3$He signals via its tracking capabilities.

The recent experiment of \citet{Ac98} employed an $^3$He ionization
chamber with a $15$~cm$^3$ active volume and a 20~Torr gas pressure.
The anode plane was segmented to permit the tracking of the incoming
muons, the isolation of the recoil tritons, and the definition of the
fiducial volume.  The anode strips were read out into flash ADCs for
measurement of the drift time and the energy loss.  Separate triggers
were employed to count the incoming muons or recoil tritons with full
efficiency.

One experimental difficulty is muon--triton pile--up, {\it i.e.}
overlapping of the prompt $\mu^-$ ionization and the delayed $^3$He
ionization.  \citet{Ac98} addressed this problem by both (i) ignoring
pile--up events and extrapolating to $t = 0$ the non pile--up
$\mu$--$^3$H time spectrum, and (ii) keeping pile--up events and using
the anode pulse--shapes for computing $\mu$--$^3$H time differences.
The consistency of (i) and (ii) was helpful in demonstrating the
correct treatment of pulse pile--up.

Another experimental difficulty is background processes.  The 1.9 MeV
triton energy--loss peak is superimposed on backgrounds that include
$^3$H break--up channels, thermal neutron capture, and $\mu$--decay
electrons.  \citet{Ac98} employed various fitting procedures in order
to separate the triton peak and continuum background and estimate the
uncertainties.  Their signal--to--background ratio was roughly 20:1.

The \citet{Ac98} experiment yielded a $^3$He $\rightarrow$ $^3$H rate,
corresponding to a statistical hyperfine initial population, of
$1496.0 \pm 4.0$~s$^{-1}$, with the largest systematic uncertainty
being the subtraction of the background.  The $\pm 0.2\%$ accuracy is
a ten--fold improvement over the earlier experiments of \citet{Fa63}
$1410 \pm 140$~s$^{-1}$, \citet{Cl65} $1467 \pm 67$~s$^{-1}$, and
\citet{Au67} $1505 \pm 40$~s$^{-1}$.  In addition \citet{Ac98}
utilized their time spectrum to establish that transfer from the $F_-$
to the $F_+$ hyperfine state as well as capture from the 2S metastable
state can be safely neglected.\footnote{For the hyperfine transition
rate \citet{Ac98} obtained $(0.006 \pm 0.008)\ \mu$s$^{-1}$ and for the
2S state lifetime they obtained $<50$ ns.}

Taken together the measurement of \citet{Ac98} and the calculation of
\citet{Co96} give $g_p(-0.954 m_\mu^2) = 8.53 \pm 1.54$ which implies
$g_p(-0.88 m_\mu^2) = 8.77 \pm 1.58 = (1.02 \pm 0.18) g_p^{PCAC}$, or
$(1.08 \pm 0.19) g_p^{PCAC}$ if we use the value of $g_p^{PCAC}$ which
includes the constant term.  These results are in nice agreement with
the PCAC prediction for the coupling $g_p$. Here the dominant
uncertainty originates from the theoretical calculation and not the
experimental data.  Clearly the key question is the exact size of the
theoretical uncertainty.  We have quoted the theoretical error of
\citet{Co96} which originates mainly from the $\Delta$ contribution to
the exchange currents.  Subsequently \citet{Ma01} have suggested this
uncertainty can be reduced by fixing some of the parameters
using results from exchange current effects in tritium beta decay.
However \citet{Ho02} found significant theoretical uncertainty arising
from different choices of the nuclear wave functions, though some of
this may be due to the fact that three-body forces were not included
and tuned to give the same binding energies for all choices. Further
efforts on establishing and reducing the theoretical uncertainty in
calculating the rate for $^3$He $\rightarrow$ $^3$H capture are
definitely worthwhile.

Lastly we  note that there have been some recent measurements of
partial capture rates and energy spectra for the breakup channels
following muon capture in $^3$He \citep{Cu92, Kuh94}. Comparisons with a
recent calculation \citep{Sk99} indicate that a good treatment of
final state effects is necessary to describe the data. However at the
present stage these channels give no information on $g_p$.

\subsubsection{$^3$He recoil asymmetry experiments}
\label{s: overview asym}
As discussed for nuclei in general in Sec. \ref{s: asymmetry} below,
the recoil asymmetry in muon capture can be very sensitive to the
coupling $g_p$.  Unfortunately production of a measurable $^3$H
asymmetry requires production of a large $\mu^-$$^3$He polarization,
which has proven to be difficult.

Early attempts to polarize $\mu^-$$^3$He, using a polarized beam on a
unpolarized target and an unpolarized beam on a polarized target, were
unsuccessful.  For example, \citet{So75} obtained a $\mu^-$
polarization of only $\sim$6\% when stopping polarized muons in
unpolarized $^3$He gas and \citet{Ne91} obtained a $\mu^-$
polarization of only $\sim$7\% when stopping unpolarized muons in
polarized $^3$He gas.\footnote{In both cases the polarizations were
smaller than expected from atomic cascade calculations \citep{Re87}.}
The breakthrough was made at LAMPF by \citet{Ba93} by repolarizing the
$\mu^-$$^3$He system with a laser--pumped Rb vapor after the atomic
cascade.

How does repolarization work?  Following the $\mu^-$$^3$He atomic
cascade a ($\mu^-$$^3$He)$^+$ positive ion is produced.  However the
($\mu^-$$^3$He)$^+$ ion is short lived, forming either 
($\mu^-$$^3$He)$^+$$^3$He molecules via collisions with surrounding He
atoms or ($\mu^-$$^3$He)$^+$e$^-$ atoms after collisions with
electron--donor impurities.  The repolarization technique developed
by \citet{Ba93} employs both spin transfer in the dissociation of the
($\mu^-$$^3$He)$^+$$^3$He molecules
\begin{equation}
\label{e: dissociation}
Rb\uparrow + He (\mu^- He)^+ \rightarrow Rb^+ + He + (\mu^- He)^+
e^{-}\uparrow
\end{equation}
and spin transfer in the collisions of the ($\mu^-$$^3$He)$^+$e$^-$ atoms 
\begin{equation}
\label{e: collision}
Rb\uparrow + (\mu^- He)^{+}e^{-}\downarrow \rightarrow Rb\downarrow +
(\mu^- He)^+ e^{-}\uparrow
\end{equation}
where the arrows denote the polarization state of transferred
electrons.  Following the dissociation process of Eq.~(\ref{e:
dissociation}) and collision process of Eq.~(\ref{e: collision}) the
polarization of the electron is shared with the muon via their
spin--spin interaction.
\citet{Ba93} discovered at high Rb$\uparrow$ densities
both processes contribute to muon repolarization on the time scale
of the muon lifetime, and produced a $\mu^-$ polarization in $^3$He
gas of 26\%.

Recently the first measurement of the recoil asymmetry in the $^3$He
$\rightarrow$ $^3$H transition was accomplished by \citet{So98} at
TRIUMF.  The experiment utilized a $^{3}$He ionization chamber to stop
the incoming muons, repolarize the $\mu^-$$^3$He system, and track the
triton recoils. For experimental details see
\citet{Bo97} and \citet{So98}. 
The chamber had an instrumented volume of $140$ cm$^3$ and a gas
pressure of 8 bar.  Anode segmentation assisted the definition of the
fiducial volume and pulse--shape read--out assisted the separation of
the $\mu$/$^3$He pulses.  Running the chamber at high voltage, high
temperature, and high Rb density was a great achievement.

\citet{So98} employed an interesting approach in
determining $\cos{ \theta }$, the direction between the $^3$H recoil
and $\mu$$^3$He polarization.  First the magnitude of $\cos{ \theta }$
was determined from the anode signal width.  If the recoil travels
perpendicular to the anode plane the drift time range and anode pulse
width are large.  If the recoil travels parallel to the anode plane
the drift time range and anode pulse width are small.  Second the sign
of $\cos{ \theta }$ was determined from the anode pulse shape.  For a
recoil traveling towards the anode the early--arriving Bragg peak
yields a fast rise time pulse.  For a recoil traveling away from the
anode the late--arriving Bragg peak yields a slow rise time pulse.

The $\mu$$^3$He polarization was determined by detecting the
$\mu$--decay electrons and using the well known correlation between
the electron momentum and the muon spin in the $\mu \rightarrow e \nu
\bar{\nu}$ decay.  The electrons were detected using a telescope
consisting of wire chambers and plastic scintillators, which permitted
ray tracing into the ionization chamber.  The asymmetry in the
electron counts, on reversing the laser polarization, gave P$_{\mu}$.

\citet{So98} achieved a muon polarization 
of P$_{\mu} = 30 \pm 4$\%.  By normalizing the observed triton
asymmetry with measured muon polarization the authors obtained a
preliminary value for the vector asymmetry for $^3$He $\rightarrow$
$^3$H capture of $A_v = 0.63 \pm 0.09 (stat.) \pm ^{0.11}_{0.14}
(syst.)$.  The largest uncertainty was the systematic error in the
experimental determination of the muon polarization.

Taking the preliminary result of \citet{So98} and the model
calculation of \citet{Co96} we obtain, at $q^2=-0.954 m_\mu^2$, $g_p /
g_p^{PCAC} = 0.40 \pm^{0.89}_{0.73}$, which implies that $g_p(-0.88
m_\mu^2)=3.3\pm ^{7.4}_{6.1}$. This result is consistent with PCAC and
the $\mu^3$He capture data, but the experimental uncertainties are
quite large.

\subsection{Radiative muon capture in $^3$He}
\label{s: RMCHe3}

\subsubsection{Theoretical calculations}

Much less has been done on RMC in $^3$He than on OMC, though there
have been calculations of the exclusive rate both in IA and
EPM. Calculations in the EPM were made by \citet{Hw78}, who however
made an incorrect assumption, and thus obtained incorrect results, and
also by \citet{Fe80}, \citet{Kl81}, \citet{Kl82} and
\citet{Ho02}. For RMC however the EPM seems to be less
reliable than for OMC. The nuclear form factors are more rapidly
varying than the nucleon form factors so that the gauge terms
involving derivatives of the form factors \citep{Ad66, Ch77, Kl85, Ho02}
are somewhat larger. Furthermore, as emphasized by
\citet{Kl84a, Kl84b}, some important terms involving intermediate state
excitations are missed in the EPM.

The RMC rate has also been calculated in IA by \citet{Kl81},
\citet{Kl82}, and \citet{Ho02}. The earlier calculations used wave
functions from the Reid soft core potential and made a number of
approximations in evaluating the nuclear matrix elements. The later
calculation of \citet{Ho02} improved on a number of these
approximations, kept some additional terms, and used a variety of
nuclear wave functions obtained from modern momentum space Faddeev
calculations. The end results for the photon spectrum however were
about the same as those of \citet{Kl82}, though the photon
polarization was somewhat different, apparently because of the higher
order terms kept.

Typical results for the RMC statistical capture rate for photons with
energies greater than 57 MeV are 0.17 s$^{-1}$ in the IA and 0.25
s$^{-1}$ in the EPM \citep{Ho02}.  Just as for OMC, the IA result is
significantly below that obtained in the EPM, and presumedly the
difference is again at least partially explained by MEC's, which will
be much more complicated here than they were for OMC, and have not yet
been calculated in detail. In view however of the problems with the
EPM applied to $^3$He, as noted above, it is not possible to simply
ascribe the difference just to MEC's. Such corrections will have to be
calculated explicitly to finally obtain a reliable rate.

In principle the various breakup channels would give additional
information, but to our knowledge there have not been any calculations
of such channels for RMC.

\subsubsection{Experimental results}

In $^3$He RMC the $^3$H--channel has a strong peaking at the recoil
energy corresponding to the kinematic limit 1.9 MeV.  In contrast the
energy spectra of charged particles from break--up channels are a
broad continua reaching to 53~MeV and therefore are straightforwardly
distinguished from $^{3}$He $\rightarrow$ $^3$H.

While the $^3$He$\rightarrow$$^3$H transition following $^3$He RMC and
the p$\rightarrow$n transition following $^1$H RMC are spin--isospin
analogs, the $^3$He $\rightarrow$ $^3$H process is experimentally
easier.  First the $^3$He rate is much larger and consequently the
background difficulties are greatly reduced.  Second the $\mu$$^3$He
atom circumvents the various chemical processes that confound muonic
hydrogen.  Of course for $^3$He RMC the presence of exchange currents
and the contribution of break--up channels have to be considered.

The $^3$H channel in $^3$He RMC has recently been measured by
\citet{Wr00} at TRIUMF.  In the experiment a $\mu^-$
beam was stopped in liquid $^3$He and gamma--triton coincidences
recorded.  The photons were detected using a pair spectrometer and the
tritons were detected via their scintillation light in the liquid
helium.  Pulse-shape read out permitted the discrimination of the
incoming muon from the recoil triton.  The detection efficiency was
calibrated via $\pi$ stops in liquid $^3$He, {\it i.e.}  with the
well--known branching ratios for $\pi^-$$^{3}$He $\rightarrow$
$\gamma$$^{3}$H and $\pi^-$$^{3}$He $\rightarrow$ $\pi^o$$^{3}$H.

In the $^3$He RMC experiment the major photon backgrounds originate
from muon decay in the target material, muon capture in the nearby
materials, and pion contamination in the muon beam.  However energy
cuts, {\it i.e.} $E_{\gamma} >$ 57 MeV, and timing cuts, {\it i.e.}
$t_{\gamma} > 440$ ns, together with the gamma--triton coincidence
requirement, were effective in discriminating the $^3$He signal from
background processes.  Note the potential background from random
coincidences of photons from RMC and tritons from OMC was negligible.

Only preliminary results are presently available for the $^3$He RMC
experiment \citep{Wr00}. The results indicate that the shape of the
measured $^3$He RMC energy spectrum is consistent with the theoretical
prediction of \citet{Kl81}, but that the overall magnitude is 20-30\%
smaller than the impulse approximation prediction. To get agreement in
magnitude \citet{Wr00} stated that one needs $g_p = 3.4$, which is
much smaller than the PCAC value.  

We stress that the $^3$He$\rightarrow$ $^3$H radiative capture rate is
potentially a very valuable data-point for the determination of $g_p$.
We therefore urge the publication of the final result from the TRIUMF
experiment and the undertaking of a modern impulse approximation plus
exchange current calculation for the process.

\section{Other few body processes}
\label{s: otherfb}

There have also been attempts to extract $g_p$ from other processes,
notably pion electroproduction. To do this \citet{Ch93} measured the
near threshold $\pi^+$ electroproduction at several momentum
transfers. The connection to $g_p$ was made via a low energy theorem
described by \citet{Va72} and reviewed by \citet{Dr92}.  The theorem
is valid in the limit of vanishing pion mass, so it had to be assumed
that higher order corrections were in fact small. Nevertheless
\citet{Ch93} obtained results which they interpreted as $g_p(t)$ over
a range of four-momentum transfer squared $0.07 < t < 0.18$
GeV$^2$. These results agreed quite well with the predicted pion pole
dominance and with the PCAC value near $t=0$.

It is clear that such an approach to `measuring' $g_p$ is quite
different than that described for OMC and RMC above. In fact there has
been a quite heated disagreement over the question of whether or not
pion electroproduction does give new information about either $g_p$ or
$g_a$ \citep{Ha00, Gu01, Be01c, Ha01, Tr01}.

In OMC $g_p$ is defined as a parameter in the most general form of the
amplitude, so that a measurement of the rate gives directly
information about $g_p$. In RMC this same amplitude appears directly,
albeit with one leg off shell and with additional gauge terms, some of
which probe the interior structure of the amplitude. To the extent
these are small, one again in RMC has an almost direct measurement of
$g_p$.

In other processes however, such as pion electroproduction, one needs
to make a number of assumptions in order to make the connection to
$g_p$. For example, the low energy theorem approach is based on chiral
symmetry, which implies the usual PCAC relation between $g_a$ and
$g_p$, which in turn allows one to eliminate one or the other from the
expressions for the amplitude. Alternatively in a Born approximation
diagram approach, which is often used, a pion pole diagram is
included. This is the diagram which is assumed to be responsible for
$g_p$. Thus in such cases is one measuring $g_p$ or just the
importance of a pion exchange diagram?

It thus appears that there is a degree of circularity in trying to use
processes such as pion electroproduction to `measure' $g_p$. To make
the interpretation, one has to assume something nearly equivalent to
the PCAC relation between $g_a$ and $g_p$, which is what is being
tested. In our view then processes such as pion electroproduction are
very interesting, and very worth measuring, but primarily as
consistency checks on our understanding of the underlying theory,
rather than as independent measurements of $g_p$.

Finally one should note that in the context of ChPT the calculations
of OMC give a relation between $g_p$ and some of the low energy
constants \citep{Fe97, Be98, An00}. Thus any process which contains
these same low energy constants could be considered as a way of
`measuring' $g_p$, but only within the context of ChPT.

\section{Exclusive OMC on complex nuclei}
\label{s: exclnuclOMC}

Determining $g_p$ from exclusive OMC on complex nuclei is not easy.
First, the reaction products are a $\sim$$100$~MeV neutrino and a
$\sim$0.1~MeV recoil, and therefore the $\beta$--rays or
$\gamma$--rays from the recoil's decay must be detected.  Second, the
effects of $g_p$ in OMC are small and subtle, and therefore
measurements of spin observables are usually required.  Last, the
observables are functions of both the weak couplings and the nuclear
wave functions, and disentangling these contributions is unavoidably
model dependent.

In allowed transitions on $J_i=0$ targets the 0p transition $^{12}$C$(
0^+ , 0 )\,\rightarrow\,^{12}$B$( 1^+ , 0 )$ and 1s--0d transition
$^{28}$Si$( 0^+ , 0 )\,\rightarrow\,^{28}$Al$( 1^+ , 2201 )$ have
attracted the most attention.  In order to isolate the coupling $g_p$,
the $^{12}$C experiments have measured recoil polarizations via the
$^{12}$B $\beta$--decay and the $^{28}$Si experiments have measured
$\gamma$--ray correlations via the $^{28}$Al $\gamma$--decay.  In
allowed transitions on $J_i\neq0$ targets the 0p nucleus $^{11}$B and
1s--0d nucleus $^{23}$Na have been studied.  These experiments
measured the hyperfine effect in $3/2^{\pm} \rightarrow 1/2^{\pm}$
transitions to extract the coupling $g_p$.  Last, the capture rate of
the $^{16}$O$( 0^+ , 0 )
\rightarrow ^{16}$N$( 0^- , 120 )$ transition
has been measured by several groups;
since the  $0^+ \rightarrow 0^-$ spin sequence offers special sensitivity 
to $g_p$.

The material on exclusive OMC is organized as follows.  In section
\ref{s: observables} we describe the physical observables in partial
transitions.  In Sec. \ref{s: helicity decomposition} we discuss the
dynamical content of the physical observables in OMC and in
Sec. \ref{s: induced currents} we describe the specific manifestations
of the coupling $g_p$ in OMC.  The experimental work on exclusive
transitions is covered in Sec. \ref{s: experiments}.  In Secs. \ref{s:
formalism} and \ref{s: nuclear models} respectively we review the
theoretical framework and nuclear models for exclusive OMC on complex
nuclei.  In Sec. \ref{s: gp results} we discuss the results for $g_p$.

One goal is a unified discussion and comparison of the different
experiments and calculations.  Another goal is a critical assessment
of the current `best values' and model uncertainties in the extraction
of the coupling $g_p$.

We don't discuss the related topic of total OMC rates on complex
nuclei.  Although a large body of accurate data is nowadays available
on total OMC rates, because the total rate is weakly dependent on
$g_p$ and the model uncertainties in total rates are large, extracting
$g_p$ isn't feasible.  We point out though that total OMC rates are
employed in analyzing the total rate of radiative capture on complex
nuclei (see Sec. \ref{s: nuclRMC}).

\subsection{Free couplings versus effective couplings}
\label{s: free}

We first wish to comment on the meaning of the coupling $g_p$ that's
extracted from partial transitions on complex nuclei, and specifically
the issue of free nucleon coupling constants versus effective nucleon
coupling constants.

Suppose one performed an exact calculation of the initial-final wave
functions and the weak transition, that is, incorporating all
components that contribute significantly to the wave functions and all
currents that contribute significantly to the transition.  Naturally
it would require a large number of basis states, a realistic
interaction between nucleons, and the contribution of nucleonic
excitations, such as the $\Delta$ resonance, and exchange currents,
such as pion exchange.  However such a calculation would involve the
free nucleon values of the weak couplings $g_a$, $g_p$, etc.

However, in practice only for few-body systems are such highly
accurate calculations feasible.  Rather in complex nuclei a common
practice is to consider only a handful of active nucleons and a
truncated model space, and to omit both exchange currents and
nucleonic excitations.  In such calculations the free coupling
constants are replaced by effective coupling constants, with their
renormalization accounting for the overall effects of the missing
components of the nuclear model.  The use of effective electric
charges and an effective axial coupling are well known examples of
this approach to the problem.

In this article our interest is in the coupling $g_p$ of the free
nucleon.  We therefore have tried to use the most sophisticated models
available, in order to extract the coupling $g_p$.  In almost every
case we discuss there are full-shell model calculations with
perturbation theory computations of core polarization effects, {\it
i.e.} the effects of truncating the model space, and exchange current
effects, {\it e.g.} the effects of pion exchange and $\Delta$
excitations.  In this way we believe we come as close as possible to
extracting the free nucleon value for the induced coupling $g_p$.

\subsection{Physical observables}
\label{s: observables}

In muon capture the initial state is the 1S ground state of the muonic
atom.  Capture then yields a muon neutrino with momentum $\vec{\nu}$
and recoil nucleus with momentum $-\vec{\nu}$, {\it i.e.}
\begin{equation}
\label{e: muon capture}
\mu^- + [A,Z] \longrightarrow [A,Z-1] + \nu .
\end{equation}
For $J_i = 0$ targets the total spin of the $\mu$--atom is $F = 1/2$,
and for $J_i > 0$ targets the possible spins of the $\mu$--atom are
$F_{\pm} = J_i \pm 1/2$.  On $J_i = 0$ targets the necessary variables
for describing capture are the $\mu$--atom orientation, the recoil
orientation and the neutrino direction.  On $J_i > 0$ targets the
capture process is further dependent on the $F$--state.

\subsubsection{Capture rates}
\label{s: rate}

The easiest observable to experimentally determine is the capture rate
of the partial transition.  For $J_i = 0$ targets we shall denote the
capture rate by $\Lambda$ and for $J_i > 0$ targets we shall denote
the hyperfine rates by $\Lambda_{\pm}$.  In addition on $J_i > 0$
targets it's useful to define the statistical rate
\begin{equation} 
\label{e: statistical rate}
\Lambda_S = (\frac{J_i + 1}{2 J_i + 1}) \Lambda_+ 
+ (\frac{J_i}{J_i + 1}) \Lambda_-
\end{equation}
where $( J_i + 1 ) / ( 2 J_i + 1 )$ and $ J_i / ( 2 J_i + 1 ) $
are the statistical populations of the $F_{\pm}$ hyperfine states.

\subsubsection{Recoil asymmetries}
\label{s: asymmetry}

Generally the direction of emission of recoils after capture is
anisotropic about the $\mu$--atom orientation axis.\footnote{An
exception is the isotropic distribution of the recoil emission from a
$F = 0$ $\mu$--atom.}  For $J_i = 0$ targets the angular distribution
for recoil emission is \citep{Mu77,Be75,Ci84}
\begin{equation}
\label{e: J=0 asymmetry}
\frac{d\Lambda}{d\Omega} 
= \frac{\Lambda}{4 \pi } 
[ 1 + \alpha ~ |\vec{P}_{\mu}| ~ P_1 ( \hat{P}_{\mu} \cdot \hat{\nu} ) ]
\end{equation}
where $\vec{P}_{\mu}$ is the muon polarization, $\hat{\nu}$ is the
neutrino direction, $P_1$ is the $\ell = 1$ Legendre polynomial, and
$\alpha$ is the asymmetry coefficient.  Note that the recoil asymmetry
is a pseudoscalar quantity, and therefore an example of parity
violation in muon capture.

For $J_i > 0$ targets the angular distribution for recoil emission is
generally more complicated.  For example a $F = 1$ $\mu$--atom may
possess both a vector polarization, {\it i.e.} non-statistical
populations of the $m_F = +1$ and $m_F = -1$ sub-states, and a tensor
polarization, {\it i.e.} non-statistical populations of the $|m_F| =
1$ and $m_F = 0$ sub-states.  Consequently for $F = 1$ capture the
angular distribution of recoil emission is \citep{Ci84,Ga68,Co93}
\begin{equation}
\label{e: J>0 asymmetry}
\frac{d\Lambda}{d\Omega} 
= \frac{\Lambda}{4 \pi } 
[ 1 + A_v ~ |\vec{P}_{\mu}| ~ P_1 ( \hat{P}_{\mu} \cdot \hat{\nu} )
+ A_t ~ |\vec{P}_{\mu}| ~ P_2 ( \hat{P}_{\mu} \cdot \hat{\nu} ) ]
\end{equation}
where $A_v$ is the vector asymmetry coefficient and $A_t$ is the
tensor asymmetry coefficient.  Note for $F > 1$ atoms even higher
ranks of $\mu$--atom orientation are possible in principle.

We stress that $\vec{P}_{\mu}$ in Eq.\ (\ref{e: J=0 asymmetry}) and
Eq.\ (\ref{e: J>0 asymmetry}) is the muon's residual polarization in
the $\mu$--atom ground state.  Unfortunately in all targets the
muon--nucleus spin--orbit interaction leads to substantial
depolarization during the atomic cascade.  For example for $J_i = 0$
targets the residual polarization is typically 17\%.  Further in $J
\neq 0$ targets the muon--nucleus spin--spin interaction leads to
additional depolarization during the atomic cascade.  The degree of
depolarization is dependent on $J_i$ and $F$.  For further details see
\citet{Fa70,Mu77,Ha75,Ha77a,Ku87}.

\subsubsection{Recoil orientations}
\label{s: orientation}

Generally after muon capture the recoil nucleus is oriented, where
orientation along both the ${\nu}$--momentum axis and the
${\mu}$--spin axis is possible.  For a $J_f = 1/2$ recoil the
orientation is described by its recoil polarization $P$, where $P = (
p_{+1/2} - p_{-1/2} ) / \sum{p_m}$ with $p_m$ denoting the population
of each magnetic sub--state $m$.  For a $J_f = 1$ recoil the
orientation is described by its polarization $P$ and its alignment
$A$, where $A = ( p_{+1} + p_{-1} - 2 p_{0} ) / \sum{p_m} $.  For $J_f
> 1$ recoils even higher ranks of orientation are possible in
principle.  Conventionally the recoil orientation along the
$\nu$--momentum axis is designated the longitudinal polarization P$_L$
and longitudinal alignment A$_L$, and the recoil orientation along the
$\mu$--spin axis is designated the average polarization P$_{av}$ and
average alignment A$_{av}$.  The recoil polarizations, {\it i.e.}
P$_L$ and P$_{av}$, are pseudoscalar quantities and additional
examples of parity violation in muon capture.  For further details see
\citet{Mu77,Be75,De72,Su76}.

Additionally triple correlations of the recoil orientation with the
$\mu$--spin direction and the $\nu$--momentum direction are possible.
Of experimental importance are the so--called forward hemisphere and
backward hemisphere polarizations.  Defining the hemispheres
relative to the $\mu$--spin axis, P$_F$ is the polarization along the
$\mu$--axis for the `forward hemisphere' recoils, and P$_B$ is the
polarization along the $\mu$--axis for the `backward hemisphere'
recoils.  Note that P$_F$ and P$_B$ are related to P$_L$ and P$_{av}$
via
\begin{equation}
\label{e: Pf}
P_F = \frac{1}{2} ( P_{av} + \frac{1}{2} P_L )
\end{equation}
\begin{equation}
\label{e: Pb}
P_B = \frac{1}{2} ( P_{av} - \frac{1}{2} P_L )
\end{equation}
which demonstrates their dependence on both the $\mu$--spin direction
and the $\nu$--momentum direction.

\subsubsection{Gamma-ray correlations}
\label{s: correlation}

Gamma-ray directional correlations with the $\nu$--momentum axis and
the $\mu$--spin axis in the sequence
\begin{equation}
\label{e: muon capture + gamma decay}
\mu^- + [Z,A] \longrightarrow [Z-1,A]^* 
+ \nu \longrightarrow [Z-1,A]^{**} +\gamma
\end{equation}
are additional observables in partial transitions.  Such correlations
originate from the recoil orientation and the recoil asymmetry in the
capture process.  For unpolarized muons a $\gamma$--$\nu$ directional
correlation is possible and for polarized muons a
$\gamma$--$\nu$-$\mu$ triple correlation is also possible.  In
discussing the correlations we shall denote the spin sequence in Eq.\
(\ref{e: muon capture + gamma decay}) by $J_i \rightarrow J_f
\rightarrow j$ where $J_i$, $J_f$ and $j$ are the angular momenta of
the three nuclear states.  Note that below we consider only
unpolarized targets and $J_f \leq 1$ recoils.  For further details see
\citet{Ci84}.
 
For unpolarized muons $\vec{P}_{\mu} = 0$ the $\gamma$--$\nu$
directional correlation is given by
\footnote{The theory of $\gamma$--ray correlations was developed
by Popov and co--workers in 
\citet{Po63,Bu64,Oz65,Bu67a,Bu67b,Bu67c,Bu70b}.
See also \citet{Pa78} and \citet{Ci84}.}
\begin{equation}
\label{e: Pmu=0 correlation}
W = 1 + a_2 P_2(\hat{k} \cdot \hat{\nu} )
\end{equation}
where $\hat{k}$ is the $\gamma$--ray direction, $\hat{\nu}$ is the
neutrino direction, $P_2$ is the $\ell = 2$ Legendre polynomial, and
$a_2$ is the $\gamma$--$\nu$ correlation coefficient.  Note that this
directional correlation $a_2$ is a consequence of the longitudinal
alignment $A_L$ of the recoil nucleus in the capture process.

For polarized muons $\vec{P}_{\mu} \neq 0$ the $\gamma$--$\nu$--$\mu$
triple correlation is given by
\begin{eqnarray}
\label{e: Pmu>0 correlation}
W = & 1 \, + \, ( \alpha + \frac{2}{3} c_1 ) \, 
\vec{P}_{\mu} \cdot \hat{\nu} 
\; \hat{k} \cdot \hat{\nu}  
\nonumber
\\
 +& ( a_2 + b_2 \; \vec{P}_{\mu} \cdot \hat{\nu} 
\; \hat{k} \cdot \hat{\nu} ) \,
P_2  ( \hat{k} \cdot \hat{\nu} )  \\
\nonumber
\end{eqnarray}
and contains three distinct correlation terms involving ($\alpha +
\frac{2}{3} c_1$), $a_2$ and $b_2$.  Note in Eq.\ (\ref{e: Pmu>0
correlation}) that (i) $\hat{k}$ enters only quadratically because of
parity conservation in $\gamma$--decays, (ii) $\vec{P}_{\mu}$ enters
only linearly because the muon is spin--$1/2$, and (iii) $\hat{\nu}$
enters in powers of two or less for $J_f \leq 1$.  In Eq.\ (\ref{e:
Pmu>0 correlation}) the correlation involving $\alpha$ is a
consequence of the recoil asymmetry and the correlation involving
$a_2$ is a consequence of the longitudinal alignment.  The remaining
terms, {\it i.e.} $c_1$ and $b_2$, originate from the triple
correlation of the recoil alignment with $\vec{P}_{\mu}$ and
$\hat{\nu}$.  If either unpolarized muons, {\it i.e.} $\vec{P}_{\mu} =
0$, or perpendicular geometry, {\it i.e.} $\vec{P}_{\mu} \cdot \hat{k}
= 0$, is employed the only non--vanishing correlation is $a_2$, and
Eq.\ (\ref{e: Pmu>0 correlation}) becomes Eq.\ (\ref{e: Pmu=0
correlation}).

It's important to recognize that $c_1$, $a_2$ and $b_2$ are functions
of both the $\mu$--capture process and the $\gamma$--decay process.
For example the coefficient $a_2$ may be written in the form of a
product $A_L B_2$, where the alignment $A_L$ is governed by the $\mu$
capture process and the parameter $B_2$ is governed by the
$\gamma$--decay process.  A handy compilation of the coefficients
$B_2$ for various $J_f \rightarrow j$ spin--parities, multipolarities,
and mixing ratios is given in Table 1 of \citet{Ci84}.  Note that in
certain cases, {\it e.g.} for M1 emission in a $1^+ \rightarrow 0^+$
decay, $B_2$ is large, but in other cases, {\it e.g.} for M1 emission
in a $1^+ \rightarrow 2^+$ decay, $B_2$ is small.  This makes the
former $1^+ \rightarrow 0^+$ case more favorable, and the latter $1^+
\rightarrow 2^+$ case less favorable, for $\gamma$--ray correlation
experiments.

\subsection{Helicity representation}
\label{s: helicity decomposition}

Herein we consider the overall dynamical content of exclusive muon
capture.  Utilizing the helicity representation we discuss the
constraints imposed on $\mu$ capture by $\nu$--handedness and
T--invariance.  We shall denote the corresponding helicity amplitudes
by $T^F_{\lambda}$, where $\lambda$ is the recoil helicity and $F$ is
the $\mu$--atom hyperfine state.  We use the superscript `$+$' for $F
= J_i + 1/2$ capture, the superscript `$-$' for $F = J_i - 1/2$
capture, and no superscript for $J_i = 0$ atoms.  Assuming the absence
of T violation in $\mu$ capture the helicity amplitudes are real
numbers.

Further details on the helicity representation in the ($\mu$,$\nu$)
reaction are given in \citet{Be75}, \citet{Mu77} and \citet{Ci84}.

\subsubsection{Capture on zero--spin targets}
\label{s: J=0 helicities}

For $J_i = 0$ targets the helicity representation depicts muon capture
as the two--body decay of a spin--1/2 $\mu$--atom into a left--handed
neutrino and a spin--$J_f$ recoil.  Choosing the z--axis along the
$\nu$--axis, the definite neutrino helicity of $\lambda_{\nu} = -1/2$
means the allowable recoil helicities are $\lambda_f$ $=$ $0$, $+1$.
The corresponding helicity amplitudes, denoted by $T_0$ and $T_1$, are
the underlying dynamical variables in $\mu$ capture on $J_i = 0$
targets.

In Table \ref{t: J=0 observables} we compile explicit formulas for
various physical observables in $0 \rightarrow J_f$ transitions in
terms of $T_0$, $T_1$ and their ratio $X = \sqrt{2} T_0 / T_1$.  The
$\gamma$--ray correlation coefficients also involve the quantity $B_2$
that is governed by the $\gamma$--decay.  With the exception of
$\Lambda$, the reaction dynamics are completely determined by
$X$, and consequently there exist numerous relations between
observables in $0 \rightarrow J_f$ transitions, for example see
\citet{Be75} and \citet{Mu77}.  Clearly although different observables
offer alternative possibilities for experimental measurements, the
dynamical content of $0 \rightarrow J_f$ transitions is somewhat
limited.

\begin{table}[htpb]
\caption{Helicity decomposition of physical observables 
in $0 \rightarrow 1$ transitions.  We give the capture rate $\Lambda$,
recoil asymmetry $\alpha$, longitudinal polarization and alignment
$P_L$ and $A_L$, average polarization and alignment $P_{av}$ and
$A_{av}$, and $\gamma$--ray correlation coefficients $a_2$, $b_2$ and
$( \alpha + \frac{2}{3} c_1)$.  The helicity amplitudes for recoil
helicities of $\lambda_f = 0, +1$ are denoted by $T_0$ and $T_1$ and
$X = \sqrt{2} T_0 / T_1$.  The parameter B$_2$ involved in the
$\gamma$--ray correlations depends on the spin-parities of the
initial-final states and the multipolarity of the gamma-radiation in
the $\gamma$-decay.}
\label{t: J=0 observables}
\begin{ruledtabular}
\begin{tabular}{cc}
 & \\
Observable & Helicity decomposition \\ 
 & \\
\hline
 & \\
$\Lambda$  & $| T_1 |^2 ( 2 + X^2 ) $ \\ 
 & \\ 
$\alpha$ & $( X^2 -2 ) / (X^2 + 2 )$ \\
 & \\
$P_L$ & $-2 / (2 + X^2)$ \\
 & \\
$A_L$ & $2 (1 - X^2) / (2 + X^2)$ \\
 & \\
$P_{av}$ & $2/3 ( 1 + 2 X ) / (2 + X^2)$ \\
 & \\
$A_{av}$ & $0$ \\
 & \\
$\alpha + \frac{2}{3} c_1$ & $B_2 \sqrt{2} ( 2 + 2X - X^2 ) 
/ ( 2 + X^2 ) $ \\
 & \\
$a_2$ & $B_2 \sqrt{2} ( 1 - X^2 ) / ( 2 + X^2 )$ \\
 & \\
$b_2$ & $B_2 \sqrt{2} ( 1 - 2 X + X^2 ) / ( 2 + X^2 )$ \\
 & \\
\end{tabular}
\end{ruledtabular}
\end{table}

Note that the $0 \rightarrow 0$ sequence is a special case.  Compared
to $J_f > 0$, where the $\lambda_f = 0, +1$ helicity states are
allowable, for $J_f = 0$, only the $\lambda_f = 0$ helicity state is
possible.  Consequently a single amplitude $T_0$ is the sole dynamical
quantity for physical observables in $0 \rightarrow 0$ transitions.

\subsubsection{Capture on nonzero--spin targets}
\label{s: J>0 helicities}

For $J_i > 0$ targets the helicity representation depicts capture as a
two--body decay of the $F = J_i \pm 1/2$ $\mu$--atom.  Based on
definite neutrino handedness and angular momentum coupling the total
number of recoil helicity states $\lambda_f$ and contributing helicity
amplitudes $T^F_{\lambda}$ is the lesser of either $ 2 J_f + 1 $ or 
$2 F + 1 $ \citep{Mu77,Ci84}.

For concreteness let's consider the example of a $1/2 \rightarrow 1/2$
transition, where the $\mu$--atom spin is either $F = 0$ or $F = 1$.
For $F = 1$ both the $\lambda_f = -1/2, +1/2$ recoil helicity states
are populated, but for $F = 0$ only the $\lambda_f = +1/2$ recoil
helicity state is populated because of the single magnetic sub--state
of the $F = 0$ $\mu$--atom.  Consequently one helicity amplitude
governs the $F = 0$ capture, two helicity amplitudes govern the $F =
1$ capture, and three dynamical variables underlie $1/2 \rightarrow
1/2$ transitions.  By comparison in $3/2 \rightarrow 1/2$ transitions
a total of four helicity amplitudes contribute, two for $F_+$ capture
and two for $F_-$ capture.

In summary for $J_i > 0$ targets many independent quantities are
experimentally accessible.  In principle the increased number of
variables allows an increased number of cross checks on model
calculations for $J_i > 0$ transitions.

\subsection{Induced currents}
\label{s: induced currents}

Most often in partial transitions the leading contributions originate
from the axial coupling $g_a$.  This subsection concerns `where to
find the coupling $g_p$?'

To answer this question we shall examine capture in (i) the $q/M
\rightarrow 0$ limit, where $q$ is the three-momentum transfer and 
$M$ is the nucleon mass, and (ii) in the
Fujii--Primakoff (FP) approximation \citep{Fu59}.  In the $q/M
\rightarrow 0$ limit the effects of $g_p$ are absent.  However in the
FP approximation, which keeps $q/M$ terms involving allowed operators,
but drops $\ell$--forbidden and gradient operators, the leading
effects of induced currents are present.  Therefore comparison of the
capture process in the $q/M \rightarrow 0$ limit and the FP
approximation is helpful in understanding the manifestation of $g_p$
in capture.

At this point it is helpful to recall the Fujii--Primakoff effective
Hamiltonian for muon capture \citep{Fu59,Pr59}.  Its form is
\begin{eqnarray}
\label{e: FP hamiltonian}
\nonumber
\tau^+ { \frac{\boldsymbol{1} - \boldsymbol{\sigma} 
\cdot \boldsymbol{\hat{\nu}}}{2}} 
\sum_{i = 1}^{A} \tau^-_i 
( 
G_V~ \boldsymbol{1} \cdot \boldsymbol{1}_i 
+ G_A~  \boldsymbol{\sigma} \cdot \boldsymbol{\sigma}_i \\ 
+ G_P~  \boldsymbol{\sigma}  \cdot \boldsymbol{\hat{\nu}} ~ 
\boldsymbol{\sigma}_i \cdot \boldsymbol{\hat{\nu}}
)
\, \delta ( \boldsymbol{r} - \boldsymbol{r}_i ) 
\end{eqnarray}
where $\boldsymbol{1}$, $\boldsymbol{1}_i$, $\boldsymbol{\sigma}$ and
$\boldsymbol{\sigma}_i$ are the $2 \times 2$ unit and spin matrices
and $\boldsymbol{r}$ and $\boldsymbol{r}_i$ are the spatial
coordinates of the lepton and the $i^{th}$ nucleon respectively,
$\boldsymbol{\hat{\nu}}$ is the $\nu$--momentum unit vector, and
$\tau^+$, $\tau^-_i$ convert the muon into a neutrino and a proton
into a neutron.  Last, $G_V$, $G_A$ and $G_P$ are the so--called
Fujii--Primakoff effective couplings
\begin{equation}
\label{e: GV}
G_V =   g_v ( 1 + \frac{ q }{ 2M }  )
\end{equation}
\begin{equation}
\label{e: GA}
G_A = - ( g_a + \frac{ q }{ 2M }  ( g_v + g_m ) )
\end{equation}
\begin{equation}
\label{e: GP}
G_P = - \frac{ q }{ 2M } ( g_p - g_a + g_v + g_m )
\end{equation}
Note that, when $q/M \rightarrow 0$, the coupling $G_P$ vanishes and
$G_V$ and $G_A$ are determined by $g_v$ and $g_a$ respectively.
Furthermore the induced currents are order $q/M$, and $g_p$ appears in
$G_P$ only.  For the canonical values of the weak couplings, as given
in Table \ref{t: parameters} (see Sec. \ref{s: gptheory}), $G_V =
1.00$, $G_A = -1.27$ and $G_P = 0.00$ in the $q/M \rightarrow 0$
limit, and $G_V = 1.03$, $G_A = -1.52$ and $G_P = -0.62$ in the FP
approximation and with $q = 100$~MeV/c.

\subsubsection{Asymmetries, orientations and correlations}
\label{s: recoil and gp}

To understand the $g_p$ sensitivity of orientations, correlations,
etc., we compare the terms with $G_A$ and $G_P$ in Eq.\ (\ref{e: FP
hamiltonian}).  Observe the operator corresponding to the $G_P$ term,
{\it i.e.} $( \boldsymbol{1} - \boldsymbol{\sigma} \cdot
\boldsymbol{\hat{\nu}} ) ~
\boldsymbol{\sigma}  \cdot \boldsymbol{\hat{\nu}} 
~ \boldsymbol{\sigma}_i \cdot \boldsymbol{\hat{\nu}}$,
cannot change the spin projection of either the lepton or the nucleon 
along the $\nu$-momentum axis.
However this limitation does not apply to
the $G_A$ term and the operator
$( \boldsymbol{1} - \boldsymbol{\sigma} \cdot \boldsymbol{\hat{\nu}} ) ~
\boldsymbol{\sigma} \cdot \boldsymbol{\sigma}_i$.
Consequently in Eq.\ (\ref{e: FP hamiltonian}) the $G_P$ term admits
only a longitudinal coupling of the weak currents whereas the $G_A$
term admits also a transverse coupling of the weak currents.  This
makes correlations, orientations, etc., sensitive to $g_p / g_a$.  For
further details see \citet{Gr85}.

To further illustrate the $g_p$ sensitivity we consider the example of
a $0^+ \rightarrow 1^+$ transition.  In the $q/M \rightarrow 0$ limit
the two helicity amplitudes, {\it i.e.} $T_0$ and T$_1$, are both
determined by the product of the coupling $g_a$ and the allowed
Gamow--Teller operator, and $X = 1$.  However in the FP approximation,
the amplitude $T_0$ is proportional to $( G_A - G_P )$ and involves a
longitudinal coupling of the weak currents, and the amplitude $T_1$ is
proportional to $ G_A $ and involves a transverse coupling of the weak
currents, and $X = ( G_A - G_P ) / G_A \simeq 0.59$ for $q =
100$~MeV/c.  Therefore the asymmetries, correlations and orientations,
which are governed by $X$, permit the determination of $g_p / g_a$.
For further details see \citet{Mu77}.

In Table \ref{t: J=0 FPA} we compile FP approximation expressions for
physical observables in $0^+ \rightarrow 1^+$ transitions in terms of
$G_A$ and $G_P$.  Note in the $q/M \rightarrow 0$ limit the recoil
polarizations are $P_L = -2/3$ and $P_{av} = +2/3$ and the recoil
alignments are $A_L = 0$ and $A_{av} = 0$, thus indicating that the
recoil is `highly polarized' but `not aligned'.  Most strikingly, in
the FP approximation the longitudinal alignment A$_L = +0.55$ is very
large and highly sensitive to $g_p / g_a$. This occurs because the
alignment measures the difference in population of the $\lambda_f = 0$
recoil sub--state, which is populated by $G_P$, and the $\lambda_f =
1$ recoil sub--state, which is not populated by $G_P$.  It therefore
represents a golden observable for the spin structure of the induced
coupling $g_p$.

\begin{table}[htpb]
\caption{Expressions and values of observables for
$0^+ \rightarrow 1^+$ transitions in the $q/M \rightarrow 0$ limit and
the Fujii-Primakoff approximation.  $G_A$ and $G_P$ are the
Fujii--Primakoff effective constants defined in Eqs.\ (\ref{e: GA})
and (\ref{e: GP}) and the common denominator is $\Gamma = (3 G_A^2 +
G_P^2 - 2 G_A G_P )$.  We tabulate the $\gamma$--ray directional
correlations for the $0^+ \rightarrow 1^+ \rightarrow 0^+$ sequence.}
\label{t: J=0 FPA}
\begin{ruledtabular}
\begin{tabular}{llll}
 & & & \\
Observable & ~~~~~~~~FPA & $q/M \rightarrow 0$ & FPA \\
 &  ~~~~~~~~Eq.\ & value & value \\
 & & & \\
\hline
 & & & \\
$\alpha$ & 
$ (3 G_A^2 + G_P^2 - 2 G_A G_P ) / \Gamma  $ & -0.33 & -0.70 \\
 & & & \\
P$_L$ &
$ -2 G_A^2 / \Gamma $ & -0.67 & -0.85 \\
 & & & \\
A$_L$ & 
$ ( -2 G_P^2 + 4 G_A G_P ) / \Gamma $ & 0.00 & -0.55  \\
 & & & \\
$P_{av}$ & 
$ ( 2 G_A^2 - \frac{4}{3} G_A G_P ) / \Gamma $ & +0.67 & +0.61 \\
 & & & \\
$A_{av}$ & $0$ & $0$ & $0$ \\
 & & & \\
$\alpha + \frac{2}{3} c_1$ & $ ( 3 G_A^2 - G_P^2 ) / \Gamma $ 
 & 1.0  & 1.21 \\
 & & & \\
$a_2$ & $ ( - G_P^2 + 2 G_P G_A  ) / \Gamma $ 
 & 0.00  & 0.28 \\
 & & & \\
$b_2$ & $ G_P^2 / \Gamma $  
 & 0.00 & 0.07 \\
 & & & \\
\end{tabular}
\end{ruledtabular}
\end{table}

\subsubsection{Hyperfine dependences}
\label{s: hyperfine and gp}

The hyperfine dependence $\Lambda_+ / \Lambda_-$ of $\Delta J^{\pi} =
J_f - J_i = \pm 1^+$ transitions in muon capture on $J_i \neq 0$
targets is also sensitive to the coupling $g_p$.  Specifically in the
multipole expansion of the Fujii-Primakoff Hamiltonian the $G_A$--term
makes allowed contributions to neutrino waves with total angular
momentum $j^{\pi} = 1/2^+$ only, whereas the $G_P$--term makes allowed
contributions to neutrino waves with total angular momentum $j^{\pi} =
3/2^+$ also.  This difference is because of $\boldsymbol{hat{\nu}}$ in
Eq. (\ref{e: FP hamiltonian}).  In $\Delta J^{\pi} = +1^+$
transitions, neutrinos with $j = 1/2^+$ may be emitted in $F_+$
capture, but neutrinos with $j = 3/2^+$ must be emitted in $F_-$
capture.  This makes $\Lambda_+ / \Lambda_- >> 1$ and a strong
function of the ratio $g_p / g_a$.  However in $\Delta J^{\pi} = -1^+$
transitions, neutrinos with $j = 1/2^+$ may be emitted in $F_-$
capture, but neutrinos with $j = 3/2^+$ must be emitted in $F_+$
capture.  This makes $\Lambda_+ / \Lambda_- << 1$ and a strong
function of the ratio $g_p / g_a$.

For example, we consider the important case of $3/2^+ \rightarrow
1/2^+$ transitions.  In the $q/M \rightarrow 0 $ limit, capture from
the $F_- = 1$ hyperfine state is governed by $g_a$ but capture from
the $F_+ = 2$ hyperfine state is forbidden.  Therefore $\Lambda_+ /
\Lambda_- = 0$.  However in the FP approximation, the two $F_+$
helicity amplitudes are proportional to $G_P$ while the two $F_-$
helicity amplitudes are proportional to either $( 4 G_A - G_P )$ or 
$(2 G_A - G_P )$.  Consequently the hyperfine dependence $\Lambda_+ /
\Lambda_-$ is highly dependent on $g_p / g_a$.

The FP approximation expressions for various observables in $3/2^+
\rightarrow 1/2^+$ transitions are given in Table \ref{t: J>0 FPA}.
In passing we mention that the asymmetries, correlations and
orientations in $F_-$ capture for $3/2^+ \rightarrow 1/2^+$
transitions are also sensitive to the coupling $g_p$, for the same
reasons as described for the $0^+ \rightarrow 1^+$ transitions in the
preceding Sec.\ \ref{s: recoil and gp}.  This is not true for the case
of asymmetries, correlations and orientations for $F_+$ capture in
$3/2^+ \rightarrow 1/2^+$ transitions, because only $j^{\pi} = 3/2^+$
neutrino emission is possible.

\begin{table}[htpb]
\caption{Equations and values for selected observables 
in $3/2^+ \rightarrow 1/2^+$ transitions in the $q/M \rightarrow 0$
limit and Fujii-Primakoff approximation.
The common denominator is $\Gamma = ( 24 G_A^2 - 16 G_A G_P + 3 G_P^2 )$.}
\label{t: J>0 FPA}
\begin{ruledtabular}
\begin{tabular}{llll}
 & & & \\
Observable & ~~~~~~~~FPA & $q/M \rightarrow 0$ & FPA \\
 & ~~~~~~~~Eq. & value & value \\
 & & & \\
\hline
 & & & \\
$\Lambda_+$/$\Lambda_-$ & $3 G_P^2 / \Gamma$
 & 0.00 & 0.027 \\
 & & & \\
$\alpha^-$ & $-\frac{1}{3} ( 4 G_A - G_P )^2 / \Gamma$
& $-0.22$  & $-0.24$ \\
 & & & \\
$\alpha^+$ &  $\frac{2}{5}$ 
& $+0.60$  &  $+0.60$ \\
 & & & \\
$P^-_L$ & $( 8 G_A^2 -G_P^2 ) / \Gamma$
 & $+0.33$ & $+0.44$ \\
 & & & \\
$P^+_L$ & $\frac{1}{5}$ & $+0.20$ & $+0.20$ \\
 & & & \\
\end{tabular}
\end{ruledtabular}
\end{table}

Related FP approximation expressions for other $\Delta J^{\pi} = \pm
1^+$ transitions are given for example by \citet{Mu77} and
\citet{Ci84}.

\subsubsection{Capture rates}
\label{s: rates and gp}

In general in $\mu$ capture the rate $\Lambda$ has a fairly weak
dependence on the coupling $g_p$.  For example in $0^+ \rightarrow
1^+$ transitions the contribution of $g_p$ is typically 10\%.  However
an exception is the capture rate of the first--forbidden $0^+
\leftrightarrow 0^-$ transition.

To understand the sensitivity it is instructive to assume the
dominance of the $\ell = 1$ retarded Gamow-Teller operator in the
first--forbidden $0^+ \rightarrow 0^-$ transition, {\it i.e.} ignoring
the contribution from the axial charge operator.  For details see
\citet{To81}.  The capture rate in $0^+ \rightarrow 0^-$ transitions
is then governed by the coupling constant combination $( G_A - G_P )$,
 which is strongly dependent on $g_p$.  For example
in going from $g_p = 0$ to $g_p = 6.7 g_a$ the rate is increased by
roughly 50\%.  In short, the quantum numbers of the $\Delta J^{\pi} =
0^-$ multipole are effective in isolating the longitudinal
contributions to weak currents, such as $g_p$.

\subsection{Experimental studies of partial transitions}
\label{s: experiments}

This subsection concerns the experimental work on exclusive OMC.  We
discuss two allowed transitions on $J_i = 0$ targets, $^{12}$C($0^+ ,
0$) $\rightarrow$ $^{12}$B$(1^+, 0)$ and $^{28}$Si($0^+ , 0$)
$\rightarrow$ $^{28}$Al($0^+, 2201$), two allowed transitions on $J_i
\neq 0$ targets, $^{11}$B$(3/2^- , 0)$ $\rightarrow$ $^{11}$Be$(1/2^-,
320)$ and $^{23}$Na($3/2^+ , 0$) $\rightarrow$ $^{23}$Ne($1/2^+,
3458)$, and the first--forbidden transition $^{16}$O($0^+ , 0$)
$\rightarrow$ $^{16}$N$(0^-, 120)$.  The experiments include
measurements of capture rates, recoil orientations, $\gamma$--ray
correlations, and hyperfine dependences.

\subsubsection{$^{12}$C($0^+ , 0$) $\rightarrow$ $^{12}$B$(1^+ , 0)$}
\label{s: 12C experiment}

The $^{12}$C($0^+ , 0$) $\rightarrow$ $^{12}$B$(1^+ , 0)$ reaction is
an allowed Gamow-Teller transition between the spin--0, isoscalar
$^{12}$C ground state and the spin--1 isovector $^{12}$B ground state.
The transition was first observed in cosmic--ray data by \citet{Go53}
via the identification of the Godfrey--Tiomino cycle, {\it i.e.} $\mu$
capture on $^{12}$C followed by $\beta$--decay of $^{12}$B.

The first investigation of the recoil polarization in the $^{12}$C$(
0^+ , 0 ) \rightarrow ^{12}$B$( 1^+ , 0 )$ transition was conducted by
\citet{Lo59} in order to measure the $\mu^-$ helicity in $\pi^-$
decay.  The application of polarization measurements to induced
currents was pioneered by \citet{Po74,Po77} at Saclay and extended by
\citet{Ro81a,Ro81b} and \citet{Tr79} at PSI.  Using ingenious
techniques these researchers have amassed impressive data on several
polarization observables in $^{12}$C$( 0^+ , 0 ) \rightarrow ^{12}$B$(
1^+ , 0 )$ capture.  More recently \citet{Ku84,Ku86} at the BOOM
facility have polished some techniques and re--measured the
polarization P$_{av}$.

The procedure for measuring the polarization P$_{av}$ of $^{12}$B
recoils from $^{12}$C capture is straightforward in principle.  First
one makes polarized $\mu^-$$^{12}$C atoms by stopping polarized muons
in carbon--containing material.  Next the polarized $\mu$$^{12}$C
atoms produce the polarized $^{12}$B recoils via muon capture.  Then
one measures the $^{12}$B $\beta$--ray asymmetry to determine the
recoil polarization P$_{av}$.  Note that the method relies on the
known correlation of the ejected $\beta$--rays with the $^{12}$B
orientation.  Also note that this method gives the average
polarization P$_{av}$, {\it i.e.} along the $\mu$--axis, not the
longitudinal polarization P$_L$, {\it i.e.} along the $\nu$--axis.

Unfortunately the measurement of P$_{av}$ is difficult in practice.
First the muon polarization is largely destroyed via the spin--orbit
interaction in the atomic cascade.  Second the $^{12}$B polarization
is easily destroyed by the spin--spin interaction in the host
material.  Third only about $1$\% of $\mu$ stops in $^{12}$C undergo the
$^{12}$C$( 0^+ , 0 ) \rightarrow ^{12}$B$( 1^+ , 0 )$ transition
\citep{Re63,Ma64}.
Consequently, the beta--ray asymmetries are small, the target choices
are limited, and backgrounds are troublesome.  Fortunately the short
$29$~ms lifetime and high $15$~MeV end point for $^{12}$B beta--decay
are ideally suited to a measurement of the polarization.

Average polarization measurements have been performed by
\citet{Po74,Po77} at Saclay and \citet{Ku84,Ku86} at BOOM.
The basic set--up comprises
a beam telescope for detecting incoming muons,
beta--ray counters for detecting $^{12}$B decays,
and Michel counters for detecting $\mu$--decays.
\citet{Po74,Po77} found
a $\geq$0.3~kG longitudinal B--field and a graphite target were
sufficient to preserve the $^{12}$B polarization for $t > 29$~ms.
Note that the distinctive lifetime/end point are used to identify the
$^{12}$B $\beta$--rays and their forward/backward count rates are used
to determine the $^{12}$B $\beta$--asymmetry.  Pulsed--beam operation
allows beta--ray detection under beam--off conditions and reduces the
backgrounds.  Transverse--field precession of the $\mu$ spin is used
to measure the muon polarization $P_{\mu}$.

One challenge is the experimental determination of a small
$\beta$--ray asymmetry ($\sim$3\%) with a reasonable accuracy
($\pm$10\%).  Consequently false asymmetries, such as geometrical and
instrumental effects, must be minimized and then measured.
\citet{Po74,Po77}
used a polarization preserving target material (graphite) and a
polarization destroying target (polyethylene) to determine false
asymmetries. \citet{Ku84,Ku86}
used a novel magnetic resonance technique to periodically destroy the
muon polarization.

In summary the resulting values of the average polarization were
$P_{av} = 0.38 \pm 0.07$ from \citet{Po74}, $P_{av} = 0.452 \pm 0.042$
from \citet{Po77}, $P_{av} = 0.462 \pm 0.053$ from \citet{Ku84,Ku86}.
We discuss the extraction of $g_p$ from $P_{av}$ in Sec. \ref{s: world
data}.

Additionally the forward (P$_{F}$) and backward (P$_{B}$)
polarizations, defined in Eqs.\ (\ref{e: Pf}) and (\ref{e: Pb}), for
the $^{12}$C$( 0^+ , 0 )$ $\rightarrow$ $^{12}$B$( 1^+ , 0 )$
transition have been measured by \citet{Tr79} and \citet{Ro81b}.
These experiments are masterpieces of ingenuity and technique.

The experiments employed a novel target with recoil direction
sensitivity.  The targets comprised a multi--layer sandwich of
`triple--foils' with each triple--foil comprising a carbon target foil
(C), a polarization preserving foil (P), and a polarization destroying
foil (D).  Arranging the P--foil upstream and D--foil downstream of
the carbon foil, {\it i.e.} DCP, permits selective retention of the
forward hemisphere recoil polarization.  Arranging the D--foil
upstream and P--foil downstream of the carbon foil, {\it i.e.} PCD,
permits selective retention of the backward hemisphere recoil
polarization.  Thereby $\beta$--ray asymmetry measurements for the DCP
configuration yield P$_F$ and the PCD configuration yield P$_B$.
Additionally a PCP sandwich enables the measurement of the average
polarization and a DCD sandwich enables the measurement of the false
asymmetries.

One advantage of measuring P$_{F/B}$ is a larger recoil polarization
and a larger beta--ray asymmetry.\footnote{Recall from Sec. \ref{s:
orientation} that P$_{F/B}$ have contributions from both P$_{av}$,
which is decreased by $\mu$ depolarization, and from P$_{L}$, which is
unchanged by $\mu$ depolarization.}  Another advantage is that
combining both P$_{F}$ and P$_{B}$ to deduce $P_{av} / P_L $ where
\begin{equation}
\label{e: expt R}
\frac{ P_{av} }{P_{L} } = 2 \frac{ P_F + P_B }{ P_F - P_B }
\end{equation}
reduces the sensitivity to false asymmetries and systematic
uncertainties.  Note that the measurements of P$_F$ and P$_B$ are
achieved by simply rotating the multi--layer target by 180 degrees.

However a major challenge in measuring P$_{F/B}$ is the small quantity
of the $^{12}$C material in the multi--layer target, since the carbon
foils must be thin enough for recoils to emerge and the P/D foils must
be thick enough for recoils to stop.  Consequently any $^{12}$B
background from $\mu^-$ stops in nearby carbon is especially perilous.
Therefore \citet{Tr79} and \citet{Ro81b} took great care to avoid
using any carbon--containing materials in the neighborhood of the
experiment.  Based on these data the authors obtained $P_{av} / P_{L}
= -0.516 \pm 0.041$.

In Table \ref{t: 12C polarizations} we summarize the various
measurements of recoil polarizations in $^{12}$C$( 0^+, 0 )$
$\rightarrow$ $^{12}$B$( 1^+, 0 )$.  Note we quote all results in
terms of the helicity amplitude ratio $X$ so as to facilitate their
comparison.  The experimental results are mutually consistent.

\begin{table*}[htpb]
\caption{Compilation of results from the recoil polarization
experiments for the transition $^{12}$C$( 0^+, 0 )$ $\rightarrow$
$^{12}$B$( 1^+, 0 )$.  The results are presented in terms of the
dynamical parameter $X = \sqrt{2} T_0 / T_1$.  They incorporate the
corrections for capture to $^{12}$B bound excited states using the
$\gamma$--ray data from \citet{Gi81} in column three and \cite{Ro81a}
in column four.  The experimental observable is listed in column two.}
\label{t: 12C polarizations}
\begin{ruledtabular}
\begin{tabular}{llll}
 & & & \\
 Ref.\ & obs.\ & 
$X$ using & $X$ using \\
      &            &
\citet{Gi81} & \citet{Ro81a} \\ 
 & & & \\
\hline
 & & & \\
\citet{Po74}  & $P_{av}$ & 
$0.10 \pm 0.11$ & $0.08 \pm 0.11$ \\ 
 & & & \\
\citet{Po77}  & $P_{av}$ & 
$0.22 \pm 0.07$ & $0.20 \pm 0.07$ \\  
 & & & \\
\citet{Ro81b} & $P_{F/B}$ & 
$0.27 \pm 0.07$ & $0.24 \pm 0.07$ \\  
 & & & \\
\citet{Ku84} & $P_{av}$ & 
$0.23_{-0.08}^{+0.10}$ & $0.21_{-0.08}^{+0.10}$ \\ 
 & & & \\
FPA & & 0.59 & 0.59 \\
 & & & \\
\end{tabular}
\end{ruledtabular}
\end{table*}

We stress the experimental results for P$_{av}$ and P$_{F/B}$ are for
muon capture to all bound states in the $^{12}$B nucleus.  Therefore a
correction is necessary to obtain the interesting ground state
polarization from the observed bound state polarization.  Measurements
of the total rate to bound states
\citep{Ma64,Re63}
and the individual rates to excited states
\citep{Bu70a,Mi72a,Gi81,Ro81a}
are employed to determine this correction.

Unfortunately the $\gamma$--ray measurements of $\mu$ capture to
individual $^{12}$B excited states are complicated by near equal
energies of several Doppler broadened
$\gamma$--rays.\footnote{Specifically (i) the $947~$keV ($2620
\rightarrow 1674$) and $953~$keV ($953 \rightarrow 0$) gamma--ray lines
and (ii) the $1668~$keV ($2620 \rightarrow 953$) and $1674~$keV ($1674
\rightarrow 0$) gamma--ray lines} Furthermore the latest results of
\citet{Ro81a} and earlier results of \citet{Bu70a}, \citet{Mi72a} and
\citet{Gi81} are in disagreement, and consequently there exists some
uncertainty in the correction for the capture to the $^{12}$B excited
states.  For a detailed discussion of the experimental data see
\citet{Me01}.  Using the $\gamma$--ray yields from \citet{Gi81} and
the model calculations of \citet{Fu87}, the world average for $X$ in
$^{12}$C$(0^+ , 0)$ $\rightarrow$ $^{12}$B$(1^+ , 0)$ is $0.23 \pm
0.06$.  However using the $\gamma$--ray yields from \citet{Ro81a} and
the model calculations of \citet{Fu87}, the world average for $X$ in
$^{12}$C$(0^+ , 0)$ $\rightarrow$ $^{12}$B$(1^+ , 0)$ is $0.20 \pm
0.06$.  Fortunately the corrections are not too large.

\subsubsection{$^{11}$B $\rightarrow$ $^{11}$Be
and $^{23}$Na $\rightarrow$ $^{23}$Ne}
\label{s: 11B experiment}

Historically studies of the hyperfine effect in nuclear muon capture
reaction were important in demonstrating the $V-A$ structure of weak
interactions.  These experiments were performed in muonic $^{19}$F by
\citet{Cu61} and \citet{Wi63}.  The first application of the hyperfine
effect to the induced coupling $g_p$ was conducted by \citet{De68} in
muonic $^{11}$B.  Recently \citet{Wi02} at PSI have improved the data
on $^{11}$B, and \citet{Jo96} at TRIUMF have extended the data to
$^{23}$Na.

Recall from Sec. \ref{s: rate} that the 1S ground state of a $J_i \neq
0$ $\mu$--atom is a hyperfine doublet with a spin $F_{\pm} = J_i \pm
1/2$.  These states are split by the spin--spin interaction of the
muon--nucleus magnetic moments.  For a positive nuclear magnetic
moment the $F_-$ state is the true atomic ground state and for a
negative nuclear magnetic moment the $F_+$ state is the true atomic
ground state.

The two hyperfine state are initially populated with statistical
weights, {\it i.e.} $n_+ = ( J_i + 1 ) / ( 2 J_i + 1 )$ and $n_- = J_i
/ ( 2 J_i + 1 )$.\footnote{The exception is the use of a polarized
target and a polarized beam \citep{Mu77,Ha77a}.}  Thereafter hyperfine
transitions from the upper F--state to the lower F--state occur by
M1--Auger emission, changing the relative occupancies of the hyperfine
states.  The rate $\Lambda_h$ of hyperfine conversion by Auger
emission is governed by (i) the wave function overlap of the electron
and the $\mu$--atom and (ii) the relative sizes of the electron
binding and the hyperfine splitting.  The wave function overlap leads
to a systematic increase in $\Lambda_h$ with Z whereas the electron
binding leads to sudden decreases in $\Lambda_h$ at $Z \sim 6$, where
K--shell emission is forbidden, and $Z \sim 18$, where L--shell
emission is forbidden.  A detailed account of hyperfine conversion by
Auger emission is given by Winston (1963) and experimental
determinations of conversion rates have generally confirmed the
calculated rates
\citep{Su87,Go93,Me01}.
Most importantly for investigating $g_p$, there exist a handful of
$\mu$--atoms with comparable hyperfine transition and muon
disappearance rates, {\it e.g.} $^{11}$B, $^{19}$F, $^{23}$Na and
$^{35}$Cl.

\citet{Te59} was first to recognize 
that the hyperfine conversion during the $\mu$--atom lifetime would
alter the muon occupancy of the $F$--states, and modify the time
spectrum of the capture products, and thus permit the determination of
the hyperfine dependence $\Lambda_+ / \Lambda_-$.  Formulas for the
time evolution of capture products in the presence of hyperfine
conversion have been published by several authors.  For the case of a
positive nuclear magnetic moment, {\it e.g.} $^{11}$B and $^{23}$Na,
\begin{eqnarray} 
\label{e: gamma times}
\nonumber
\frac{d N }{ d t} \propto &
e^{- \Lambda^-_D t}  [ 
(\frac{n^- }{ n^+} + \frac{\Lambda_h }{ \Lambda_h + \Delta \Lambda_D}) \\
 & +(\frac{\Lambda^+ }{ \Lambda^-} - 
\frac{\Lambda_h }{\Lambda_h - \Delta \Lambda_D } ) 
e^{-( \Lambda_h + \Delta \Lambda_D  ) t} ] \\
\nonumber
\end{eqnarray}
where $n^+$, $n^-$ and $\Lambda_+$, $\Lambda_-$ are the initial
populations and the capture rates of the $F_+$, $F_-$ hyperfine
states, $\Lambda_D^-$ is the $F_-$ state disappearance rate, $\Delta
\Lambda_D = \Lambda^+_D - \Lambda^-_D$ is the hyperfine disappearance
increment, and $\Lambda_h$ the hyperfine conversion rate.  In almost
all cases of interest $\Lambda_h >> \Delta \Lambda_D$ and therefore
the factor $\Lambda_h / ( \Lambda_h \pm \Delta \Lambda_D )$ is close
to unity.

As discussed in Sec.\ \ref{s: hyperfine and gp} the hyperfine
dependence $\Lambda_+ / \Lambda_-$ of partial transitions with $\Delta
J^{\pi} = \pm 1^+$ spin--sequences is especially sensitive to $g_p$.
Eq.\ (\ref{e: gamma times}) demonstrates clearly that the hyperfine
dependence $\Lambda_+ / \Lambda_-$ in such capture is encoded in the
time spectrum of the reaction products.

The $^{11}$B$( 3/2^- , 0 )$$\rightarrow$$^{11}$Be$( 1/2^- , 320 ) $
reaction is a $\Delta J^{\pi} = 1^+$ allowed transition from the
$^{11}$B ground state to the $320~$keV $^{11}$B first--excited state.
The $320~$keV, $1/2^-$ $^{11}$Be excited state decays by
$\gamma$--emission ($\tau = 115 \pm 10 $~fs) to the $^{11}$Be ground
state, and the $^{11}$Be $1/2^+$ ground state decays by
$\beta$--emission ($\tau = 13.81 \pm 0.08 $~s) to the $^{11}$B ground
state.  For reference, the $^{11}$B disappearance rate is $\Lambda_D =
( 0.4787 \pm 0.0008 ) \times 10^{6} s^{-1}$ \citep{Wi02} and the
$^{11}$B hyperfine transition rate is $\Lambda_h = ( 0.181 \pm 0.016 )
\times 10^{6} s^{-1}$ \citep{Wi02}.

One nice feature in the experimental study of the hyperfine effect in
the $^{11}$B$( 3/2^- , 0 )$$\rightarrow$$^{11}$Be$( 1/2^- , 320 )$
transition is the level structure of the $^{11}$Be nucleus.  Since the
only particle--stable states are the $1/2^+$ ground state and the
$1/2^-$ excited state this minimizes any concerns of cascade feeding from
muon capture to higher--lying $^{11}$Be states.  However one
difficulty is the very low rate for the $^{11}$B$( 3/2^- , 0
)$$\rightarrow$$^{11}$Be$( 1/2^- , 320 )$ transition, with only
about $0.2$\% of $\mu$ stops in $^{11}$B undergoing this reaction.

Two $^{11}$B$( 3/2^- , 0 )$$\rightarrow$$^{11}$Be$( 1/2^- , 320 )$
measurements, the pioneering work by \citet{De68} at CERN and the
recent work by \citet{Wi02} at PSI, have been performed.  Both counted
incoming muons in a plastic scintillator beam telescope, detected
outgoing $\gamma$--rays in a high resolution Ge detector, and used a
natural isotopic abundance boron target.  The $320~$keV $\gamma$--ray
signal-to-noise ratio was 1:2 in the earlier \citet{De68} experiment
and 5:1 in the later \citet{Wi02} experiment.

Both experiments fit their $320~$keV $\gamma$--ray time spectra to the
time dependence of Eq.\ (\ref{e: gamma times}) in order to extract
$\Lambda_+ / \Lambda_-$.  The instrumental time resolutions were
determined via muonic x--rays and the continuum backgrounds were
subtracted via neighboring energy windows.
\citet{Wi02} analyzed
both the $320~$keV $\gamma$--ray time spectrum and the Michel time
spectrum in order to reduce the correlations between $\Lambda_D^-$,
$\Lambda_h$ and $\Lambda_+ / \Lambda_-$.  For $^{11}$B$( 3/2^- , 0 )$
$\rightarrow$ $^{11}$Be$( 1/2^- , 320 )$ the \citet{De68} experiment
yielded $\Lambda_+ / \Lambda_- < 0.26$ and the \citet{Wi02} experiment
yielded $\Lambda_+ / \Lambda_- = 0.028 \pm 0.021$.  These experiments
clearly demonstrate the strong dependence of the capture process on
the hyperfine state.

In allowed capture on $^{23}$Na nuclei a large fraction of
Gamow--Teller strength is to several levels in the energy region
1--4~MeV \citep{Si95}.  In particular two $\Delta J^{\pi} = -1^+$
reactions, $^{23}$Na$( 3/2^+ , 0 )$$\rightarrow$$^{23}$Ne$( 1/2^+ ,
1017 )$ and $^{23}$Na$( 3/2^+ , 0 )$$\rightarrow$$^{23}$Ne$( 1/2^+ ,
3458 )$, exhaust a sizable fraction of GT strength.  For reference,
the $^{23}$Na disappearance rate is $\Lambda_D = ( 0.831 \pm 0.002 )
\times 10^6 s^{-1}$ \citep{Su87} and the $^{23}$Na hyperfine
conversion rate is $\Lambda_h = ( 15.5 \pm 1.1 ) \times 10^6 s^{-1}$
\citep{Go94}.

Compared to $^{11}$B, in $^{23}$Na (i) the capture rates are
considerably larger and (ii) the hyperfine rate and disappearance rate
are more readily distinguished, thus making the measurement more
straightforward.  However the large number of $^{23}$Ne states, and
greater fragmentation of Gamow--Teller strength, means cascade feeding
from higher--lying levels to lower--lying levels is a worry.

The study of the hyperfine effect in the $\mu$$^{23}$Na atoms was
conducted at TRIUMF \citep{Go94,Jo96}.  Incident muons were counted in
a plastic scintillator telescope and stopped in a sodium metal target.
Emerging $\gamma$--rays were detected in high purity Ge detectors with
surrounding NaI Compton suppressors.  For $^{23}$Na$( 3/2^+ , 0
)$$\rightarrow$$^{23}$Ne$( 1/2^+ , 1017 )$ and $^{23}$Na$( 3/2^+ , 0
)$$\rightarrow$$^{23}$Ne$( 1/2^+ , 3458 )$ the measurement yielded
$\Lambda_+ / \Lambda_- = 0.18 \pm 0.03$ and $\Lambda_+ / \Lambda_-
\leq 0.19$ and revealed a very large hyperfine effect.  Note that in
analyzing their data the authors had to account for the direct
production and the indirect production of the $1017~$keV $\gamma$--rays.
For more details on the interpretation of their data see \citet{Jo96}.

\subsubsection{$^{16}$O$(0^+ , 0)$ $\rightarrow$ $^{16}$N$( 0^- , 120)$}
\label{s: 16O experiment}

The $^{16}$O$( 0^+ , 0 ) \rightarrow ^{16}$N$( 0^- , 120 )$ transition
is a first forbidden transition from the $J_i^{\pi} = 0^+$, $^{16}$O
ground state to the $J_f^{\pi} = 0^-$, $^{16}$N metastable state.  As
discussed in Sec.\ \ref{s: rates and gp} the capture rates of $\Delta
J^{\pi} = 0^-$ transitions are especially sensitive to the
longitudinal component of the axial current, and hence to $g_p$.  The
$^{16}$N level structure comprises four particle--stable bound states:
the ($2^-$, 0) ground state and ($0^-$, 120), ($3^-$, 298), and ($1^-$,
397) excited states.  Note that the $^{16}$N$(0^- , 120)$ state
decays, with a half-life $t_{1/2} = 5.25 \mu$s, both by $\gamma$--ray
emission to the $^{16}$N$( 2^-, 0 )$ ground state and by $\beta$--ray
emission to the $^{16}$O$( 0^+ , 0)$ ground state.  Also note that the
dominant decay of the $1^-$, $^{16}$N state is via $\gamma$--ray
cascade through the $0^-$, $^{16}$N state.

One experimental difficulty is the low rate of the $^{16}$O$( 0^+ , 0
)$ $\rightarrow$ $^{16}$N$( 0^- , 120 )$ transition.  This difficulty
is compounded by the low energy of the de--excitation $\gamma$--ray,
resulting in large backgrounds from Michel bremsstrahlung and Compton
scattering.  Furthermore because of the meta--stability of the
$^{16}$N$( 0^- , 120 )$ state the $120~$keV $\gamma$--ray time
spectrum is a convolution of the $1.8 \mu s$ $\mu$$^{16}$O lifetime
and the $5.2 \mu s$ $^{16}$N$( 0^- , 120 )$ lifetime.  Consequently
some care is needed in applying $\mu$--stop timing gates as the
$120~$keV $^{16}$N $\gamma$--ray and other $^{16}$N $\gamma$--rays are
impacted differently.  Last the feeding of the $^{16}$N$(0^- , 120)$
level from the $^{16}$N$(1^- , 398)$ level also complicates the
extraction of $\Lambda$.

The first experimental studies of $^{16}$0 $\rightarrow$ $^{16}$N
partial transitions were conducted by \citet{Co63}, \citet{Coh64} and
\citet{As64} in the early sixties.  The experiments detected the
de--excitation $\gamma$--rays from $^{16}$N excited states using NaI
detectors.  Unfortunately, because of the limited resolution of the
NaI detectors, they suffered from poor signal--to--noise and
unidentified background lines.  Consequently from the late sixties to
the late seventies a further series of $\gamma$--ray experiments were
conducted using Ge detectors by \citet{De69}, \citet{Ka73} and
\citet{Gu79}.

The experimental results for $\mu$ capture rates to $^{16}$N bound
states are summarized in Table \ref{t: 16O results}.  All experiments
indicate substantial capture to the $2^-$ ground state, some capture
to the $0^-$ and $1^-$ excited states, and negligible capture to the
$3^-$ excited state.  Unfortunately measurement--to--measurement
discrepancies of a factor of two are apparent for $^{16}$O($0^+ , 0)$
$\rightarrow$ $^{16}$N($0^- , 120)$.  Probably one should reject the
early measurements with NaI detectors due to poor resolution and thus
uncertain backgrounds.  Further the peculiar time dependence of the
interesting $120~$keV $\gamma$--ray is probably the origin of
disagreement between the early studies of \citet{De69} and later
studies of \citet{Ka73} and \citet{Gu79}.  In the later experiments
special efforts was made to tackle this difficulty.
\citet{Ka73}  employed 
a continuous beam without a $\mu$--stop time gate and \citet{Gu79}
employed a pulsed beam with a long $\mu$--stop time gate.  These
experiments, using different timing techniques and yielding consistent
results, give a mean value for the $^{16}$O($0^+ , 0)$ $\rightarrow$
$^{16}$N($0^- , 0)$ rate of $\Lambda = 1520 \pm 100$s$^{-1}$.

\begin{table}[htpb]
\caption{Muon capture rates in $^{16}$O to the $( 0^- , 120 )$ and
$( 1^- , 397 )$ excited states of $^{16}$N from the published results
of \citet{Co63}, \citet{As64}, \citet{De69}, \citet{Ka73} and
\citet{Gu79} in units of $\times$10$^3$s$^{-1}$. No evidence was found
for production of the $^{16}$N$( 3^- , 298 )$ excited state.}
\label{t: 16O results}
\begin{ruledtabular}
\begin{tabular}{lll}
 & & \\
 Ref.\ & $0^+ \rightarrow 0^-$ & $0^+ \rightarrow 1^-$ \\
 & & \\
\hline
 & & \\
\citet{Co63}  & $1.1 \pm 0.2$           & $1.7 \pm 0.1$          \\
 & & \\
\citet{As64}  & $1.6 \pm 0.2$           & $1.4 \pm 0.2$          \\
 & & \\
\citet{De69}  & $0.85^{+0.15}_{-0.060}$ & $1.85^{+0.36}_{-0.17}$ \\
 & & \\
\citet{Ka73}  & $1.56 \pm 0.03$         & $1.31 \pm 0.11$        \\
 & & \\
\citet{Gu79} & $1.50 \pm 0.11$         & $1.27 \pm 0.09$        \\
 & & \\
\end{tabular}
\end{ruledtabular}
\end{table}

\subsubsection{$^{28}$Si($0^+ , 0$) $\rightarrow$ $^{28}$Al($1^+ , 2201$)}
\label{s: 28Si experiment}

The study of $g_p$ by measurement of $\gamma$--ray angular
correlations in $\mu^-$ [A,Z] $\rightarrow$ $\nu$ [A,Z-1]$^*$
$\rightarrow$ $\gamma$ [A,Z-1]$^{**}$ transitions was originally
proposed by Popov and co--workers.\footnote{See
\citet{Po63,Bu64,Oz65,Bu67a,Bu67b,Bu67c,Bu70b}.}  The authors examined
$\gamma$--ray correlations in allowed and forbidden transitions on $J
= 0$ and $J \neq 0$ targets, and emphasized the $g_p$ sensitivity of
the $0^+ \rightarrow 1^+ \rightarrow 0^+$ spin sequence.

An experimental method for $\gamma$--ray correlation measurements was
subsequently suggested by \citet{Gr68}.  The method exploits the
Doppler energy shift of nuclear $\gamma$--ray from in--flight decay.
To illustrate the method we consider a recoil nucleus with a
$\gamma$--decay lifetime denoted by $\tau$ and a stopping time denoted
by $\tau_S$ (the stopping time is typically $\sim 0.5$~ps in
medium-weight nuclei).  If $\tau << \tau_{S}$ the recoil is in motion
as it decays and consequently the $\gamma$--ray energy is Doppler
shifted by
\begin{equation}
\label{e: doppler}
\frac{\Delta E }{ E_o }  = \frac{ E - E_o }{ E_o } =  \beta cos{ \theta } 
\end{equation}
where $E_o$ is the $\gamma$--ray energy in the recoil reference frame,
$E$ is the $\gamma$--ray energy in the laboratory reference frame,
$\beta = v / c$ is the velocity of the recoil in the laboratory, and
$\theta$ is the angle between the $\gamma$--ray and the recoil
momentum.  As a consequence of Eq.\ (\ref{e: doppler}) the energy
spectra of nuclear $\gamma$--rays from muon capture are Doppler
broadened when $\tau << \tau_{S}$.  Further the exact lineshape of the
Doppler spectrum is a function of the correlations between the
$\gamma$--ray, recoil and $\mu$--spin directions.  As recognized by
\citet{Gr68}, this permits a determination of the $\gamma$--ray
correlation coefficients by measurement of the $\gamma$--ray Doppler
energy spectrum.

Two experimental configurations for correlation measurements have
special significance.  In the first arrangement, the $\gamma$--$\nu$
configuration, the experiment is conducted with either unpolarized
muons or perpendicular geometry, so that $\vec{P}_{\mu} \cdot \hat{k}
= 0$ where $\hat{k}$ is the $\gamma$-ray direction.  {}From Eq.\
(\ref{e: Pmu>0 correlation}) this configuration yields sensitivity to
the correlation coefficient $a_2$ only.  In the second arrangement,
the $\gamma$--$\nu$--$\mu$ configuration, the experiment is conducted
with both polarized muons and non--perpendicular geometry, so that
$\vec{P}_{\mu} \cdot \hat{k} \neq 0$.  {}From Eq.\ (\ref{e: Pmu>0
correlation}) this configuration yields sensitivity to the correlation
coefficients $\alpha + \frac{2}{3} c_1$ and $b_2$ also.  Note that,
due to different powers of $\hat{k}$ in Eq.\ (\ref{e: Pmu>0
correlation}), the Doppler lineshape arising from $a_2$ is symmetric
about $E_o$ whereas the Doppler lineshapes arising from $\alpha +
\frac{2}{3} c_1$ and $b_2$ are asymmetric about $E_o$.

Measurements of $\gamma$--$\nu$ and $\gamma$--$\nu$--$\mu$
correlations have pros and cons.  One disadvantage of
$\gamma$--$\nu$--$\mu$ correlation measurements is the small $\mu$
polarization in the $\mu$--atom ground state.  However one advantage
is that the Doppler lineshape may be manipulated by varying the
$\mu$--spin direction or $\gamma$--detector position, which is helpful
in separating the Doppler signal from the continuum background.
Furthermore distortion of the $\gamma$--ray lineshape due to
slowing--down of the recoil nucleus is more straightforwardly
separated from the asymmetric effects of $\gamma$--$\nu$--$\mu$
correlations than from the symmetric effects of $\gamma$--$\nu$
correlations.

Two $\mu^-$ [A,Z] $\rightarrow$ $\nu$ [A,Z-1]$^*$ $\rightarrow$
$\gamma$ [A,Z-1]$^{**}$ transitions have attracted the most attention,
\begin{equation}
\label{e: 28Si capture 1}
^{28}{\rm Si}( 0^+ , 0 ) \; \longrightarrow \; 
^{28}\!\!{\rm Al}( 1^+ , 2201 ) \; \longrightarrow \; 
^{28}\!\!{\rm Al}( 0^+ ,973 ) 
\end{equation}
\begin{equation}
\label{e: 28Si capture 2}
^{28}{\rm Si}( 0^+ , 0 ) \; \longrightarrow \; 
^{28}\!\!{\rm Al}( 1^+ , 2201 ) \; \longrightarrow \; 
^{28}\!\!{\rm Al}( 2^+ , 37 ) .
\end{equation}
They involve a common GT transition from the $^{28}$Si($0^+$, 0)
ground state to the $^{28}$Al($1^+$, 2201 ) excited state.  Note that
the $0^+ \rightarrow 1^+ \rightarrow 0^+$ sequence involves a pure M1
$\gamma$--decay whereas the $0^+ \rightarrow 1^+ \rightarrow 2^+$
sequence involves a mixed E2/M1 $\gamma$--decay with mixing ratio
$\delta (E2/M1) = 0.37 \pm 0.11$ \citep{Ku98}.  Additionally the
$2201~$keV state lifetime is $59 \pm 6$~fs and slowing--down effects are
non--negligible.  Although at first glance the two sequences involve
six correlations, {\it i.e.}  $(\alpha + \frac{2}{3} c_1 )^{1229}$,
$a_2^{1229}$ and $b_2^{1229}$ for the $1229~$keV $\gamma$--ray and $(
\alpha + \frac{2}{3} c_1 )^{2170}$, $a_2^{2170}$ and $b_2^{2170}$ for
the $2170~$keV $\gamma$--ray, these coefficients are related to a
single underlying dynamical parameter, the helicity amplitude ratio
$X$, as described in Sec.\ \ref{s: J=0 helicities}.

The ground-breaking work on the $\gamma$--ray correlations in these
spin sequences was conducted at SREL by \citet{Mi72b}.  They measured
the $\gamma$--ray lineshapes from $\mu$ stops in both natural Si and
enriched $^{28}$SiO$_2$ targets.  Unfortunately the statistics were
limited and the authors were forced to assume the absence of
distortions of the Doppler lineshape due to slowing--down of the
recoil ion.

More recently an improved measurement of the $a_2$ coefficient was
conducted at TRIUMF by \citet{Mo97}.  The experiment was performed in
perpendicular geometry, {\it i.e.}  with $\vec{P}_{\mu} \cdot \hat{k}
= 0$, and utilized a coincidence technique with Compton suppression to
improve the $1229~$keV $\gamma$--ray signal--to--noise.  In analyzing
the data the authors treated the $a_2$ coefficients for $1229~$keV and
$2170~$keV $\gamma$--rays as independent, since the $2170~$keV
multipolarity was not measured at the time.  However they included
recoil slowing--down effects in the fit of the lineshapes.

Also recently a new measurement of all the coefficients has been
performed at the Dubna phasotron by \citet{Br95} and \citet{Br00}.  To
identify both $( \alpha + \frac{2}{3} c_1 )$ and $b_2$ the authors
recorded the Doppler spectra for different values of $\vec{P}_{\mu}
\cdot \hat{k}$.  In \citet{Br95} they used forward/backward positioned
Ge detectors to vary $\vec{P}_{\mu} \cdot \hat{k}$ and in \citet{Br00}
they used $\mu$ spin precession to vary $\vec{P}_{\mu} \cdot \hat{k}$.
In analyzing their data the authors enforced the dynamical relations
between correlation coefficients and fit the $2201$~keV lifetime and
$2170$~keV mixing ratio.

Note that a concern in all experiments is the production of the
$^{28}$Al$( 1^+ , 2201 )$ state by either (i) ($\mu$, n$\nu$) or
($\mu$, nn$\nu$) capture on $^{29}$Si or $^{30}$Si isotopes, or (ii)
($\mu$,$\nu$) capture to higher--lying $^{28}$Si levels.  Such
contributions would distort the lineshapes and impact the extraction
of $(\alpha + \frac{2}{3} c_1 )$, $a_2$ and $b_2$.  Note that
\citet{Mi72a} obtained constraints on contributions from
$^{29}$Si($\mu$, n$\nu$) and \citet{Mo97} and \citet{Br00} obtained
limits on cascade feeding from higher levels.  However a small
contribution from (i) or (ii) is not excluded.

In Table \ref{t: 28Si results} we summarize the various measurements
of $\gamma$--ray correlations for $^{28}$Si$( 0^+, 0 )$ $\rightarrow$
$^{28}$Al$( 1^+, 2201 )$.  The experimental configurations are
denoted by $\vec{P}_{\mu} \cdot \hat{k} = 0$ or $\vec{P}_{\mu} \cdot
\hat{k} \neq 0$ and indicate the absence or presence of sensitivity to
$(\alpha + \frac{2}{3} c_1 )$ and $b_2$.  To assist the comparison of
experiments we quote all results in terms of the helicity amplitude
ratio $X$.  The `world data' weighted mean is $X = 0.554 \pm 0.042$.

\begin{table}[htpb]
\caption{Compilation of recent results from the $\gamma$--ray correlation
experiments for the transition $^{28}$Si$( 0^+, 0 )$ $\rightarrow$
$^{28}$Al$( 1^+, 2201 )$.  The results are presented in terms of the
dynamical parameter $X = \sqrt{2} T_0 / T_1$.  The analyzed gamma-rays
are listed in column two and the experimental technique is listed in
column three.  FPA is the Fujii-Primakoff approximation value.}
\label{t: 28Si results}
\begin{ruledtabular}
\begin{tabular}{llll}
 & & & \\
 Ref. & $\gamma$--ray & $\gamma$--ray & 
$X = \sqrt{2} T_0 / T_1$ \\
 & trans. & corr. & \\
 & & & \\
\hline
 & & & \\
\citet{Mo97} & 1229 &  `$\gamma$--$\nu$' & $0.454^{+0.12}_{-0.11}$ \\
 & & & \\
\citet{Br95} & 1229, 2171 & `$\gamma$--$\nu$--$\mu$' & $0.543 \pm 0.052$ \\
 & & & \\
\citet{Br00} & 1229, 2171 & `$\gamma$--$\nu$--$\mu$' & $0.566 \pm 0.045$ \\
 & & & \\
 FPA & & & 0.59 \\
  & & & \\
\end{tabular}
\end{ruledtabular}
\end{table}

\subsection{Theoretical framework for exclusive OMC}
\label{s: formalism}

Herein we describe the theoretical treatment of physical observables
in $(\mu$, $\nu$) reactions.  Our main goals are to outline the steps
and assumptions in calculating the observables and provide some
details of work on $^{11}$B, $^{12}$C, $^{16}$O, $^{23}$Na and
$^{28}$Si.  In Sec.\ \ref{s: operators} we describe the operators that
contribute to $\mu$ capture.  In Sec.\ \ref{s: impulse approximation}
we discuss the application of the impulse approximation and in Sec.\
\ref{s: exchange currents} we discuss the effects due to exchange
currents.  The detailed discussion of nuclear models is left to
Sec. \ref{s: nuclear models}.

\subsubsection{Multipole operators}
\label{s: operators}

The model calculation of partial transitions on complex nuclei is
generally conducted via a multipole expansion of the Fujii--Primakoff
effective Hamiltonian of Eq. (\ref{e: FP hamiltonian}).  We refer the
reader who is interested in the details of the formalism to the
articles by
\citet{Pr59}, \citet{Mo60}, \citet{Lu63},
\citet{Wa75}, \citet{Mu77} and  \citet{Ci84}.
Herein we briefly describe the specific operators that contribute to
transitions of interest for $g_p$, {\it i.e.} the transitions $0^+
\rightarrow 0^-$, $0^+ \rightarrow 1^+$ and $3/2^{\pm} \rightarrow
1/2^{\pm}$.  We employ the notation of \citet{Wa75} for the
contributing electroweak operators, (${\cal L}_J\!-\!{\cal M}_J$ and
${\cal T}^{el}_J\!-\!{\cal T}^{mag}_J$), and notation of
\citet{Do79,Do80} for the contributing multipole operators, $M_J(qx)$,
$\boldsymbol{M}_{JL}(qx) \cdot \boldsymbol{\sigma}$,
$\boldsymbol{M}_{JL}(qx) \cdot \boldsymbol{\nabla}$ and $M_{J}(qx) ~
\boldsymbol{\sigma} \cdot \boldsymbol{\nabla}$.  Note that ${\cal
L}_J\!-\!{\cal M}_J$ involves a longitudinal coupling to the lepton
field and is dependent on $g_p$, while ${\cal T}^{el}_J\!-\!{\cal
T}^{mag}_J$ involves a transverse coupling to the lepton field and is
independent of $g_p$.

$0^+ \rightarrow 0^-$ transitions, {\it e.g.}\ $^{16}$O, involve a
unique $J^{\pi} = 0^-$ multipole, a single electroweak weak operator,
$L^5_0 - M^5_0$, and two multipole operators, $\boldsymbol{M}_{01}
\cdot \boldsymbol{\sigma}$ and $M_0 \boldsymbol{\sigma} \cdot
\boldsymbol{\nabla}$.  The $\boldsymbol{M}_{01} \cdot
\boldsymbol{\sigma}$ operator is the $\ell = 1$ retarded Gamow--Teller
operator originating from the space component of the axial current.
The $M_0 \boldsymbol{\sigma} \cdot \boldsymbol{\nabla}$ operator is
the axial charge operator originating from the time component of the
axial current.  Note that the contribution of $g_p$ in $0^+
\rightarrow 0^-$ transitions is via $\boldsymbol{M}_{01} \cdot
\boldsymbol{\sigma}$.

$0^+ \rightarrow 1^+$ transitions, {\it e.g.}\ $^{12}$C and $^{28}$Si,
involve a unique $J^{\pi} = 1^+$ multipole, two electroweak weak
operators, $L^5_1 - M^5_1$ and $T^{el5}_1 - T^{mag}_1$, and four
multipole operators, $\boldsymbol{M}_{10} \cdot \boldsymbol{\sigma}$,
$\boldsymbol{M}_{12} \cdot \boldsymbol{\sigma}$, $M_1
\boldsymbol{\sigma} \cdot \boldsymbol{\nabla}$ and
$\boldsymbol{M}_{11} \cdot \boldsymbol{\nabla}$.  The
$\boldsymbol{M}_{10} \cdot \boldsymbol{\sigma}$ operator is the
allowed GT operator.  The remaining contributions include the axial
current's time--component, {\it i.e.} $M_1 \boldsymbol{\sigma} \cdot
\boldsymbol{\nabla}$, and second--forbidden corrections, {\it i.e.}
$\boldsymbol{M}_{12} \cdot \boldsymbol{\sigma}$.  Note that the
leading contribution of $g_p$ in $0^+ \rightarrow 1^+$ transitions is
via $\boldsymbol{M}_{10} \cdot \boldsymbol{\sigma}$.
 
For transitions on $J_i \neq 0$ targets a range of multipoles are
involved, {\it i.e.} $| J_i - J_f |$ to $( J_i + J_f )$.  For example,
a $1/2^+ \rightarrow 1/2^+$ transition, {\it e.g.} $^1$H or $^3$He,
involves $J^{\pi} = 0^+, 1^+$ multipoles and a $3/2^+ \rightarrow
1/2^+$ transition, {\it e.g.} $^{11}$B or $^{23}$Na, involves $J^{\pi}
= 1^+, 2^+$ multipoles.  For $1/2^+ \rightarrow 1/2^+$ transitions the
$L_0 - M_0$ operator yields an additional contribution from the
allowed Fermi operator.  For $3/2^+ \rightarrow 1/2^+$ transitions the
$L_2 - M_2$ and $T^{el}_2 - T^{mag5}_2$ operators yield additional
contributions from the $\ell$-forbidden multipole operators $M_{2}$
and $\boldsymbol{M}_{22} \cdot \boldsymbol{\sigma}$ and the gradient
multipole operators $\boldsymbol{M}_{21} \cdot \boldsymbol{\nabla}$
and $\boldsymbol{M}_{23} \cdot \boldsymbol{\nabla}$.  However the
leading contribution of $g_p$ in $1/2^+ \rightarrow 1/2^+$ transitions
and $3/2^+ \rightarrow 1/2^+$ transitions is still via
$\boldsymbol{M}_{10} \cdot \boldsymbol{\sigma}$.  Note that the
multipoles with $J^{\pi} = 0^+$ and $J^{\pi} = 2^+$ are independent of
$g_p$.

\subsubsection{Impulse approximation}
\label{s: impulse approximation}

In principle the weak amplitudes in muon capture have one--, two-- and
many--body contributions.  However in practice the starting point for
most calculations is to approximate the weak nuclear amplitude by a
summation of A one--body amplitudes, {\it i.e.} the impulse
approximation.  This amounts to ignoring the effects of pion exchange
currents, $\Delta$--hole excitations, etc.

Assuming a one--body form for nuclear currents, the required
multi--particle weak matrix elements
\mbox{$<J_f||O^{J}||J_i>$}
between an initial state $| J_i >$ and final state $| J_f >$ may be
written in terms of single--particle weak matrix elements $<
\alpha^{\prime} || O^{J} || \alpha >$ between single--particle states
labeled by $| \alpha >$ and $| \alpha^{\prime} >$ as
\citep{Do79}
\begin{equation}
\label{e: transition densities}
< J_f || O^{J} || J_i > =  
\sum_{\alpha , \alpha^{\prime}} C(J,  \alpha , \alpha^{\prime}, J_f, J_i) 
 < \alpha' || O^{J} || \alpha >  
\end{equation}
where the states $\alpha \equiv [ n, j, \ell ]$ and $\alpha^{\prime}
\equiv [ n^{\prime}, j^{\prime}, \ell^{\prime} ]$ and in $\mu$ capture
the operator $O^{J}$ is ${\cal M}_J - {\cal L}_J$ or ${\cal T}^{el}_J
- {\cal T}^{mag}_J$.  The coefficients $C( J , \alpha ,
\alpha^{\prime}, J_f, J_i)$ are called one--body transition densities
and determine the contributions of each single--particle matrix
element $< \alpha^{\prime} || O^{J} || \alpha >$ to the
multi--particle matrix element $< J_f || O^{J} || J_i >$.  Note that
the one--body transition densities are determined by the nuclear
structure while the single--particle matrix elements are functions of
the weak couplings.  Eq.\ (\ref{e: transition densities}) therefore
represents a convenient separation of nuclear structure and weak
dynamics.\footnote{Strictly the radial form of the nuclear wave
functions also enters the computation of the single--particle matrix
elements.}  We stress that in practice in applying Eq.\ (\ref{e:
transition densities}) the summation is truncated to a finite number
of the single particle transitions.

\subsubsection{Exchange currents}
\label{s: exchange currents}

At some level the impulse approximation will break down, and
consequently the evaluation of contributions from pion exchange, delta
excitation, etc., is important in extracting $g_p$.  One approach to
computing exchange currents is to use low energy theorems to constrain
soft pion contributions.  Another approach involves enumerating a
plausible set of Feynman diagrams that incorporate $\pi$'s,
$\Delta$'s, etc.  We shall not attempt to cover in detail the broad
topic of exchange currents, but rather we summarize their application
to muon capture on complex nuclei.

The work of \citet{Ku78} was pivotal in establishing the importance of
soft pion exchange currents in various electroweak processes.  The
authors observed that soft pions produce large effects in the time
component of the axial current and the space component of the vector
current.  Further they recognized that arguments based on chiral
symmetry fix the size of these effects, yielding a powerful tool in
determining the corrections from 2--body currents.  In particular for
muon capture the soft--pion corrections to axial charge operators,
while substantial, are well determined.  Such $M_{J}
\boldsymbol{\sigma} \cdot \boldsymbol{\nabla}$ operators compete with
the leading contributions of $g_p$ in both allowed and $0^+
\rightarrow 0^-$ transitions.  For further details see
\citet{Gu78,GuS79} and \citet{To86}.

Unfortunately, for the space component of the axial current and the
time component of the vector current the constraints dictated by
chiral symmetry are ineffective in determining the contributions of
2--body currents.  For example for the allowed Gamow-Teller operator
and the allowed Fermi operator this approach is not helpful.  Instead
the modifying effects of exchange currents must be addressed by
explicitly evaluating a specific set of Feynman graphs.  An example of
such an approach is the N$\pi$$\rho$$A_1$ phenomenological Lagrangian
model of \citet{Iv79a}.  Note that the Feynman graphs and
corresponding operators are obviously not unique and the couplings
constants and form factors are frequently not well known.
Consequently the calculation of 2--body corrections to GT matrix
elements and Fermi matrix elements have significant uncertainties.  In
particular the $\Delta$'s contribution is poorly determined.  For
examples of applications to muon capture see \citet{Ad90}.

\subsection{Nuclear models for partial transitions}
\label{s: nuclear models}

Next we consider the specific structure and nuclear models for partial
transitions on A$= 11, 12, 16, 23$ and $28$ nuclei.  Our focus is on
the elements of the models that impact the determination of $g_p$.

Nowadays full--space shell model calculations are routinely performed
for $4 < A < 16$, {\it i.e.} 0p shell nuclei, and $16 < A < 40$, {\it
i.e.} 1s--0d shell nuclei, and the empirical determination of
effective interactions from least--squares--fits to nuclear data is
well established.  For example, the parameters of the 0p shell
interaction were obtained by \citet{Co67} by fitting $4 < A < 16$
level energies and the parameters of the 1s--0d interaction were
obtained by \citet{Wi84} by fitting $16 < A < 40$ level energies.  In
addition semi--empirical interactions, which incorporate assumptions
for the particular form of the effective interaction, are available
\citep{Br88}.
Both empirical and semi--empirical interactions are capable of
reproducing many features and phenomena in 0p and 1s--0d nuclei.

Note that an alternative approach is the microscopic derivation of the
effective NN interaction from the free NN interaction.  The derivation
involves a power--series relating the effective interaction and free
interaction, but unfortunately the question of convergence is tricky.
In general such interactions either give less satisfactory model--data
agreement than empirical interactions, {\it e.g.} the interaction of
\citet{Ku66}, or need some ad--hoc tuning of model parameters, {\it
e.g.} the interaction of \citet{Ha73}.  However the comparison of
results from empirical interactions and realistic interactions is
helpful in understanding and evaluating the model uncertainties.

\subsubsection{$^{11}$B$( 3/2^- , 0 )$ $\rightarrow$ 
$^{11}$Be$( 1/2^- , 320 )$}
\label{s: 11B theory}

Several calculations of capture rate and hyperfine dependences have
been performed for the $^{11}$B$( 3/2^- , 0 )$ $\rightarrow$
$^{11}$Be$( 1/2^- , 320 )$ transition
\citep{Be71,Ko82,Ko84,Ku94,Su97}.
\citet{Be71} first discussed
the relatively high sensitivity to $g_p$ and relatively low
sensitivity to nuclear structure of $\Lambda_+ / \Lambda_-$.  More
recently \citet{Ku94} assessed the effects of different 0p--shell
effective interactions while \citet{Su97} assessed the effects of a
$^{11}$Be neutron halo.

The simplest picture of $^{11}$B$( 3/2^- , 0 )$ $\rightarrow$ 
$^{11}$Be$( 1/2^- , 320 )$ consists of a $0s_{1/2}^4$$0p_{3/2}^7$
initial state, $0s_{1/2}^4$$0p_{3/2}^6$$0p_{1/2}^1$ final state, and
$0p_{3/2} \rightarrow 0p_{1/2}$ single particle transition.  However
in reality the $^{11}$B$( 3/2^- , 0 )$ initial state has substantial
contributions from both $0s_{1/2}^4$$0p_{3/2}^5$$0p_{1/2}^2$ and
$0s_{1/2}^4$$0p_{3/2}^3$$0p_{1/2}^4$ configurations, and interference
between the $0p_{1/2} \rightarrow 0p_{3/2}$ single particle transition
and the $0p_{3/2} \rightarrow 0p_{1/2}$ single particle transition is
important.  For example the capture rate is grossly over--estimated in
a simple $0p_{3/2} \rightarrow 0p_{1/2}$ picture.

Full 0p--shell model calculations for $^{11}$B$( 3/2^- , 0 )$
$\rightarrow$ $^{11}$Be$( 1/2^- , 320 )$ with well established
effective interactions confirm the leading transition is $0p_{3/2}
\rightarrow 0p_{1/2}$, with a substantial contribution from $0p_{1/2}
\rightarrow 0p_{3/2}$ and a significant contribution from $0p_{3/2}
\rightarrow 0p_{3/2}$.  Since the Gamow--Teller matrix elements for
$0p_{3/2} \rightarrow 0p_{1/2}$ and $0p_{1/2} \rightarrow 0p_{3/2}$
have opposite signs they interfere destructively and dramatically
decrease both $\Lambda_+$ and $\Lambda_-$.  Typical model--to--model
variations in the $0p_{3/2} \rightarrow 0p_{1/2}$ and $0p_{1/2}
\rightarrow 0p_{3/2}$ densities are roughly $\pm 10\%$.  Note that the
remaining $0p_{1/2} \rightarrow 0p_{1/2}$ and $0p_{3/2} \rightarrow
0p_{3/2}$ densities show larger model--to--model variations, however
their contributions to capture are smaller.

A unique complication for the $^{11}$B$( 3/2^- , 0 )$ $\rightarrow$
$^{11}$Be$( 1/2^- , 320 )$ transition is the $^{11}$Be neutron halo.
The halo is interesting in the context of nuclear structure studies
but worrisome in the context of extracting $g_p$.  Recently
\citet{Su97} has assessed the impact of the neutron halo on the
capture rate and its hyperfine dependence.
\citet{Su97} reported it tends to reduce both $\Lambda_+$ and $\Lambda_-$
but barely changes $\Lambda_+ / \Lambda_-$.

\subsubsection{$^{12}$C$( 0^+ , 0 )$ $\rightarrow$ $^{12}$B$( 1^+ , 0 )$}
\label{s: 12C theory}

A large number of model calculation are available for $^{12}$C$( 0^+ ,
0 )$ $\rightarrow$ $^{12}$B$( 1^+ , 0 )$.  Early studies of $\mu$
capture rates and $\beta$--decay rates were performed by \citet{Fl59},
\citet{Oc72}, \citet{Mu71}, \citet{Im75}, and \citet{Mu78}.  More
comprehensive studies of rates and polarizations in $\mu$ capture were
published by Morita and co--workers,\footnote{See
\citet{Ko78,Am81,Fu83a,Fu83b,Ko85,Fu87,Mo94}.}
and by \citet{Su76}, \citet{Ro79}, \citet{Ci81},
\citet{Hay00} and \citet{Au02}.
They include investigations of two--body currents and core
polarization effects.

The simplest picture of $^{12}$C$( 0^+ , 0 )$ $\rightarrow$ $^{12}$B$(
1^+ , 0 )$ consists of a $0s_{1/2}^4$$0p_{3/2}^8$ initial state,
$0s_{1/2}^4$$0p_{3/2}^7$$0p_{1/2}^1$ final state, and a $0p_{3/2}
\rightarrow 0p_{1/2}$ single particle transition.  Like the $^{11}$B
g.s., the $^{12}$C g.s.  has substantial contributions from both
$0s_{1/2}^4$$0p_{3/2}^6$$0p_{1/2}^2$ and
$0s_{1/2}^4$$0p_{3/2}^4$$0p_{1/2}^4$ configurations, and the
interference of amplitudes from $0p_{3/2} \rightarrow 0p_{3/2}$
transitions and $0p_{1/2} \rightarrow 0p_{3/2}$ transitons is
important.  Again the capture rate is grossly over--estimated by a
simple $0p_{3/2} \rightarrow 0p_{1/2}$ picture.

Note that the overall pattern of the transition densities obtained
from full 0p--shell model calculations with well established effective
interactions is quite similar for $^{11}$B$( 3/2^- , 0 )$
$\rightarrow$ $^{11}$Be$( 1/2^- , 320 )$ and $^{12}$C$( 0^+ , 0 )$
$\rightarrow$ $^{12}$B$( 1^+ , 0 )$.  Specifically for $^{12}$C the
largest density is $0p_{3/2} \rightarrow 0p_{1/2}$, the
next--to--largest density is $0p_{1/2} \rightarrow 0p_{3/2}$, and the
contributions from $0p_{3/2} \rightarrow 0p_{3/2}$ and $0p_{1/2}
\rightarrow 0p_{1/2}$ are small.  As in $^{11}$B, in $^{12}$C the
interference of $0p_{3/2} \rightarrow 0p_{1/2}$ with $0p_{1/2}
\rightarrow 0p_{3/2}$ is important in reducing the Gamow--Teller
matrix element.

A nice feature of $^{12}$C$( 0^+ , 0 )$ $\rightarrow$ $^{12}$B$( 1^+ ,
0 )$ is that related data on other electroweak processes are
available, {\it e.g.}  $^{12}$B $\beta$--decay, $^{12}$C(e,e')
scattering and $^{12}$C$( 1^+ , 15.1~$MeV$ )$ $\gamma$--decay, and are
helpful in testing the model calculations.  For example the one-body
transition densities have been extracted from these data by
\citet{Do95} and \citet{Ha78} and generally support the model
calculations.

\subsubsection{$^{23}$Na$(3/2^+ , 0 ) \rightarrow 
^{23}$Ne$(1/2^+, 3458 )$}
\label{s: 23Na theory}

Calculations of capture rates and hyperfine dependencies in $^{23}$Na
$\rightarrow$ $^{23}$Ne transitions have been performed by
\citet{Jo96}, \citet{Ko97} and \citet{Si98}.  The calculations have
been conducted in the full 1s--0d model space using the Wildenthal
empirical interaction \citep{Wi84} and the Kuo--Brown realistic
interaction \citep{Ku66}.  A study of 2--body currents and core
polarization was made by \citet{Ko97}.

The $^{23}$Na ground state is $J^{\pi} = 3/2^+$, indicating the
simplest picture of a single unpaired proton in a $0d_{5/2}$ orbital
is wrong.  Rather the $A = 20$--$24$ mass region is well known for
examples of light deformed nuclei and rotational band spectra.
Consequently the spherical shell model representation of these nuclei
is quite complex, with $A = 23$ wave functions having small
occupancies for the (d$_{5/2}$)$^7$ configuration and large
occupancies of the 1s$_{1/2}$, 1d$_{3/2}$ orbitals.

For concreteness we describe the $^{23}$Na$( 3/2^+ , 0 )$
$\rightarrow$ $^{23}$Ne$( 1/2^+ , 3458)$ transition, which is the
strongest transition in the $\mu^-$$^{23}$Na experiment.  Full 1s--0d
shell model calculations with well tested effective interactions show
$0d_{5/2}$ $\rightarrow$ $0d_{3/2}$ is the strongest single particle
transition and $0d_{3/2}$ $\rightarrow$ $0d_{3/2}$ is the next
strongest single particle transition.  Other contributions are
typically 10--20\% of $0d_{5/2}$ $\rightarrow$ $0d_{3/2}$.  The
variations of the one-body transition densities between models are
$\leq 10\%$ for $0d_{5/2}$ $\rightarrow$ $0d_{3/2}$ densities and
$\leq 25\%$ for $0d_{3/2}$ $\rightarrow$ $0d_{3/2}$ densities.

\subsubsection{$^{28}$Si$(0^+ , 0 ) \rightarrow ^{28}$Al$(1^+, 2201 )$}
\label{s: 28Si theory}

Several authors have calculated the rates and correlations for
$^{28}$Si$( 0^+ , 0 )$ $\rightarrow$ $^{28}$Al$( 1^+ , 2201 )$.  The
first efforts were made by \citet{Ci76} and \citet{Pa78,Pa81}, but
employed relatively crude nuclear wave functions, the 1p--1h wave
function of \citet{Do70} and the truncated 1s--0d wave function of
\citet{Mc71}.  More recently \citet{Ku00}, \citet{Ku01}, \citet{Si99},
and \citet{Ci98} have conducted full 1s--0d calculations.
\citet{Si99} and \citet{Ci98}
also considered both 2--body currents and  core polarization
in $^{28}$Si$( 0^+ , 0 )$ $\rightarrow$ $^{28}$Al$( 1^+ , 2201 )$.

In the simplest picture $^{28}$Si$( 0^+ , 0 )$ $\rightarrow$
$^{28}$Al$( 1^+ , 2201 )$ consists of a full 0d$_{5/2}^{12}$ initial
state, $J^{\pi} = 1^+$ ($0d_{5/2}^{-1}$, $0d_{3/2}^1$) final state,
and 0d$_{5/2}$ $\rightarrow$ 0d$_{3/2}$ single particle transition.
However the model calculations show the simple picture is insufficient
and configurations with several particles in 1s$_{1/2}$--0d$_{3/2}$
orbitals are important.
 
A special remark on the one--body transition densities in the
$^{28}$Si$( 0^+ , 0 )$ $\rightarrow$ $^{28}$Al$( 1^+ , 2201 )$
transition is worthwhile.  Unlike the previous examples of $^{11}$B,
$^{12}$C and $^{23}$Na, in $^{28}$Si the variations from model to
model are large, {\it e.g.}  the densities from \citet{Ku66} and
\citet{Wi84} are quite different.  Also no particular single particle
transition is dominant, {\it e.g.} the interaction of \citet{Wi84}
shows $0d_{5/2}$ $\rightarrow$ $0d_{3/2}$,
$0d_{5/2}$ $\rightarrow$ $0d_{5/2}$, and $1s_{1/2}$ $\rightarrow$
$0d_{3/2}$ with similar magnitudes.  Therefore, as discussed by
\citet{Ku00}, the model calculations are especially sensitive to
interference effects.

\subsubsection{$^{16}$O$(0^+ , 0 ) \rightarrow ^{16}$N$(0^-, 120 )$}
\label{s: 16O theory}

Because the nucleus $^{16}$O is doubly magic and the transition
$^{16}$O($0^+ , 0$) $\rightarrow$ $^{16}$N($0^- , 0$) is first
forbidden this case is special.  The simple model for $^{16}$O($0^+ ,
0$) $\rightarrow$ $^{16}$N($0^- , 0$) comprises a 0p closed shell
initial state, and a 1p--1h $J^{\pi} = 0^-$ final state, and involves
$1p_{1/2}$ $\rightarrow$ $2s_{1/2}$ and $1p_{3/2}$ $\rightarrow$
$1d_{3/2}$ single particle transitions.  Several authors have computed
the rates of $\mu$ capture ($\Lambda_{\mu}$) and $\beta$--decay
($\Lambda_{\beta}$) for $^{16}$O($0^+ , 0$) $\leftrightarrow$
$^{16}$N($0^- , 0$) within this scheme.  For example see \citet{Gu78} 
and references therein.
They found the rates to be strongly dependent on the
$0p_{3/2}$$^{-1}$--$0d_{3/2}$ admixture in the $^{16}$N wave function,
but the ratio $\Lambda_{\mu} / \Lambda_{\beta}$ to be near--model
independent.

Since the work of \citet{Br66} we know the above picture is not
complete, and that 2p--2h configurations in the $^{16}$O ground state
are important.  The effects of 2p--2h configurations on $\mu$ capture
and $\beta$-decay were first investigated by \citet{GuS79} who
considered $(2s_{1/2})^{2}(1p_{1/2})^{-2}$ and
$(1d_{3/2})^{2}(1p_{3/2})^{-2}$ admixtures.  They found the effects of
2p--2h configurations on the axial charge matrix element $M_0
\boldsymbol{\sigma} \cdot \boldsymbol{\nabla}$ and the retarded
Gamow--Teller matrix element $\boldsymbol{M}_{10} \cdot
\boldsymbol{\sigma}$ were opposite in sign, and consequently the
near--cancellation of model uncertainties in $\Lambda_{\mu} /
\Lambda_{\beta}$ broke down.  Subsequently \citet{To81} evaluated the
effects of 2p--2h configurations in $^{16}$N$(0^- , 120 )
\leftrightarrow ^{16}$O$(0^+, 0 )$ transitions with more extensive 
configurations and various effective interactions.  They
concluded that 2p--2h configurations decrease the $\beta$--decay rate
by factors of $2$ to $4$, decrease the $\mu$ capture rate by factors
of $1.5$ to $2$, and increase $\Lambda_{\mu} / \Lambda_{\beta}$ by
factors of $1.5$ to $2$, thus confirming the model dependence observed
by \citet{GuS79}.

Recently \citet{Ha90} and \citet{Wa94} have performed full
4~$\hbar$$\omega$ (3~$\hbar$$\omega$) calculation for $^{16}$O
($^{16}$N) low--lying levels.  These calculations nicely reproduce the
excitation energies of the $^{16}$O isoscalar positive parity states
and the $^{16}$N isovector negative parity states, and indicate
significant 4p--4h probabilities in $^{16}$O.  In both works the
authors stress the large destructive interference between $M_0
\boldsymbol{\sigma} \cdot \boldsymbol{\nabla}$ and
$\boldsymbol{M}_{10} \cdot \boldsymbol{\sigma}$ in $\beta$--decay, and
therefore substantial model dependences in $\Lambda_{\beta}$ and
$\Lambda_{\mu} / \Lambda_{\beta}$.  Consequently the authors argue
that $\Lambda_{\mu}$, not $\Lambda_{\mu} / \Lambda_{\beta}$, is
preferable for extracting the coupling $g_p$.

The foregoing discussions of $^{16}$O$(0^+ , 0 )$ $\leftrightarrow$
$^{16}$N$(0^-, 120 )$ transitions show a worrisome sensitivity to the
multi--particle wave functions.  In addition the matrix element $M_0
\boldsymbol{\sigma} \cdot \boldsymbol{\nabla}$ is modified
considerably by 2--body currents from soft--pion exchange.  These
effects are discussed in detail by \citet{Gu78} and \citet{To81}.  The
calculations indicate that they increase the $\beta$--decay rate by a
factor of about four and increase the $\mu$ capture rate by a factor
of about two.  The contribution of $M_0 \boldsymbol{\sigma} \cdot
\boldsymbol{\nabla}$ is larger in $\beta$--decay than $\mu$ capture.
This further complicates the extraction of $g_p$ from capture on
$^{16}$O.

\subsection{Coupling $g_p$ from partial transitions}
\label{s: gp results}

\subsubsection{Recommended values of $g_p / g_a$}
\label{s: world data}

In Table \ref{t: gp recommended} we list our recommended values for
$g_p / g_a$ from exclusive OMC on complex nuclei.  By `recommended values for $g_p / g_a$' we
mean our assessment of the best values 
from the current world experimental data-set
and the most complete model calculations. In quoting these
values we combined the experimental results of \citet{Po74},
\citet{Po77}, \citet{Ro81b} and \citet{Ku84} to obtain a world average
of the helicity amplitude ratio $X = 0.23 \pm 0.06$ for $^{12}$C$( 0^+
, 0 )$$\rightarrow$$^{12}$B$( 1^+ , 0 )$ and combined the experimental
results of \citet{Br95}, \citet{Mo97} and \citet{Br00} to obtain a
world average of the helicity amplitude ratio $X = 0.55 \pm 0.04$
for $^{28}$Si$(0^+ , 0 )$$\rightarrow$$^{28}$Al$(1^+, 2201
)$.\footnote{In the literature both the helicity amplitude ratio,
denoted $X$, and the neutrino--wave amplitude ratio, denoted $x$, have
been used in this context.  Although both $X$ and $x$ are suitable for
representing the dynamical content of $\Delta J^{\pi} = 1^+$
transitions, they are different, {\it i.e.} $X = ( - 2x + \sqrt{2} ) /
( x + \sqrt{2} ) $.  Therefore it is important not to confuse the two
variables.    See Sec.\ \ref{s: operators}.}  For the hyperfine
dependence on $^{11}$B we employ the results of \citet{Wi02} and for
the hyperfine dependence on $^{23}$Na we employ the results of
\citet{Jo96}.  For the capture rate of the $^{16}$O$(0^+ , 0
)$$\rightarrow$$^{16}$N$(0^-, 120 )$ transition we averaged the
experimental results of \citet{Ka73} and \citet{Gu79}.  Note that in
Table \ref{t: gp recommended} the quoted errors include only
experimental uncertainties.  Also note we quote the results in Table
\ref{t: gp recommended} in terms of $g_p / g_a$ not $g_p$.  In most 
cases the measured quantities are recoil polarizations, $\gamma$-ray
correlations or hyperfine dependences, and therefore are governed by
ratios of nuclear matrix elements and of weak coupling constants.
Consequently quoting $g_p / g_a$ is more natural and more appropriate.

In order to extract the coupling $g_p / g_a$ from experimental data a
model is necessary.  Our model choices, and arguments for selecting
them, are given below.

For $^{11}$B we took the results of \citet{Ku94} yielding $g_p / g_a =
4.3 \pm^{2.8}_{4.3}$.  These authors used the full 0p space with
Cohen--Kurath interaction but omitted the effects of core polarization
and exchange currents.  Note the earlier calculation of \citet{Be71}
yields a similar value of $g_p /g_a = 4 \pm^{3}_{3}$.  According to
\citet{Su97} the effects of the $^{11}$Be neutron halo on the
hyperfine dependence are small.

For $^{12}$C we used the model calculations of \citet{Fu87} and
excited state yields of \citet{Gi81} to obtain $g_p / g_a = 9.8 \pm
1.8$.  These authors used the full 0p space with Hauge--Maripuu
interaction and accounted for core polarization and exchange currents.
The core polarization effects were computed to second--order in
perturbation theory and the exchange currents were computed with
contributions from pair currents, pionic currents and $\Delta$
excitations.  Earlier calculations with more rudimentary wave
functions and less sophisticated treatments of core polarization and
exchange currents gave similar results.

For $^{16}$O a number of determinations of $g_p / g_a$ are published.
We took the value $g_p / g_a = 6.0 \pm 0.4$ from \citet{Wa94}, which
incorporates both 4p--4h $^{16}$O configurations and 3p--3h $^{16}$N
configurations and reproduces the $^{16}$N $\beta$--decay rate.  The
4p--4h calculation of \citet{Ha90} yields a similar value of $g_p /
g_a = 5$--$7$.  We note, however, that the earlier calculations which
omit 4p--4h configurations have generally preferred higher values for
$g_p / g_a$.

\begin{table}[htpb]
\caption{Recommended values of $g_p / g_a$ from
ordinary capture on complex nuclei. The quoted errors are experimental
uncertainties, and do not include model uncertainties.}
\label{t: gp recommended}
\begin{ruledtabular}
\begin{tabular}{ll}
 & \\
transition  & $g_p / g_a$ \\
 & \\
\hline
 & \\
 $^{11}$B$( 3/2^- , 0 )$ $\rightarrow$ $^{11}$Be$( 1/2^- , 320 )$
 & $4.3^{+2.8}_{-4.3}$ \\
 & \\
 $^{12}$C$( 0^+ , 0 )$ $\rightarrow$ $^{12}$B$( 1^+ , 0 )$ 
 & $9.8 \pm 1.8$ \\
 & \\
 $^{16}$O$(0^+ , 0 ) \rightarrow ^{16}$N$(0^-, 120 )$
 & $6.0 \pm 0.4$ \\
 & \\
 $^{23}$Na$(3/2^+ , 0 ) \rightarrow ^{23}$Ne$(1/2^+, 3458 )$
 & $6.6^{+2.6}_{-2.4}$ \\
 & \\
 $^{28}$Si$(0^+ , 0 ) \rightarrow ^{28}$Al$(1^+, 2201 )$ 
 & $1.0^{+1.1}_{-1.2} $ \\
 & \\
\end{tabular}
\end{ruledtabular}
\end{table}

For $^{23}$Na we took the results of \citet{Ko97} yielding $g_p / g_a
= 6.6 \pm^{2.6}_{2.4}$.  The authors used the full 1s--0d space with
the Brown--Wildenthal interaction, and accounted for core polarization
to first--order and exchange currents from soft pions and $\Delta$
excitations.  The results of \citet{Jo96} are similar to \citet{Ko97}.

For $^{28}$Si we took the results of \citet{Si99} yielding $g_p / g_a
= 1.0 \pm^{1.1}_{1.2}$.  The authors used the full 1s--0d space with
the Brown--Wildenthal interaction.  They included core polarization
corrections but omitted exchange current corrections.  Note that
\citet{Ci98} found the effects of soft--pion exchange were negligible.
For comparison, the earlier calculations of \citet{Ci76} and
\citet{Pa81} gave values for $g_p / g_a$ of $3.3 \pm 1.0$ and $1.5
\pm^{0.9}_{1.1}$ respectively.

\subsubsection{Model sensitivities of $g_p / g_a$}
\label{s: model sensitivites}

The interesting observables in allowed transitions are governed by
ratios of nuclear matrix elements.  More specifically, in $0^+
\rightarrow 1^+$ transitions the observables are completely determined
by, and in $3/2^{\pm} \rightarrow 1/2^{\pm}$ transitions the
observables are strongly dependent on, the helicity amplitude ratio $X
= \sqrt{2} T_0 / T_1$.\footnote{Note for $\Delta J^{\pi} = 1^+$
multipoles the helicity amplitude ratio $\sqrt{2} T_0 / T_1$ and
multipole amplitude ratio $\sqrt{2} ( L^5_1 -M^5_1 ) / ( T^{el5}_1 -
M^{mag}_1 ) $ are identical} Therefore understanding the model
uncertainties in computing $X$ is central to tracing the model
dependences in recoil polarizations, $\gamma$--ray correlations and
hyperfine dependences.  For example see \citet{Ju00}.

Tables \ref{t:me1} and \ref{t:me2} show the relative contributions of
$\boldsymbol{M}_{10} \cdot \boldsymbol{\sigma}$, etc., to $L^5_1 -
M^5_1$ (or $T_0$) and $T^{el5}_1 - M^{mag}_1$ (or $T_1$) for the
relevant transitions.  The tables show that $T^{el5}_1 -M^{mag}_1$ is
entirely dominated by $\boldsymbol{M}_{10} \cdot \boldsymbol{\sigma}$.
However, while $\boldsymbol{M}_{10} \cdot \boldsymbol{\sigma}$ is the
leading piece in the $L^5_1 -M^5_1$ term, the contributions from
$M_{1} \boldsymbol{\sigma} \cdot \boldsymbol{\nabla}$ of about
30--40\% and $\boldsymbol{M}_{12} \cdot \boldsymbol{\sigma}$ of up to
25\% are important.  Clearly the leading source of model
dependence in computing $X$ is therefore uncertainties in the ratios
of $M_{1} \boldsymbol{\sigma} \cdot \boldsymbol{\nabla} /
\boldsymbol{M}_{10} \cdot \boldsymbol{\sigma}$ and
$\boldsymbol{M}_{12} \cdot \boldsymbol{\sigma} /
\boldsymbol{M}_{10} \cdot \boldsymbol{\sigma}$,
{\it i.e.} corrections arising from 
axial charge and 2$^{nd}$--forbidden effects.

\begin{table}[htpb]
\caption{Comparison of the corrections from the terms involving
$M_1 \boldsymbol{\sigma} \cdot \boldsymbol{\nabla}$,
$\boldsymbol{M}_{12} \cdot \boldsymbol{\sigma}$, and
$\boldsymbol{M}_{11} \cdot \boldsymbol{\nabla}$ to the multipole
amplitude of $ L^5_1 - M^5_1 $ for $A = 11, 12, 23$ and $28$ and
several effective interactions denoted CKPOT \citep{Co67}, PKUO
\citep{Ku66}, USD \citep{Wi84} and KUOSD \citep{Ku66}.  The values in
columns 3-5 correspond to the percentage change in the matrix elements
as the correction terms are successively included. Note that
$\boldsymbol{M}_{11} \cdot \boldsymbol{\nabla}$ doesn't contribute to
$ L^5_1 - M^5_1 $.}
\label{t:me1}
\begin{ruledtabular}
\begin{tabular}{llclcl}
 & & & & & \\
A & 
Int. &
$L^5_1$-$M^5_1$ &
$M_1 \boldsymbol{\sigma} \cdot \boldsymbol{\nabla}$  &
$\boldsymbol{M}_{12} \cdot \boldsymbol{\sigma}$  &
$\boldsymbol{M}_{11} \cdot \boldsymbol{\nabla}$  \\
 & & & corr. ( \% ) & corr. ( \% ) & corr. ( \% ) \\
 & & & & & \\
\hline
 & & & & & \\
 11 & CKPOT  &  -0.103  &    -30.6  &     -2.6  &      0.0 \\ 
 11 & PKUO   &  -0.094  &    -32.4  &      0.6  &      0.0 \\ 
 12 & CKPOT  &   0.107  &    -34.5  &     -9.4  &      0.0 \\ 
 12 & PKUO   &  -0.071  &    -41.6  &    -18.4  &      0.0 \\ 
 23 & USD    &   0.090  &    -31.8  &      4.1  &      0.0 \\ 
 23 & KUOSD  &  -0.071  &    -44.4  &     13.1  &      0.0 \\ 
 28 & USD    &   0.070  &    -37.2  &    -27.5  &      0.0 \\ 
 28 & KUOSD  &  -0.071  &    -54.5  &    -18.4  &      0.0 \\ 
 & & & & & \\
\end{tabular}
\end{ruledtabular}
\end{table}

\begin{table}[htpb]
\caption{Comparison of the corrections from the terms involving
$M_1 \boldsymbol{\sigma} \cdot \boldsymbol{\nabla}$,
$\boldsymbol{M}_{12} \cdot \boldsymbol{\sigma}$, and
$\boldsymbol{M}_{11} \cdot \boldsymbol{\nabla}$ to the multipole
amplitude of $T^{el5}_1 - T^{mag}_1$ for $A = 11, 12, 23$ and $28$ and
several effective interactions denoted CKPOT
\citep{Co67}, PKUO \citep{Ku66}, USD \citep{Wi84} and KUOSD
\citep{Ku66}.  The values in columns 3-5 correspond to the percentage
change in the matrix elements as the correction terms are successively
included. Note that $M_1 \boldsymbol{\sigma} \cdot
\boldsymbol{\nabla}$ doesn't contribute to $T^{el5}_1 - T^{mag}_1$.}
\label{t:me2}
\begin{ruledtabular}
\begin{tabular}{llclcl}
 & & & & & \\
A & 
Int. &
$T^{el5}_1$-$T^{mag}_1$ &
$M_1 \boldsymbol{\sigma} \cdot \boldsymbol{\nabla}$  &
$\boldsymbol{M}_{12} \cdot \boldsymbol{\sigma}$  &
$\boldsymbol{M}_{11} \cdot \boldsymbol{\nabla}$  \\
 & & & corr. ( \% ) & corr. ( \% ) & corr. ( \% ) \\
 & & & & & \\
\hline
 & & & & & \\
 11 & CKPOT  &  -0.243  &      0.0  &      0.9  &      0.6 \\ 
 11 & PKUO   &  -0.222  &      0.0  &     -0.2  &      1.7 \\ 
 12 & CKPOT  &   0.249  &      0.0  &      3.1  &     -0.4 \\ 
 12 & PKUO   &  -0.164  &      0.0  &      5.4  &      2.1 \\ 
 23 & USD    &   0.213  &      0.0  &     -1.4  &     -2.2 \\ 
 23 & KUOSD  &  -0.170  &      0.0  &     -3.6  &     -4.8 \\ 
 28 & USD    &   0.171  &      0.0  &      8.6  &      4.7 \\ 
 28 & KUOSD  &   0.106  &      0.0  &      4.2  &      0.7 \\ 
 & & & & & \\
\end{tabular}
\end{ruledtabular}
\end{table}

Uncertainties in one--body transition densities are important sources
of model dependencies in $M_{1} \boldsymbol{\sigma} \cdot
\boldsymbol{\nabla} / \boldsymbol{M}_{10} \cdot \boldsymbol{\sigma}$
and $\boldsymbol{M}_{12} \cdot \boldsymbol{\sigma} /
\boldsymbol{M}_{10} \cdot \boldsymbol{\sigma}$.
For example let's consider the transitions on boron and carbon which
involve the interference of a $0p_{3/2} \rightarrow 0p_{1/2}$ single
particle transition and a $0p_{1/2} \rightarrow 0p_{3/2}$ single
particle transition.  Under the interchange of initial and final
states the matrix element $\boldsymbol{M}_{10} \cdot
\boldsymbol{\sigma}$ changes sign but the matrix element $M_{1}
\boldsymbol{\sigma} \cdot \boldsymbol{\nabla}$ does not.  Consequently
$M_{1} \boldsymbol{\sigma} \cdot \boldsymbol{\nabla} /
\boldsymbol{M}_{10} \cdot \boldsymbol{\sigma}$ 
is quite sensitive to the $0p_{1/2} \rightarrow 0p_{3/2}$ admixture in
these $A = 11, 12$ transitions.  Fortunately the destructive
interference of $0p_{1/2} \leftrightarrow 0p_{3/2}$ amplitudes is also
reflected in, and thus calibrated by, the capture rates of $^{11}$B$(
3/2^- , 0 )$ $\rightarrow$ $^{11}$Be$( 1/2^- , 320 )$ and $^{12}$C$(
0^+ , 0 )$ $\rightarrow$ $^{12}$B$( 1^+ , 0 )$.

A special comment is worthwhile for $^{28}$Si$( 0^+ , 0 )$
$\rightarrow$ $^{28}$Al$( 1^+ , 2201 )$.  The calculations indicate
this transition involves the destructive interference of numerous
single particle transitions with comparable one--body densities.  See
Sec. \ref{s: 28Si theory} for details.  Consequently for $^{28}$Si the
calculation of $X$ may be especially sensitive to the model
uncertainties.

The inevitable truncation of model spaces, {\it e.g.} 0p orbitals for
$A = 11, 12$ and 1s--0d orbitals for $A = 23, 28$, is another source
of model dependence in $M_{1} \boldsymbol{\sigma} \cdot
\boldsymbol{\nabla} /
\boldsymbol{M}_{10} \cdot \boldsymbol{\sigma}$
and $\boldsymbol{M}_{12} \cdot \boldsymbol{\sigma} /
\boldsymbol{M}_{10}
\cdot \boldsymbol{\sigma}$.
Such core polarization effects have been studied by
\citet{Fu87} for $^{12}$C,
\citet{Si99} for $^{28}$Si,
and \citet{Ko97} for $^{23}$Na.  For $^{12}$C \citet{Fu87} found 
downward renormalizations of 13\% for $\boldsymbol{M}_{10} \cdot
\boldsymbol{\sigma}$, 33\% for $\boldsymbol{M}_{12} \cdot
\boldsymbol{\sigma}$, and 22\% for $M_{1} \boldsymbol{\sigma} \cdot
\boldsymbol{\nabla}$, and for $^{28}$Si \citet{Si99} found downward
renormalizations of 11\% for $\boldsymbol{M}_{10} \cdot
\boldsymbol{\sigma}$, 30\% for $\boldsymbol{M}_{12} \cdot
\boldsymbol{\sigma}$, and 39\% for $M_{1} \boldsymbol{\sigma} \cdot
\boldsymbol{\nabla}$.  The small effect of core polarization on 
$\boldsymbol{M}_{10} \cdot
\boldsymbol{\sigma}$ is because the model spaces are `complete spaces'
for this operator.  Note that the renormalization of
$\boldsymbol{M}_{10} \cdot \boldsymbol{\sigma}$ has strong support
from experimental data on allowed $\beta$--decay and (p,n)/(n,p)
reactions, and the renormalization of $\boldsymbol{M}_{12} \cdot
\boldsymbol{\sigma}$ has some support from experimental data on second
forbidden $\beta$--decay.  For example see \citet{Wa92} and
\citet{Ma98}.

The contributions arising from exchange currents in allowed
transitions have been studied by \citet{Fu87} for $^{12}$C,
\citet{Ci98} for $^{28}$Si,
and \citet{Ko97} for $^{23}$Na.  For $^{12}$C \citet{Fu87} found
corrections of $-$4\% to $\boldsymbol{M}_{10} \cdot
\boldsymbol{\sigma}$, $+$10\% to $\boldsymbol{M}_{12} \cdot
\boldsymbol{\sigma}$, and $+$41\% to $M_{1} \boldsymbol{\sigma} \cdot
\boldsymbol{\nabla}$.  Recall the large renormalization of the axial
charge operator $M_{1} \boldsymbol{\sigma} \cdot \boldsymbol{\nabla}$
arises from large soft--pion contributions in the axial current's time
component.  This renormalization is supported by experimental data on
first--forbidden $\beta$--decay.

The assumed form for the radial dependence of the nuclear wave
functions is another source of model dependence, {\it e.g.} at the
surface of the nucleus the difference in harmonic oscillator and
Wood--Saxon wave functions are large.  Such effects were investigated
for $^{12}$C, $^{23}$Na and $^{28}$Si by \citet{Ko00}.  They found the
sensitivity of $X$ to the radial form of the nuclear wave function was
typically 5\% or less.

Lastly we consider the determination of $g_p / g_a$ from the capture
rate of the $^{16}$O($0^+ , 0 $) $\rightarrow$ $^{16}$N($0^- , 120 $)
transition.  Note that this requires knowing the absolute values of
$\boldsymbol{M}_{01} \cdot \boldsymbol{\sigma}$ and $M_1
\boldsymbol{\sigma} \cdot \boldsymbol{\nabla}$, {\it i.e.} not ratios
like $M_{1} \boldsymbol{\sigma} \cdot \boldsymbol{\nabla} /
\boldsymbol{M}_{10} \cdot \boldsymbol{\sigma}$.
Also the capture rate is highly sensitive to the 2p--2h, 4p--4h
structure of the $^{16}$O ground state and the 2--body contributions
due to soft--pion exchange.  Consequently, as discussed by
\citet{Wa94} and \citet{Ha90}, the extraction of $g_p /g_a$ from
$^{16}$O($0^+ , 0 $) $\rightarrow$ $^{16}$N($0^- , 120)$ is a
formidable challenge.  Indeed \citet{Wa94} have cautioned their result
for $g_p / g_a$ is highly model dependent.

\subsubsection{Conclusions and outlook for $g_p / g_a$}
\label{s: conclusion}

The results in Table \ref{t: gp recommended} are largely consistent
with PCAC and $g_p / g_a =6.42$.  Specifically the values obtained
from $^{11}$B, $^{16}$O and $^{23}$Na all support the prediction of
PCAC.  The situation for $^{12}$C is more borderline, with theory and
experiment just under $2 \sigma$ apart, but not enough to cause
concern.  In contrast however for $^{28}$Si the experiment
determination and theoretical prediction are in obvious disagreement.

Ignoring the puzzle of $^{28}$Si for now, we believe the theoretical
uncertainties in extracting $g_p / g_a$ from these data are likely
about $\pm$2 or so.  For example for recoil polarizations and
$\gamma$--ray correlations the largest contribution to model
uncertainties arises via corrections at the 30--40\% level from the
$M_1 \boldsymbol{\sigma} \cdot \boldsymbol{\nabla}$ operator.
Although the matrix elements for this gradient operator are rather
difficult to compute accurately, we note even a 50\% uncertainty in
$M_1 \boldsymbol{\sigma} \cdot \boldsymbol{\nabla}$ will produce only
a 10-20\% uncertainty in extracting $g_p / g_a$.  A similar situation
arises in extracting the coupling $g_p / g_a$ from hyperfine
dependences, although here the additional contributions from $\Delta J
= 2^+$ multipoles may increase somewhat the model uncertainty.
However the model uncertainty is probably larger for $^{16}$O($0^+ , 0
$) $\rightarrow$ $^{16}$N($0^- , 120)$, since absolute values and not
ratios of matrix elements are needed, and those required are quite
difficult to accurately compute.  See Sec. \ref{s: 16O theory} for
details.

The value of $g_p / g_a = 1.0 \pm^{1.1}_{1.2}$ from $^{28}$Si is
rather puzzling.  Interestingly the experimental results for $^{12}$C
of $X = 0.26 \pm 0.06$ and $^{28}$Si of $X = 0.55 \pm 0.04$ are quite
different, while the theoretical predictions for $^{12}$C and
$^{28}$Si are not, {\it i.e.} the calculated corrections arising from
axial charge and 2$^{nd}$--forbidden operators are similar in $^{12}$C
and $^{28}$Si.  Inspection of the one--body transition densities for
$^{12}$C and $^{28}$Si does indicate a difference in the two cases.
Whereas for $^{12}$C the transition is mainly $0p_{3/2} \rightarrow
0p_{1/2}$ with some $0p_{1/2} \rightarrow 0p_{3/2}$, for $^{28}$Si the
transition involves numerous single particle transitions of similar
magnitudes.  Therefore sensitivity to the nuclear model for the
$^{28}$Si$(0^+ , 0 ) \rightarrow ^{28}$Al$(1^+, 2201 )$ transition is
likely larger.  In addition we observe that the agreement of theory
and experiment is good for the $^{12}$C$( 0^+ , 0 ) \rightarrow
^{12}$B$( 1^+ , 0 )$ capture rate but poor for the $^{28}$Si$(0^+ , 0
) \rightarrow ^{28}$Al$(1^+, 2201 )$ capture rate, indicating the GT
matrix element is reproduced well for $^{12}$C but reproduced poorly
for $^{28}$Si.  For further details see \citet{Gor99}.

Finally what new experimental and theoretical work is worthwhile?  For
partial transitions on complex nuclei a limiting factor is the small
number of the available transitions, {\it i.e.}  four allowed
transitions and one first--forbidden transition.  Given the
unavoidable sensitivity to nuclear structure a larger data--set of
partial transitions would be helpful.  With more data an improved
understanding of contributions from $M_{1} \boldsymbol{\sigma} \cdot
\boldsymbol{\nabla}$ and $\boldsymbol{M}_{12} \cdot
\boldsymbol{\sigma}$ is presumably possible.  Unfortunately of course
the experimental techniques for measuring recoil polarizations,
gamma--ray correlations and hyperfine dependences are often limited to
special cases.  However new experimental studies of muon capture on
$^{32}$S and $^{35}$Cl are under way. Additionally a measurement 
of the hyperfine dependence
in the $^6$Li$( 1^+ , 0 )$ $\rightarrow$ $^6$He$( 0^+ , 0 )$ reaction,
where accurate nuclear wavefunctions and related nuclear data
are available, is definitely interesting.

\section{Inclusive RMC on complex nuclei}
\label{s: nuclRMC}

Several factors have motivated investigations of inclusive RMC on
complex nuclei.  First the radiative rate on complex nuclei is highly
sensitive to $g_p$.  Second the branching ratios for nuclear RMC are
comparatively large, {\it e.g.} the $^{12}$C rate is about 100 times
the $^{1}$H rate and the $^{40}$Ca rate is about 1000 times the
$^{1}$H rate.  Third the early theoretical studies implied the ratio
of radiative capture to ordinary capture was only mildly model
dependent.

We note comprehensive reviews that cover nuclear RMC were published by
\citet{Mu77} and \citet{Gm87}.  These authors have discussed in detail
the formalism and methods for RMC calculations on complex nuclei.
Herein we simply outline the major approaches, referring the reader to
these reviews for more detail, and focus mainly on recent
developments.  We update the status of model calculations in
Sec. \ref{s: nuclearRMCtheory} and experimental data in Sec. \ref{s:
nuclearRMCexpt}.  We discuss the interpretation of the RMC branching
ratio data in Sec. \ref{s: nuclearRMCinter} and describe the situation
for the RMC photon asymmetry data in Sec. \ref{s: nuclearRMCcorr}.

\subsection{Theory of nuclear RMC}
\label{s: nuclearRMCtheory}

The theory of RMC in nuclei has a long history, dating back to
\citet{Ro65}. Although they were not the first to consider the
problem, they carefully laid out the approach which has become the
standard in subsequent calculations.  The standard approach to RMC in
nuclei is to develop a non-relativistic Hamiltonian for the process
which is then evaluated in impulse approximation (IA) between nuclear
states. There are thus two main ingredients, the Hamiltonian and
nuclear structure.

In the usual approach the Hamiltonian is derived from the same five
diagrams, Fig. \ref{f: RMCdiag}(a)-(e), used to describe capture on
the nucleon. This Hamiltonian, which originates as a relativistic
amplitude, is then expanded in powers of the nucleon momentum either
directly, or using a Foldy-Wouthuysen procedure. The leading order is
independent of the nucleon momentum and the linear order correction is
typically 10-20\% and is neglected in many calculations. The
correction to the non-relativistic Hamiltonian from the terms
quadratic in the nucleon momentum was considered by \citet{Sl80} and
found to be very small.

This single particle Hamiltonian is then summed over all nucleons and
matrix elements of the result are taken between nuclear states. In
early calculations very crude nuclear states were used, {\it e.g.} a
simple Slater determinant of harmonic oscillator wave functions. In
recent calculations however more modern shell model wave functions
have been used which are derived using realistic interactions.

The muon deposits a lot of energy in the nucleus and thus there can be
many levels excited in the final nucleus. However all measurements so
far are of the inclusive rate, and so a technique must be developed to
sum over all final states. In early calculations closure was
used. However this introduces a new parameter $k_{max}$ corresponding
to the average maximum photon energy, or equivalently the average
nuclear excitation. Unfortunately the rate is just as sensitive to
$k_{max}$ as to $g_p$. Furthermore, as pointed out by \citet{Ch81},
the closure sum includes many excited states which are not allowed in
the radiative process due to energy conservation.

There have been several approaches to attempt to bypass this
problem. One approach involves obtaining the spectrum of excited
states from some other source. \citet{Fo64} did this for OMC, by
recognizing that much of the strength went to the giant dipole
resonance state and that one could get this strength from empirical
photoabsorption cross sections.\footnote{Later calculations extended
this to consider SU$_4$ breaking and the excitation of the spin dipole
state by the axial current. For example see  \citet{Ca70}.}  This idea
was generalized to RMC by
\citet{Fe66} and further generalized by \citet{Ch81} who added
additional phenomenological components to account for transitions to
quadrupole states. This approach lessened the dependence on $k_{max}$
but did not totally eliminate it, as there were still matrix elements,
e. g. the leading $p/m$ corrections which could not be obtained this
way.

Another approach involved sum rules, in which various matrix elements
were evaluated using energy weighted sum rules. This approach was
considered for example by \citet{Sl78} and more recently applied to
OMC by \citet{Na86}. It was developed further and used more recently
for RMC by \citet{Ro90}.  While still somewhat phenomenological, and
still depending to some extent on the closure approximation, this
approach is much less sensitive to average excitation energies than
the closure approximation.

In principle the best method is to calculate explicitly transitions to
all possible excited states and sum the results. This takes the energy
dependence into account properly, which can be important
\citep{Fe92}. One must always truncate the sum somewhere however,
which in principle introduces errors. Modern shell model codes are
good enough however to produce enough of the excited states so that
for at least some nuclei the important transitions can be calculated
and summed. Most recent calculations \citep{Gm91, Gm86, Gm90, Er98}
have used this approach.

One can also attempt to modify or improve the basic
Hamiltonian. Recall that the standard Hamiltonian \citep{Ro65} is a
single particle operator coming from the same basic five diagrams,
Fig. \ref{f: RMCdiag}(a)-(e), used for RMC on the proton. In
principle one should include various meson exchange corrections, which
lead to two body operators, in the same fashion as has been done for
OMC in light nuclei (See Sec. \ref{s: exchange currents}). To our
knowledge this has not been done, at least recently, for heavy nuclei.

An alternative approach is to look at these meson exchange corrections
as effects of the intermediate pion rescattering in the nuclear
medium. This leads to a renormalization of the effective couplings in
the nuclear medium. This approach has a long an involved history, and
applies to a number of processes involving axial current matrix
elements. It is a bit outside the scope of this review however. The
reader interested in pursuing this further can consider the recent
papers of \citet{Ki94} or \citet{Kol00} or older studies such as that
of \citet{Ak85}.

Still another attempt to modify the basic Hamiltonian has been
proposed by \citet{Gm86}. The idea here is to use current conservation
to evaluate parts of the matrix element. This is analogous to the
Siegert theorem approach which has been used for other processes, and
originates in the observation that the nuclear matrix element of the
impulse approximation Hamiltonian, unlike the nucleon matrix element,
does not correspond to a conserved current. An attempt is made to fix
this by expanding the photon field and using the continuity equation
to eliminate parts of the three-vector current in favor of the charge
distribution.  This leads to some different terms and to what the
authors call a modified impulse approximation (MIA).

Calculations using this MIA suggest that it is extremely important
\citep{Gm86,Gm90,Gm91,Er98}. It reduces the RMC rate by a factor of
two or more and thus increases the value of $g_p$ needed to fit a
given experimental result rather dramatically. The approach also seems
to produce rates which are much less sensitive to $g_p$ than the
IA. It does however suppress usual impulse approximation results so
that they are in better agreement with phenomenological
approaches. However there are some caveats. The very fact that what is
effectively enforcing gauge invariance makes such a huge difference is
worrisome. The results also depend, though not strongly, on an
arbitrary choice among various ways to make the original expansion of
the photon field. It also appears that one is using the continuity
equation on only part of the current, still leaving some reference to
the three vector current in the problem. Clearly this needs to be
looked at more carefully to determine if it is indeed correct since it
makes such a large difference in the final results.

Finally we should mention one other modern calculation \citep{Fe92}
which uses a relativistic Fermi gas model together with relativistic
mean field theory to examine the A and Z dependence of the RMC and OMC
rates. The Fermi gas model was used in very early calculations, but is
clearly too crude a model to give good results for specific details
for individual nuclei. It does however allow one to elucidate general
trends and select out features which are important. This will be
discussed further in the next section.

\subsection{Measurement of nuclear RMC}
\label{s: nuclearRMCexpt}

The first observation of radiative capture was made at CERN by
\citet{Co62}.  Muons were stopped in Fe and photons were detected by
gamma--ray conversion in a iron sheet/spark chamber sandwich and
energy deposition in a large volume NaI crystal.  After subtraction of
backgrounds a total of five photons from RMC on Fe were identified and
yielded a Fe RMC branching ratio of roughly $10^{-4}$.  During the
following ten years a few more studies of nuclear RMC were conducted
by \citet{Con64}, \citet{Ch65} and \citet{Ro73}.  However the
measurements remained extremely difficult because of the low signal
rate and the high background rates.  In particular under--counted
neutron backgrounds most likely corrupted these early experiments.

More recently the availability of higher quality muon beams and higher
performance pair spectrometers has dramatically improved the
experimental situation.  The new era for nuclear RMC was pioneered by
\citet{Ha77b} for RMC on Ca.  Later experimental programs at PSI in
the late 1980's and TRIUMF in the early 1990's have produced an
extensive body of nuclear RMC data.  Today's beams and detection
systems allow data collection on $Z > 12$ targets at count rates of
$\sim 1000$ RMC photons/day with nearly background--free conditions.

In principle the method for determining the branching ratio for
nuclear RMC is straightforward.  It simply requires counting the
incoming muons and outgoing photons and determining their detection
efficiencies.  However as discussed in detail in Sec. \ref{s: radcapt}
the small branching ratio means troublesome $\gamma$--ray backgrounds
arise from $\mu$ decay in the target material, radiative $\mu$ capture
in the neighboring materials, and the pion contamination in the muon
beam.  Additionally neutron backgrounds from ordinary capture in the
target material and other sources on the accelerator site are
dangerous if the n/$\gamma$ discrimination is not highly effective.

Two basic types of pair spectrometers have been widely employed in the
RMC studies of recent years.  In one approach, as used by
\citet{Fri88}, the photons are converted in a relatively thin passive
converter and the e$^+$e$^-$ pair is tracked in a multiwire chamber
arrangement.  In another approach, as used by \citet{Do88}, the
photons are converted in a relatively thick active converter and the
e$^+$e$^-$ pair is measured by combinations of Cerenkov detectors and
NaI crystals.  The former approach offers good resolution, typically
1--2\%, but at the expense of a low efficiency, typically $10^{-5}$.
The latter approach offers high efficiency, typically 0.5\%, but at
the expense of a poor resolution, typically 20\%.  Such experimental
set--ups have yielded excellent n/$\gamma$ discrimination and provided
nearly background--free RMC spectra.

Most recently a novel large solid--angle pair spectrometer was
developed at TRIUMF by \citet{Wr92} for RMC on $^{1}$H.  The
spectrometer offers both a relatively high detection efficiency and a
relatively good energy resolution.  It has permitted quick and
straightforward measurements of nuclear RMC for numerous targets, and
significantly extended the RMC data--set
\citep{Ar92,Go98,Be99}. 

\subsection{Interpretation of nuclear RMC}
\label{s: nuclearRMCinter}

\begin{figure}
\begin{center}
\epsfig{file=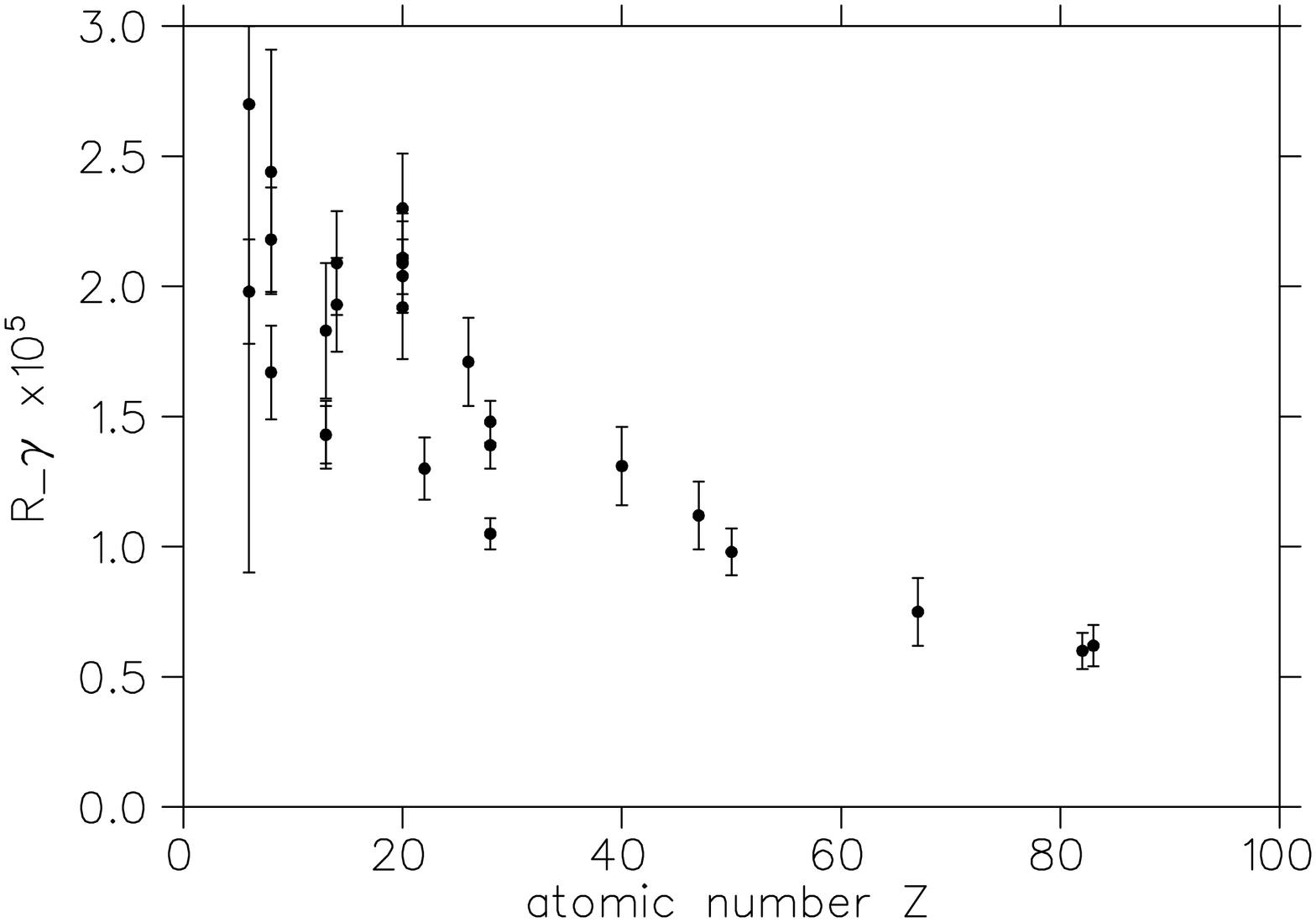,bbllx=0pt,bblly=0pt,
bburx=800pt,bbury=800pt,width=7.5cm,angle=90,clip=}
\end{center}
\vspace{0.0cm}
\caption{The world data for $R_{\gamma}$ versus $Z$ on $A > 3$ nuclei.}
\label{f: R v Z}
\end{figure}

\begin{figure}
\begin{center}
\epsfig{file=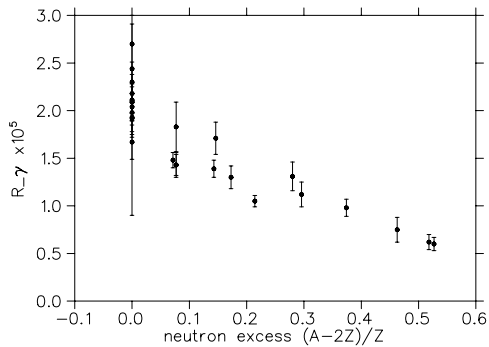,bbllx=0pt,bblly=0pt,
bburx=800pt,bbury=800pt,width=7.5cm,angle=90,clip=}
\end{center}
\vspace{0.0cm}
\caption{The world data for $R_{\gamma}$ versus 
$\alpha = ( A - 2 Z ) / Z$ on $A > 3$ nuclei.}
\label{f: R v alpha}
\end{figure}

The world data--set\footnote{We omit the results from the early
experiments of \protect\citet{Con64},
\protect\citet{Ch65}, and \protect\citet{Ro73} due to
large neutron backgrounds.} for nuclear RMC, consisting of targets
from carbon to bismuth, is summarized in Table \ref{t: rmc results} in
which we tabulate the ratio $R_{\gamma}$ of the $E_{\gamma} > 57$ MeV
partial radiative rate to the total ordinary rate.  In addition we
plot the data versus atomic number Z in Fig. \ref{f: R v Z} and versus
neutron excess $\alpha = ( A - 2Z ) / Z$ in Fig. \ref{f: R v alpha}.
The figures illustrate some intriguing trends, {\it i.e.}  that
$R_{\gamma}$ is observed to decrease from $\sim 2 $ to $\sim 0.6$ with
increasing $Z$ and increasing $\alpha$.\footnote{Note that in this
section we will quote all values of $R_{\gamma}$ in units of
$10^{-5}$.}  Note the overall trend is somewhat smoother with neutron
excess than atomic number. Specifically, the isotope effect in the
mass $^{58, 60, 62}$Ni isotopes and the odd/even--A effect in the
Al--Si, Ca--Ti pairs fit the $\alpha$--dependence but not the
Z--dependence.  Below we discuss the $R_{\gamma}$--data in the context
of the determination of the coupling $g_p$.

For $^{40}$Ca the value of $R_{\gamma}$ is well--established by
numerous experiments. The average of those included in Table \ref{t:
rmc results} is $R_{\gamma} = 2.08 \pm 0.11$.  A number of
calculations are available including the phenomenological models of
\citet{Fe66} and \citet{Ch81}, the microscopic shell model
calculations of \citet{Gm86} and the sum rule calculations of
\citet{Ro90}.  Unfortunately as discussed earlier the calculation 
of inclusive RMC on complex nuclei is notoriously difficult.
It is often
difficult to quantify the effect of the various approximations and
assumptions and so to set limits on the uncertainty in the theoretical
predictions.

\begingroup
\begin{table}[tpb]
\caption{Summary of the world data on the quantity $R_{\gamma}$
on $A > 3$ targets, where $R_{\gamma}$ is the ratio of the RMC rate
with $E_{\gamma} > 57$~MeV to the OMC rate in units of $10^{-5}$.  In
order to assist comparisons in most cases the quoted results are for
the closure approximation spectra shape where the corresponding value
of the parameter $k_{max}$ is given in column four.  The quantity
$\alpha$ is the neutron excess $( A - 2Z ) / Z$.}
\label{t: rmc results}
\begin{ruledtabular}
\begin{tabular}{llll} 
  & & & \\
 Target & $\alpha$ & R$_{\gamma}$ ( $\times 10^{-5}$ ) & 
$k_{max}$ ( MeV ) \\
  & & & \\
 \hline
  & & & \\
\citet{Be99} \\
   $^{16}_{8}$O   & 0.000 &  1.67 $\pm$   0.18 &  88.4 $\pm$    2.3 \\ 
   $^{27}_{13}$Al & 0.077 &  1.43 $\pm$   0.11 &  90.1 $\pm$    1.8 \\ 
   $^{28}_{14}$Si & 0.000 &  2.09 $\pm$   0.20 &  89.4 $\pm$    1.8 \\ 
   $^{nat}_{22}$Ti & 0.173 &  1.30 $\pm$   0.12 &  89.2 $\pm$    2.0 \\ 
   $^{nat}_{40}$Zr & 0.280 &  1.31 $\pm$   0.15 &  89.2 $\pm$    3.4 \\ 
   $^{nat}_{47}$Ag & 0.296 &  1.12 $\pm$   0.13 &  89.0 $\pm$    3.2 \\
\citet{Go98} \\
   $^{58}_{28}$Ni & 0.071 &  1.48 $\pm$   0.08 &  92.0 $\pm$    2.0 \\ 
   $^{60}_{28}$Ni & 0.143 &  1.39 $\pm$   0.09 &  90.0 $\pm$    2.0 \\ 
   $^{62}_{28}$Ni & 0.214 &  1.05 $\pm$   0.06 &  89.0 $\pm$    2.0 \\ 
\citet{Ar92} \\
   $^{27}_{13}$Al & 0.077 &  1.43 $\pm$   0.13 &  90.0 $\pm$    2.0 \\ 
   $^{28}_{14}$Si & 0.000 &  1.93 $\pm$   0.18 &  92.0 $\pm$    2.0 \\ 
   $^{40}_{20}$Ca & 0.000 &  2.09 $\pm$   0.19 &  93.0 $\pm$    2.0 \\ 
   $^{nat}_{42}$Mo  & 0.283  &  1.11 $\pm$   0.11 &  90.0 $\pm$  2.0 \\ 
   $^{nat}_{50}$Sn  & 0.374 &  0.98 $\pm$   0.09 &  87.0 $\pm$   2.0 \\ 
  $^{nat}_{82}$Pb & 0.527 &  0.60 $\pm$   0.07 &  84.0 $\pm$    3.0 \\ 
\citet{Ar91} \\
   $^{12}_{6}$C & 0.000 &  1.98 $\pm$   0.20 &                    \\ 
   $^{16}_{8}$O & 0.000 &  2.18 $\pm$   0.20 &                    \\ 
   $^{40}_{20}$Ca & 0.000 &  2.04 $\pm$   0.14 &                    \\ 
\citet{Fri88} \\
    $^{16}_{8}$O & 0.000 &  3.80 $\pm$   0.40 &                    \\ 
\citet{Do86} \\
    $^{12}_{6}$C & 0.000 &  2.70 $\pm$   1.80 &                    \\ 
    $^{16}_{8}$O & 0.000 &  2.44 $\pm$   0.47 &  89.9 $\pm$    5.0 \\ 
    $^{27}_{13}$Al & 0.077 &  1.83 $\pm$   0.26 &  88.8 $\pm$    1.8 \\ 
   $^{40}_{20}$Ca & 0.000 &  2.30 $\pm$   0.21 &  92.5 $\pm$    0.7 \\ 
   $^{nat}_{26}$Fe & 0.146 &  1.71 $\pm$   0.17 &  90.2 $\pm$    1.1 \\ 
   $^{165}_{67}$Ho & 0.463 &  0.75 $\pm$   0.13 &  84.1 $\pm$    5.1 \\ 
   $^{209}_{83}$Bi & 0.518 &  0.62 $\pm$   0.08 &  88.2 $\pm$    0.6 \\ 
\citet{Fr85} \\
   $^{40}_{20}$Ca & 0.000 &  1.92 $\pm$   0.20 &  90.8 $\pm$    0.9 \\ 
\citet{Ha77b} \\
   $^{40}_{20}$Ca & 0.000 &  2.11 $\pm$   0.14 &  86.5 $\pm$    1.9 \\ 
  & & & \\
\end{tabular}
\end{ruledtabular}
\end{table}
\endgroup

The phenomenological calculations of \citet{Ch81}, based on the Giant
Dipole Resonance model \citep{Fo64,Fe66} with parameters determined
from electromagnetic data but including effects of a quadrupole
resonance with parameters fit to the OMC rate, gave $R_{\gamma}=2.4$
using the canonical value $g_p / g_a \simeq 6.7$.  The calculation of
\citet{Ro90} used fairly realistic RPA wave functions but with the
phenomenological aspects of a sum rule calculation to obtain
$R_{\gamma}=1.87$.  The microscopic shell model calculation of
\citet{Gm86} used simple 1p-1h wave functions, with some corrections
to obtain $R_{\gamma}=4.25$ in the standard IA and $R_{\gamma}=2.28$
in the MIA. Thus the MIA makes a large difference, and brings
the results in closer agreement with the more phenomenological
approaches. Note that all of these calculations, except the standard
IA approach, apparently neglected the velocity or $p/m$ terms which
could be up to 10-20\%.

On a superficial level all of these results, except perhaps for the
standard IA result, are reasonably consistent with the experimental
result. However if one turns the question around and asks what limits
are set on $g_p / g_a$ things become much less clear. A theoretical
value of $R_{\gamma}$ less than the experiment implies a larger value
of $g_p / g_a$ is needed to fit the data. Thus the sum rule
calculation of \citet{Ro90} requires $g_p / g_a \sim 8$. On the other
hand the phenomenological calculations of \citet{Ch81} and the MIA
results of \citet{Gm86} require $g_p / g_a \sim 4-5$ and the IA result
implies that $g_p / g_a$ is much smaller.  Thus one has to conclude
that it is unreasonable to claim that $g_p / g_a \sim 6.7$ is
well--established by RMC on $^{40}$Ca.

For $^{16}$O a number of measurements and calculations are also
available.  Unfortunately measurement of RMC on $^{16}$O is more
difficult, since the $\gamma$--ray yield per $\mu^-$ stop is $\sim 2
\times 10^{-5}$ for $^{40}$Ca and $\sim 0.4 \times 10^{-5}$ for
$^{16}$O, and most probably the earlier experiments have
under--subtracted contamination from $\gamma$--ray backgrounds.
Therefore we employ the recent result from \citet{Be99} of $R_{\gamma}
= 1.67 \pm 0.18$.  Like $^{40}$Ca, both phenomenological calculations
and microscopic calculations are available, but similarly the
sensitivities of results to approximations are difficult to estimate.
Taking $g_p / g_a \simeq 6.7$ the phenomenological calculation of
\citet{Ch85} gave $R_{\gamma} = 2.08$, the shell model calculation of
\citet{Gm86} gave $R_{\gamma} = 1.61$ in MIA and $R_{\gamma} = 3.10$
in IA and sum rule calculation of \citet{Ro90} gave $R_{\gamma} =
1.73$.  Again, with the exception of the IA result, these are
reasonably consistent with the experimental result \citep{Be99}
$R_{\gamma} = 1.67 \pm 0.18$ with the sum rule and MIA approaches
implying roughly the canonical value of $g_p / g_a$ and the approach
of \citet{Ch85} requiring a somewhat smaller value. The IA result
would require a significantly smaller value to fit the data.  However,
like RMC on $^{40}$Ca, due to the approximations in the calculations
and the differences among the results of different calculations the
case for $g_p / g_a \sim 6.7$ in RMC on $^{16}$O is not firmly
established.

There are also both theoretical and experimental results for
$^{12}$C. The most recent result, and the one with by far the smallest
uncertainty is that of \citet{Ar91}, $R_{\gamma} = 1.98 \pm
0.20$. This is to be compared with the sum rule result
\citep{Ro90} of $R_{\gamma} = 1.42$, and the IA result, $R_{\gamma} =
3.60$, and the MIA result, $R_{\gamma} = 1.48$ \citep{Gm90}. In this
case both MIA and sum rule results are too low, implying $g_p / g_a
\sim 10-13$, to fit the data whereas the IA is too high, implying a
very low value of $g_p / g_a$.

Finally we should mention the calculations for $^{58,60,62}$Ni of
\citet{Er98} carried out in a microscopic model using the
quasiparticle RPA. Again the IA results for $R_{\gamma}$ are much
higher than those in MIA, but even the MIA results are significantly
higher than the experiment \citep{Go98}.

Thus by now there have been a number of experiments and enough
calculations that we can make comparisons of $R_{\gamma}$ for several
different nuclei. The situation can only be described as confused. The
standard impulse approximation calculations are consistently too high,
implying a value of $g_p / g_a $ much smaller than the canonical
value. The MIA start out too low for $^{12}C$, implying a large value
of $g_p / g_a $, and rise with increasing A to become significantly
too high for the Ni isotopes, implying there a small value of $g_p /
g_a $.  The phenomenological calculations of \citet{Ch81} and
\citet{Ch85} for $^{16}$O and $^{40}$Ca are both too high and the sum
rule calculations show no consistent pattern. It appears clear that we
are not yet at a stage where theoretical uncertainties are
sufficiently under control to consistently reproduce experimental
results.

In addition to discussing the detailed calculations for specific
nuclei it is worthwhile to consider the systematics of RMC data versus
atomic number and neutron excess.  Such systematics were examined in
the non--relativistic Fermi--gas calculation of \citet{Ch80}  and the
relativistic Fermi--gas calculation of \citet{Fe92}.  These models
clearly oversimplify the nuclear structure, for example omitting the
important effects of giant resonances in muon capture, or the effects
of shell closures. However they demonstrate a number of interesting
dependences of muon capture on Z and $\alpha$.

For concreteness we consider the calculation of \citet{Fe92}.  These
authors have carefully studied the dependence of inclusive OMC and
inclusive RMC on the input parameters and the model assumptions.  They
stress that the OMC rate, the RMC rate, and their ratio, are highly
sensitive to phase--space effects, {\it i.e.} things that alter the
available energy for the neutrino and the photon.  For example
including the $\mu^-$ atomic binding energy, which increases from
$\sim 0.1$~MeV in light nuclei to $\sim 10$~MeV in heavy nuclei,
decreases the OMC rate by a factor of two and the RMC rate by a factor
of eight for the heaviest nuclei.  Other parameters, {\it e.g.}  for
Fermi--gas models the Coulomb energy, symmetry energy, etc., also have
large effects on the available energy and therefore the rates.
Consequently the authors caution that reliably extracting the coupling
$g_p$ from inclusive RMC on complex nuclei is difficult.

Despite such concerns the calculation of \citet{Fe92} does reproduce
the overall dependence of $R_{\gamma}$ data with Z and $\alpha$.  This
model--data agreement however suggests no reason to invoke a
large--scale medium--modification of $g_p / g_a$ to explain the
Z--dependence of $R_{\gamma}$ data.

Earlier suggestions for an A--dependence of the coupling $g_p$, as
discussed by \citet{Gm87} and \citet{Do88}, were largely grounded in
large values of $g_p / g_a$ obtained from shell model calculations on
light nuclei and small values of $g_p / g_a$ obtained from Fermi--gas
calculations on heavy nuclei.  Given the different systematics of the
various models and the difficulty of all models in consistently
fitting the data, such differences do not provide real evidence for an
A--dependence of the coupling $g_p$.

In summary, although the RMC rate on complex nuclei is undoubtedly
$g_p$--dependent, the problem of separating the coupling constant and
nuclear structure is extraordinarily tricky.  While the overall
features of nuclear RMC are generally accommodated by $g_p / g_a =
6.7$ there is a model uncertainty of at least 50\%. Thus RMC in heavy
nuclei is not yet competitive with RMC or OMC in hydrogen or light
nuclei as a source of information on $g_p / g_a $.

For inclusive RMC on complex nuclei we find the progress on model
calculations is lagging the progress on experimental data, and a
breakthrough on the theoretical side is needed before this particular
tool for studying the coupling is competitive.  However, exclusive RMC
on complex nuclei, which is so far unmeasured, might offer
interesting possibilities.

\subsection{Comments on photon asymmetries in nuclear RMC}
\label{s: nuclearRMCcorr}

In RMC the gamma--ray direction and muon polarization are correlated
according to
\begin{equation}
N( \theta ) = N ( 1 + \alpha P_{\mu} \cos{ \theta } )
\end{equation}
where $P_{\mu}$ is the muon polarization, $\alpha$ is the asymmetry
coefficient, and $\theta$ is the angle between the $\mu$--spin axis
and the $\gamma$--ray momentum axis.  The left handedness of the
neutrino leads to $\alpha \simeq +1$, and $g_p$ is manifest as
departures of $\alpha$ from unity
\citep{Fe75}. The specific dependence of $\alpha$ on $g_p / g_a$ for 
RMC on $^{40}$Ca was computed by \citet{Ro65}, \citet{GmK81},
\citet{Ch81} and \citet{Gm87b}.

To measure the asymmetry one stops the incoming muons in a suitable
target, precesses the muon spin in a magnetic field, and measures the
time spectrum of the outgoing photons.  The resulting time spectrum
consists of an exponential decay with a sinusoidal modulation, where
the amplitude of the sine--wave is governed by the product $P_{\mu}
\alpha$.  Note that the time spectrum of the Michel electrons is
employed to measure $P_{\mu}$ and isolate $\alpha$.  In the first
measurement by \citet{Di71} the authors used a single NaI crystal for
$\gamma$--ray detection.  In the later measurements by \citet{Ha77b},
\citet{Do86} and \citet{Vi90} the authors used a separate $\gamma$-ray
converter and NaI calorimeter.  A $^{40}$Ca target was used in each
experiment.

The experimental difficulties originate from the tiny RMC branching
ratio and the small $\mu$ residual polarization.  Consequently the
backgrounds are severe.  They include $\gamma$--rays from pion capture
in the target, $\gamma$--rays from Michel bremsstrahlung in the
target, neutrons from muon capture in the target, and n/$\gamma$
backgrounds from cosmic--ray and accelerator sources.  Note that a
prompt cut reduces the $\gamma$--rays following $\pi$--capture and a
$E > 57$~MeV cut reduces the bremsstrahlung following $\mu$--decay.
The neutron background is reduced via the converter--calorimeter
set--up and cosmic--ray background is reduced via combined passive and
active shielding.

In Table \ref{t: asym results} we summarize the results of
\citet{Ha77b}, \citet{Do86} and \citet{Vi90}.
We omit the earliest results of \citet{Di71}, which suffered severe
neutron backgrounds.  The various experiments are mutually consistent
and yield a world average value $\alpha = 1.02 \pm 0.25$.

\begin{table}[tpb]
\caption{Summary of the world data on the photon asymmetry
in radiative capture on calcium.}
\label{t: asym results}
\begin{ruledtabular}
\begin{tabular}{ll} 
  & \\
 Ref.  & asymmetry $\alpha$ \\
  & \\
 \hline
  & \\
\citet{Ha77b}\   & $0.90 \pm 0.50$ \\
  & \\
\citet{Do86}\ & $0.90 \pm 0.43$ \\
  & \\
\citet{Vi90}\ & $1.32^{+0.54}_{-0.47}$ \\
  & \\
world average        & $1.02 \pm 0.25$ \\
  & \\
\end{tabular}
\end{ruledtabular}
\end{table}

Unfortunately the effect of $g_p$ on $\alpha$ is (i) not large and
(ii) model dependent.  For example with $g_p / g_a \simeq 6.7$ the
phenomenological model of \citet{Ch81} gives $\alpha \simeq 0.80$ and
the shell model of \citet{Gm87b} gives $\alpha \simeq 0.90$.
Conservatively the measurements yield $g_p / g_a < 20$.

\section{Summary}
\label{s: summary}

The determination of $g_p$ is important for several reasons.  First,
whereas the values of the proton's other weak couplings are nowadays
well determined, the induced pseudoscalar coupling is still poorly
determined.  Second, based on chiral symmetry arguments a solid
theoretical prediction for $g_p$ with 2--3\% accuracy is available.
Third, the prediction is founded on some elementary properties of the
strong interaction, and determining the coupling is therefore an
important test of quantum chromodynamics at low energies.

Since the classic review of \citet{Mu77} an extensive body of
experimental data on the coupling $g_p$ has been accumulated.  This
work spans the elementary processes of ordinary muon capture and
radiative muon capture on hydrogen, muon capture on few--body systems,
and exclusive OMC and inclusive RMC on complex nuclei.  The
experimental approaches have ranged from ultra--high precision
measurements to ultra--rare process measurements, and include some
novel studies of spin phenomena in complex nuclei.  The disentangling
of the coupling constant from the physical observables has involved
diverse fields from $\mu$--molecular chemistry to traditional nuclear
structure and exchange current effects.

One would expect muon capture on hydrogen to be the most
straightforward and most easily interpreted of the muon capture
reactions.  This is the situation from the nuclear perspective, but
unfortunately there exist atomic and molecular complications which
also must be thoroughly understood. For OMC there are a several older
experiments together with the most recent and precise measurement of
\citet{Ba81a} whereas for RMC there is only the TRIUMF experiment
\citep{Wr98}. To extract $g_p$ from these results we updated the
theoretical calculations to include the current best values of the
other weak couplings and their $q^2$ dependences.  Additionally we
updated the capture rate of \citet{Ba81a} for the present
world-average value of the positive muon lifetime.  We found, for the
standard values of the $\mu$ chemistry parameters and specifically
with $\Lambda_{op} = 4.1 \times 10^4$~s$^{-1}$, values of $g_p = 12.2
\pm 1.1$ from the TRIUMF RMC experiment \citep{Wr98}, $g_p = 10.6 \pm
2.7$ from the Saclay OMC experiment \citep{Ba81a}, and $g_p = 10.5 \pm
1.8$ for the world average of all OMC experiments.  These updates
increased the $g_p$ extracted from OMC, as compared to the original
analysis, so that now both OMC and RMC give results larger than
predicted by PCAC and in fact agree better with each other than with
PCAC. When uncertainties are taken into account however, only the
TRIUMF result is clearly inconsistent with PCAC, while the OMC results
are only marginally inconsistent with PCAC.

Our strong prejudice that PCAC is not violated at such a level, makes
the situation for $\mu$ capture in hydrogen very puzzling.  We
therefore have examined some suggestions for solving the puzzle,
such as modifications to the $\mu$ chemistry and the role of
the $\Delta$ resonance.  By `tuning parameters' the discrepancy
between the hydrogen data and the PCAC prediction can be reduced
somewhat, but no clear-cut solution, which makes OMC and RMC
simultaneously agree with PCAC, has emerged.

New work on $\mu$ capture in liquid H$_2$ and gaseous H$_2$ is  in
progress.  At TRIUMF an investigation of the $\mu$ chemistry in liquid
hydrogen has taken data, and at PSI a measurement of $\mu^-$
lifetime in gaseous hydrogen is now under way.  Hopefully these
experiments will help to clarify the situation on $g_p$.  Looking
further ahead, perhaps both new facilities, {\it e.g.} an intense muon
source at a neutrino factory, and new techniques, {\it e.g.} neutron
polarizations or hyperfine effects, will permit a determination of
$g_p$ to an accuracy of 2-3\% or better.

The situation for muon capture on pure deuterium is inconclusive.
Recall that in deuterium the $g_p$ sensitivity is smaller than in
hydrogen, the neutron experiments are harder, and nuclear models are
needed.  At present the world data for muon capture in
pure deuterium comprises the experiment of \citet{Ca89}, which appears
consistent with PCAC, and the experiment of \citet{Ba86}, which
may be inconsistent with PCAC. Also the results on
$\mu$d capture by \citet{Be73} using a H$_2$/D$_2$ target
are even more puzzling.

Further studies of $\mu$d capture are clearly worthwhile.  Obviously
resolving the possible discrepancy between the doublet rates obtained by
\citet{Ca89} and \citet{Ba86} is important.  Additionally we note the
chemistry of muons in pure H$_2$ and pure D$_2$ is quite different in
several ways, including a slower rate for hyperfine depopulation and
near absence of muonic molecules in pure D$_2$.  Such features may
permit the study of $g_p$ via alternative approaches, such the
hyperfine dependence of the capture reaction.
 
In $\mu$+$^{3}$He $\rightarrow$ $^{3}$H+$\nu$ capture, the recent
precision measurement of the statistical capture rate by \citet{Ac98}
and ground--breaking measurement of the recoil angular correlation by
\citet{So98}, were major achievements.  Preliminary data on the
radiative capture rate in the $^{3}$He $\rightarrow$ $^{3}$H channel
are also available \citep{Wr00}.  Further modern treatments of $A = 3$
wave functions and 2--body exchange currents have been applied to the
process by \citet{Co96} and others.  The OMC results for $g_p$ are
completely consistent with PCAC, and the value of $g_p = 8.53 \pm
1.54$ from \citet{Ac98} is probably the best individual determination
of $g_p$.

The extraction of the coupling from the measurement of the $\mu$+
$^{3}$He $\rightarrow$ $^{3}$H+$\nu$ statistical rate is unfortunately
limited by theoretical uncertainties in calculating contributions from
exchange currents.  Further theoretical work is definitely worthwhile
to quantify the contributions arising from radiative corrections and
nail-down the uncertainties arising from exchange currents, etc, but
improvements in extracting the coupling may be difficult.  A precision
measurement of the recoil correlation, which has enhanced sensitivity
to $g_p$ and reduced sensitivity to exchange currents, would be
extremely interesting.

A significant body of new data on exclusive transitions in ordinary
capture on complex nuclei has been collected recently.  This includes
the measurement of $\gamma$--ray correlations in $\mu$$^{28}$Si and
hyperfine dependences in $\mu$$^{11}$B and $\mu$$^{23}$Na.  These data
complement earlier investigations of polarizations in $\mu$$^{12}$C
and rates in $\mu$$^{16}$O.  Furthermore, modern well--tested models
of nuclear structure in $0p$, $1s$--$0d$ nuclei offer improved
multi--particle wave functions for interpreting these experiments.  In
most cases, with the exception of $^{28}$Si, the values of
$g_p$ that are extracted from the experiments with such wave functions
are consistent with PCAC.  Generally the major uncertainty in
extracting $g_p$ originates in the interplay of the contributions from
the pseudoscalar coupling, arising from the space part of the axial
current, and the axial charge, arising from the time part of the axial
current.  Unfortunately it is difficult to precisely quantify such
theoretical uncertainties.

Additional experiments on exclusive OMC could be helpful to our
understanding of the interplay of the induced pseudoscalar
contribution and the axial charge contribution.  Clearly if a wealth
of data were available for nuclear OMC a better assessment of model
uncertainties would be possible.  However, the experiments generally
involve the use of methods and observables that cannot be applied to
large numbers of exclusive transitions.  Alternatively a
measurement of observables in a transition such as $^{6}$Li$( 1^+ , 0
)$ $\rightarrow$ $^6$He$( 0^+ , 0 )$, where highly accurate
wave functions and related nuclear data are nowadays available, is
definitely interesting.

In inclusive RMC on complex nuclei the application of pair
spectrometers has yielded data with good statistics and little
background, and enabled the systematics of the RMC rate across the
periodic table to be mapped--out.  Unfortunately the situation in
regards to the model calculation of the inclusive rate is less
satisfactory.  In general the models employed for inclusive RMC have
too many assumptions and too many parameters in order to reliably
extract the coupling $g_p$.  Furthermore questions remain regarding
the effective Hamiltonian for radiative capture on complex nucleus.
Therefore we believe that earlier claims for large renormalizations of
$g_p$ in inclusive RMC on complex nuclei were premature and that, at
the level of the uncertainties, the data are consistent with PCAC.

In conclusion we hope we have convinced the reader of the importance
of the coupling $g_p$, both as a fundamental parameter in nucleon weak
interactions and as an important test of low energy quantum
chromodynamics.  In recent years an impressive body of experimental
data has been accumulated and most data are consistent, within
sometimes large experimental and theoretical uncertainties, with the
PCAC prediction for the coupling $g_p$.  Unfortunately the results
from $\mu$ capture in hydrogen, which should be the simplest and
cleanest process, are very puzzling.  We believe that the resolution
of this puzzle should be a high priority, and that, until the
situation is clarified, the accurate determination of $g_p$, and
implied testing of low energy QCD, remains an important task but an
elusive goal.

\section*{Acknowledgments}

The authors would like to thank David Armstrong, Jules Deutsch, Mike
Hasinoff, and David Measday for valuable discussions and Jean-Michel
Poutissou for encouraging us to prepare this review and for supporting
the sabbatical at TRIUMF of one of us (T.~G.). We would also like to
thank David Measday for a careful reading of the manuscript. This work
was supported in part by grants from the Natural Sciences and
Engineering Research Council of Canada and the U. S. National Science
Foundation.
 

\end{document}